\documentclass[12pt,a4paper]{article}
\usepackage{feynmp,amsmath,amssymb}
\usepackage{url}
\usepackage{epsfig}
\usepackage[dvips]{color}
\usepackage{hangcaption} 
\usepackage{tabls}
\usepackage{multicol}
\usepackage{portland}
\usepackage{graphicx}
\makeatletter
\newif\if@preliminary
\@preliminaryfalse
\def\preliminary{\@preliminarytrue}

\def\preprintno#1{\def\@preprintno{#1}}
\def\address#1{\def\@address{#1}}
\def\email#1#2{\thanks{\tt #1@{}#2}}
\def\abstract#1{\def\@abstract{#1}}
\renewcommand\abstractname{ABSTRACT}
\newlength\preprintnoskip
\newlength\abstractwidth
\@titlepagetrue
\renewcommand\maketitle{\begin{titlepage}%
  \let\footnotesize\small
  \hfill\parbox{\preprintnoskip}{%
  \begin{flushright}\@preprintno\end{flushright}}\hspace*{1cm}
  \vskip 60\p@
  \begin{center}%
    {\Large\bf\boldmath \@title \par}\vskip 1cm%
    {\sc\@author \par}\vskip 3mm%
    {\@address \par}%
    \if@preliminary
      \vskip 2cm {\large\sf FINAL DRAFT \par \@date}%
    \fi
  \end{center}\par
  \@thanks
  \vfill
  \begin{center}%
    \parbox{\abstractwidth}{\centerline{\abstractname}%
    \vskip 3mm%
    \@abstract}
  \end{center}
  \end{titlepage}%
  \setcounter{footnote}{0}%
  \let\thanks\relax\let\maketitle\relax
  \gdef\@thanks{}\gdef\@author{}\gdef\@address{}%
  \gdef\@title{}\gdef\@abstract{}\gdef\@preprintno{}
}%
\def\@citex[#1]#2{\if@filesw\immediate\write\@auxout{\string\citation{#2}}\fi
  \def\@citea{}\@cite{\@for\@citeb:=#2\do
    {\@citea\def\@citea{,\penalty\@m}\@ifundefined
       {b@\@citeb}{{\bf ?}\@warning
       {Citation `\@citeb' on page \thepage \space undefined}}%
\hbox{\csname b@\@citeb\endcsname}}}{#1}}
\def\citerange{\@ifnextchar [{\@tempswatrue\@citexr}{\@tempswafalse\@citexr[]}}
\def\@citexr[#1]#2{\if@filesw\immediate\write\@auxout{\string\citation{#2}}\fi
  \def\@citea{}\@cite{\@for\@citeb:=#2\do
    {\@citea\def\@citea{--\penalty\@m}\@ifundefined
       {b@\@citeb}{{\bf ?}\@warning
       {Citation `\@citeb' on page \thepage \space undefined}}%
\hbox{\csname b@\@citeb\endcsname}}}{#1}}
\long\def\@makecaption#1#2{%
  \vskip\abovecaptionskip
  \sbox\@tempboxa{#1: \emph{#2}}%
  \ifdim \wd\@tempboxa >\hsize
    #1: \emph{#2}\par
  \else
    \hbox to\hsize{\hfil\box\@tempboxa\hfil}%
  \fi
  \vskip\belowcaptionskip}
\def\fmslash{\@ifnextchar[{\fmsl@sh}{\fmsl@sh[0mu]}}
\def\fmsl@sh[#1]#2{%
  \mathchoice
    {\@fmsl@sh\displaystyle{#1}{#2}}%
    {\@fmsl@sh\textstyle{#1}{#2}}%
    {\@fmsl@sh\scriptstyle{#1}{#2}}%
    {\@fmsl@sh\scriptscriptstyle{#1}{#2}}}
\def\@fmsl@sh#1#2#3{\m@th\ooalign{$\hfil#1\mkern#2/\hfil$\crcr$#1#3$}}
\makeatother

\newcommand{\pd}{\partial}
\newcommand{\LL}{\mathcal{L}}

\newcommand{\tr}[1]{\mathop{\rm tr}\left\{#1\right\}}
\newcommand{\mfrac}[2]{#1 / #2}

\newcommand{\ii}{\mathrm{i}}

\newcommand{\pp}{{\prime\,2}}

\newcommand{\sw}{s_w}
\newcommand{\cw}{c_w}

\newcommand{\hc}{\text{h.c.}}

\newcommand{\eff}{{\rm eff}}
\newcommand{\Pp}{\ensuremath\mathcal{P}_+}

\newcommand{\Ppm}{\ensuremath\mathcal{P}_\pm}
\newcommand{\pS}{$S$}
\newcommand{\pT}{$T$}
\newcommand{\pU}{$U$}
\newcommand{\pSp}{$S\,$}
\newcommand{\pTp}{$T\,$}
\newcommand{\pUp}{$U\,$}

\newcommand{\vB}{\mathbf{B}}

\newcommand{\vD}{\mathbf{D}}
\newcommand{\vV}{\mathbf{V}}
\newcommand{\vT}{\mathbf{T}}
\newcommand{\vW}{\mathbf{W}}
\newcommand{\va}{\mathbf{a}}
\newcommand{\vj}{\mathbf{j}}
\newcommand{\vt}{\mathbf{t}}
\newcommand{\vw}{\mathbf{w}}
\newcommand{\vpi}{\boldsymbol{\pi}}
\newcommand{\vphi}{\boldsymbol{\phi}}
\newcommand{\vrho}{\boldsymbol{\rho}}

\newcommand{\MeV}{{\ensuremath\rm MeV}}
\newcommand{\GeV}{{\ensuremath\rm GeV}}
\newcommand{\TeV}{{\ensuremath\rm TeV}}

\newcommand{\fb}{{\ensuremath\rm fb}}
\newcommand{\ab}{{\ensuremath\rm ab}}

\newcommand{\z}{\phantom{0}}

 \newcommand{\ga}{\alpha}
 \newcommand{\gs}{\sigma}
 \newcommand{\gD}{\Delta}
 \newcommand{\CL}{{\cal L}}

\newcommand{\whizard}{\texttt{WHIZARD}}

\newcommand{\oMega}{\texttt{O'Mega}}
\newcommand{\pythia}{\texttt{PYTHIA}}
\newcommand{\simdet}{\texttt{SIMDET}}
\newcommand{\minuit}{\texttt{MINUIT}}
\begin{document}
\preprintno{DESY 05--067\\hep-ph/0604048\\[0.5\baselineskip] April~2006}
\title{%
 Determination of New Electroweak Parameters\\ at the ILC --- 
Sensitivity to New Physics
}
\author{%
 M.~Beyer\email{michael.beyer}{uni-rostock.de}$^a$,
 W.~Kilian\email{wolfgang.kilian}{desy.de}$^{b,c}$,
 P.~Krstono\v{s}i\'c\email{krstonos}{mail.desy.de}$^c$,
 K.~M\"onig\email{klaus.moenig}{desy.de}$^d$,
 J.~Reuter\email{juergen.reuter}{desy.de}$^c$, 
 E.~Schmidt\email{erik.schmidt}{uni-rostock.de}$^a$,
 H.~Schr\"oder\email{h.schroeder}{uni-rostock.de}$^a$
}
\address{\it%
$^a$Institute of Physics, University of Rostock, D--18051 Rostock, Germany\\
$^b$Fachbereich Physik, University of Siegen, D--57068 Siegen, Germany\\
$^c$Deutsches Elektronen-Synchrotron DESY, D--22603 Hamburg, Germany\\
$^d$DESY, Platanenallee 6, D--15738 Zeuthen, Germany
\\[.5\baselineskip]
}
\abstract{%
We present a study of the sensitivity of an International Linear
Collider (ILC) to electroweak parameters in the absence of a light
Higgs boson.  In particular, we consider those parameters that have
been inaccessible at previous colliders, quartic gauge couplings.
Within a generic effective-field theory context we analyze all
processes that contain quasi-elastic weak-boson scattering, using
complete six-fermion matrix elements in unweighted event samples, fast
simulation of the ILC detector, and a multidimensional parameter fit
of the set of anomalous couplings.  The analysis does not rely on
simplifying assumptions such as custodial symmetry or approximations
such as the equivalence theorem. We supplement this by a similar new
study of triple weak-boson production, which is sensitive to the same
set of anomalous couplings.  Including the known results on triple
gauge couplings and oblique corrections, we thus quantitatively
determine the indirect sensitivity of the ILC to new physics in the
electroweak symmetry-breaking sector, conveniently parameterized by
real or fictitious resonances in each accessible spin/isospin channel.
}
\maketitle


\section{Introduction}

Uncovering the mechanism of electroweak symmetry breaking (EWSB) is a
central issue for the next generation of particle colliders, the LHC
and the ILC.  The previous generation of precision experiments, in
particular data from LEP and SLC, have established the description of
electroweak interactions as a spontaneously broken gauge theory, but
the underlying physics that triggers the formation of a scalar (Higgs)
condensate and thus breaks the electroweak $SU(2)_L\times U(1)_Y$
symmetry is still unknown.  All possible scenarios necessarily involve
yet unseen degrees of freedom and their interactions.  They range from
purely weakly interacting models, such as the minimal Standard Model
(SM) with a light Higgs boson and its supersymmetric generalizations
(e.g., the MSSM), to strongly-interacting settings that could indicate
the opening-up of further gauge sectors or extra
dimensions~\cite{reviews,CPNSH}.

In any case, the Higgs condensate induces masses and longitudinal
polarization components for the weak gauge bosons $W^+,W^-$, and $Z$.
Therefore, a precise study of weak-boson interactions is a nontrivial
measurement of parameters that are related to the unknown
symmetry-breaking sector.  

It may happen that this new physics involves resonances in the elastic
scattering of vector bosons and, in analogy with the form factors of
QCD, in the form factors of vector-boson production.  As a special
case, the SM Higgs boson is a scalar resonance in the $VV\to VV$
($V=W,Z$) elastic scattering amplitude (below the physical region if
the Higgs is light).  Other possible resonances include vector or
tensor states.  Alternatively, the weak-boson scattering amplitudes
and form factors might be featureless while saturating the unitarity
limit at high energies.

Narrow resonances such as the MSSM Higgs may be understood as
elementary particles.  The renormalizability of some
weakly-interacting models supports this view and allows us to
extrapolate the theory up to very high scales and small distances such
as the Planck scale, before any four-dimensional field-theoretic
understanding breaks down.  On the other hand, if resonances are
broad, and if in the absence of light Higgs states renormalizability
is lost, the distinction between elementary and composite states is
meaningless.  For instance, the underlying theory may be a QCD-like
confining gauge theory like
technicolor~\cite{technicolor,technicolor_reviews}, extended and
walking technicolor~\cite{ETC,tumblingtc,walkingtc},
topcolor~\cite{topcolor,topcolorasstc}, a Little-Higgs
model~\cite{little}, deconstructed dimensions~\cite{deconstruction}
or an extra-dimensional Higgs-less theory with Kaluza-Klein towers of
vector resonances~\cite{higgsless}.  A
phenomenological analysis of electroweak symmetry breaking should
therefore account for all of these possibilities.

This can be done in a model-dependent way by predicting observables
within some definite framework and comparing with data.  In
weakly-interacting models where precision calculations are possible,
this is straightforward.  Unfortunately, if the EWSB mechanism
involves strong interactions, our current knowledge is far too limited
to do this.  In minimal technicolor as the classic strong-interaction
theory, the QCD analogy has been exploited to predict some
vector-boson interactions, only to rule out the simplest class of
models by the detailed comparison with LEP data.  While there are many
ways to overcome these constraints, the possibility to accomodate data
in qualitatively different models is usually paid for by a loss of
predictivity.  Since we cannot discard the scenario of strong
electroweak symmetry breaking altogether, the accumulation of more
data in a new energy range is the only path to a significant
improvement in our understanding.

Nevertheless, a phenomenological approach should be able, at least, to
give quantitative information on the sensitivity of new collider
experiments, even if nothing is known or assumed about the underlying
theory.  This is possible, and results are often expressed in terms of
limits on 'new-physics' scales~$\Lambda$.  Unfortunately, the meaning
of such a scale is rather unclear, since it usually depends on
arbitrary normalization factors in effective operators.  Furthermore,
our experimental understanding of the signatures and analysis
possibilities at the next generation of colliders, LHC and ILC (for a
recent overview see \cite{Allanach:2006fy}), so far did not allow us
to accomplish this task in full generality.

In the present paper we present a new analysis of electroweak
observables at the ILC that, together with previous results from
LEP/SLC, should complete the picture.  (We expect that similar
results will become available soon for the LHC
environment~\cite{CPNSH,LHC-TGC} and thus enable us to exploit the
LHC/ILC complementarity~\cite{Weiglein:2004hn}.)  We express the
results on weak-boson interactions in a generic effective-theory
language and transform this into transparent sensitivity estimates by
rephrasing results in terms of would-be resonance mass
parameters~\cite{Krst,Kilian:2005bz}. This allows for a unique and
precise definition of the accessible scale $\Lambda$ in each distinct
interaction channel, that does not depend on arbitrary operator
normalizations.

\subsection{Weak-Boson Interactions at Colliders}

At high-energy colliders there are several processes that probe the
electroweak symmetry-breaking sector.  Vector-boson form factors are
accessible in single and double production of electroweak gauge
bosons.  In a more direct way, we can address the mechanism of
symmetry breaking by measuring the quasi-elastic scattering of vector
bosons that are radiated from incoming fermions.  This is supplemented
by data on triple vector-boson production in fermion
annihilation~\cite{WWZ,Beyer:2004km}. Furthermore, new degrees of freedom
in the symmetry-breaking sector can directly interact with fermions or
manifest themselves in four-fermion interactions via ``oblique''
corrections to gauge-boson propagators.

New effects in fermion pair production, i.e., contact interactions and
oblique corrections, are already constrained by the combination of
low-energy data with the $Z$-peak results of LEP~I and SLC.  The
current status of these measurements is summarized in \cite{:2005em}.
Since the LEP~II collider did produce on-shell $W^+W^-$ pairs, we also
have experimental constraints on the low-energy tail of $W$ form
factors, encoded in the set of triple-gauge couplings (TGC).

The quality of all these data will greatly improve at future
colliders.  The higher energy that is probed in the current Tevatron
run and later at the LHC gives a much better lever arm on four-fermion
data, and we also expect a more precise TGC determination~\cite{TGC}.
Further significant improvements in accuracy are foreseen for the
ILC~\cite{ILCres}. If this machine is run on the $Z$ peak again (GigaZ
option), it will replace the existing data on oblique corrections.  We
give a brief account of this in Sec.~\ref{sec:LEP}. 

A measurement of quasi-elastic vector boson scattering is clearly the
most direct probe of the Higgs mechanism.  Without the Higgs boson the
amplitude matrix of this class of processes saturates the
tree-unitarity bound at $1.2\;\TeV$~\cite{Uni}.  With a Higgs boson,
unitarity is restored (for heavy Higgs bosons
see~\cite{heavy_sm_higgs}).  Actually measuring this has been
considered since the planning of the SSC.  

For a phenomenological description, we have to distinguish two
complementary approaches.  In the high-energy range much beyond
$1\;\TeV$ that would have been covered by the SSC if it had been
built, unitarity saturation invalidates any low-energy expansions, so
the processes are described by arbitrary amplitude functions.  There
is no way to find a finite set of parameters that accounts for all
possibilities.  To simplify the discussion, previous studies therefore
concentrated on a small set of reference models, e.g., a single scalar
or vector resonance, and estimated the perspective of observing them
in data.

With less energy being available at the LHC, the prospects for
discovering, e.g., resonances in the high-energy range is clearly
worse~\cite{LHCres,WW-LHC}.  However, low-energy expansions become
more appropriate, and thus we have a well-defined framework of
interpreting future data in terms of few parameters.  This is even
more true for the current ILC proposal.  There, the high $e^+e^-$
luminosity and the clean environment allow for precision analyses, but
the c.m.\ energy is limited to $500$--$1000\;\GeV$.  This does not
reach into the energy range where perturbative unitarity becomes an
issue.  However, measurements are foreseen to be rather precise and
lead us to the unambiguous parameter determinations that we describe
in the current paper.

\subsection{The Layout of the Paper}

In this paper we present a new, improved estimate of the ILC
sensitivity to the amplitudes of quasi-elastic vector-boson
scattering, using both triple vector-boson production and vector-boson
scattering as complementary processes.  We describe the analyses and
results in Sec.~\ref{sec:TVB} (triple weak-boson production)
and~\ref{sec:VBS} (weak-boson scattering).  Other experimental
constraints are briefly reviewed in Sec.~\ref{sec:LEP}.  As one should
expect, the results are expressed in terms of sensitivity ranges for a
set of low-energy parameters, the anomalous couplings
$\alpha_{4,5,6,7,10}$.  The necessary definitions are collected in
Sec.~\ref{sec:chiral}.  In the simulation and numerical analysis of
scattering processes, we put particular emphasis on model
independence, so we do not assume custodial symmetry, and we refrain
from calculational simplifications such as the effective
$W$~approximation or the Goldstone Equivalence Theorem that have
proven numerically unreliable.

In Sec.~\ref{sec:resonances}, we discuss resonances and their relation
to the measurable low-energy parameters.  As mentioned above, this is
not because resonances have to be present in weak-boson scattering,
but the idea is to give an unambiguous meaning to the notion of a
sensitivity reach in terms of a scale~$\Lambda$.  This language is
then used for the interpretation of our numerical results, as given in
Sec.~\ref{sec:combine}.  If resonances turn out to be actually
present, and accessible at LHC, this way of interpreting data
furthermore allows for a straightforward relation of high-energy and
low-energy measurements, as they can be provided by the combination of
LHC and ILC.


\section{Anomalous Couplings and the Chiral Lagrangian}
\label{sec:chiral}

Below the energy range where new degrees of freedom become visible or
non-perturbative models have to be used, electroweak interactions are
described in terms of an effective field theory: the chiral
Lagrangian~\cite{chL}.  The particular formulation of this Lagrangian
in terms of elementary fields is not unique, but any two different
formulations are related by reparameterizations that do not affect the
$S$~matrix. The physical results depend just on the symmetries and on
the content of asymptotic fields, i.e., the known
particles~\cite{WWrev}.  

The amplitudes derived from this Lagrangian are organized in a
perturbative expansion in powers of $1/(4\pi v)$, where $v$ is the
electroweak scale, set by the Fermi constant as $v=(\sqrt2
G_F)^{-1/2}=246\;\GeV$.  To be precise, the perturbative series
involves the parameters $g$ and $g'$ (the electroweak couplings) and
$E/(4\pi v)$, where $E$ is some combination of the typical process
energies and external-particle masses~\cite{NDA}.

The lowest order in this expansion gives rise to an exact low-energy
theorem~\cite{LET} for the amplitudes of weak-boson scattering, that
depends only on the known value of the electroweak scale~$v$. The
next-to-leading order (NLO) introduces transversally polarized gauge
bosons, one-loop corrections, and a set of new parameters that govern
the second order in the energy expansion, known as anomalous
couplings.  These encode information on the unknown physics that we
are interested in.  While higher orders (two-loop corrections,
one-loop effects of anomalous couplings, and further new free
parameters) are interesting as well, the limited precision of actual
experiments lets us truncate the series at NLO.  In some cases,
higher-order effects may be important, however.

As mentioned before, any such an effective-field theory description is
limited in scope.  It fails at the threshold of the first resonance,
e.g., a Higgs boson or a (``techni-$\rho$'') vector resonance.
However, this can always be remedied by coupling such resonances in a
generic way, introducing their coupling constants as free parameters.
The framework thus retains its generality beyond the threshold.  A
more important limitation comes from the fact that, in the absence of
a SM-like Higgs boson, scattering amplitudes of vector bosons saturate
the unitarity bound at high energy, such that a perturbative expansion
is no longer possible.  Naively, we would expect this bound to be at
$E=4\pi v=3\;\TeV$, but a more precise estimate~\cite{Uni} sets this
scale at $E=1.2\;\TeV$ if all anomalous couplings vanish.

The formal setup of the electroweak chiral Lagrangian is well-known
and has been described in several papers and textbooks.  In order to
introduce the framework and notation for the later sections, we list
the relevant definitions and relations here.

In a generic gauge, the degrees of freedom consist of the
usual fermions, the gauge bosons $W^1,W^2,W^3,B$ (in the gauge basis)
or $W^+,W^-,Z,A$ (in the physical basis), and the scalar Goldstone
bosons $w^+,w^-,z$ that, after symmetry breaking, provide the
longitudinal polarization states of the massive gauge bosons.
Without oblique corrections,
the relation of the gauge and physical bases is given by
\begin{subequations}
\begin{align}
  W^1 &= \tfrac1{\sqrt2}(W^+ + W^-), &
  W^3 &= \cw Z + \sw A, \\
  W^2 &= \tfrac\ii{\sqrt2}(W^+ - W^-), &
  B   &= -\sw Z + \cw A,
\end{align}
\end{subequations}
where $\sw$ and $\cw$ are the sine and cosine of the weak mixing
angle, respectively.  Contracting the $W$ field with Pauli matrices,
$\vW = W^k\tfrac{\tau^k}{2}$, we introduce the field strength tensors
\begin{align}
  \vW_{\mu\nu} &= \pd_\mu\vW_\nu - \pd_\nu\vW_\mu + ig[\vW_\mu, \vW_\nu], \\
  \vB_{\mu\nu} &= \Sigma\left(\pd_\mu B_\nu - \pd_\nu B_\mu\right)
                  \frac{\tau^3}{2}\Sigma^\dagger.
\end{align}
The Goldstone bosons $\vw \equiv w^k\tau^k$ are labeled analogously,
\begin{align}
  w^1 &= \tfrac1{\sqrt2}(w^+ + w^-), &
  w^2 &= \tfrac\ii{\sqrt2}(w^+ - w^-), &
  w^3 &= z,
\end{align}
and enter only via the Goldstone (or Higgs) field matrix,
\begin{equation}
  \Sigma = \exp\left(-\frac{\ii}{v}\vw\right).
\end{equation}
The covariant derivative of the Higgs field is
\begin{equation}
  \vD\Sigma = \pd\Sigma + \ii g\vW\Sigma 
                        - \ii g'\Sigma \left(B\frac{\tau^3}{2}\right),
\end{equation}
where the gauge couplings, again in the absence of anomalous
couplings, are given by their usual definitions $g = \mfrac{e}{\sw}$
and $g'= \mfrac{e}{\cw}$.

It is customary to introduce further, related fields, that allow us
to write all terms in the Lagrangian in a manifestly $SU(2)_L$
gauge-invariant way.  These are
\begin{align}
  \vV &= \Sigma(\vD\Sigma)^\dagger = -(\vD\Sigma)\Sigma^\dagger,
&
  \vT &= \Sigma\tau^3\Sigma^\dagger.
\end{align}

All expressions may be much simplified by adopting the unitarity gauge
where $\vw\equiv 0$.  In this gauge, the latter two fields reduce to
\begin{align}
  \vV &\Rightarrow -\frac{\ii g}{2}\left[\sqrt2(W^+\tau^+ + W^-\tau^-)
                           + \frac1\cw Z \tau^3\right],
&
  \vT &\Rightarrow \tau^3
\end{align}
i.e., the vector field $\vV$ is composed of those components of the
gauge fields that acquire masses.  $\vT$ projects onto the
electrically neutral component; in particular, in unitarity gauge
we have $\tr{\vT\vV}= -\frac{\ii g}{\cw} Z$.  

However, there are good reasons to retain the gauge-invariant form of
the Lagrangian.  In particular, at high energies the leading behavior
of vector boson scattering amplitudes is related to Goldstone
scattering amplitudes~\cite{EQT}, so we may consider the opposite
limit and omit the gauge fields while keeping only the Goldstone
bosons in the Lagrangian.  In this case, we obtain
\begin{align}
  \vV &= \frac{\ii}{v}
  \left(\pd w^k + \frac1v \epsilon^{ijk} w^i\pd w^j\right)\tau^k + O(v^{-3}),
\\
  \vT &= \tau^3 
  + 2\sqrt2\frac{\ii}{v}\left(w^+\tau^+ - w^-\tau^-\right) + O(v^{-2}).
\end{align}

The bosonic part of the lowest-order chiral Lagrangian reads
\begin{equation}
  \LL_0 = -\frac{1}{2}\tr{\vW_{\mu\nu}\vW^{\mu\nu}}
        -\frac{1}{2}\tr{\vB_{\mu\nu}\vB^{\mu\nu}}
        -\frac{v^2}{4}\tr{\vV_\mu\vV^\mu}
	+\beta_1\LL_0'+\sum_i\alpha_i\LL_i
\end{equation}
At NLO, we have to include anomalous couplings.  The purely bosonic, C
and CP invariant interactions that appear are~\cite{chL}
\begin{subequations}
  \label{eq:efflagrange}
  \begin{align}
    \LL_0' &= \tfrac{v^2}{4}\tr{\vT\vV_\mu}\tr{\vT\vV^\mu}
    \\
    \LL_1 &= gg' \tr{\vB_{\mu\nu}\vW^{\mu\nu}}
    \\
    \LL_2 &= \ii g' \tr{\vB_{\mu\nu}[\vV^\mu,\vV^\nu]}
    \\
    \LL_3 &= \ii g \tr{\vW_{\mu\nu}[\vV^\mu,\vV^\nu]}
    \\
    \LL_4 &= \left(\tr{\vV_\mu\vV_\nu}\right)^2
    \\
    \LL_5 &= \left(\tr{\vV_\mu\vV^\mu}\right)^2
    \\
    \LL_6 &= \tr{\vV_\mu\vV_\nu} \tr{\vT\vV^\mu} \tr{\vT\vV^\nu}
    \\
    \LL_7 &= \tr{\vV_\mu\vV^\mu} \left(\tr{\vT\vV_\nu}\right)^2
    \\
    \LL_8 &= \tfrac14 g^2 \left(\tr{\vT\vW_{\mu\nu}}\right)^2
    \\
    \LL_9 &= \tfrac12 \ii g
    \tr{\vT\vW_{\mu\nu}} \tr{\vT[\vV^\mu,\vV^\nu]}
    \\
    \LL_{10} &= \tfrac12
    \left(\tr{\vT\vV_\mu}\right)^2 \left(\tr{\vT\vV_\nu}\right)^2
  \end{align}
\end{subequations}
In this list, there are three operators ($\LL_0',\LL_1,\LL_8$) that
affect gauge-boson propagators directly (oblique corrections).  Three
additional operators ($\LL_2,\LL_3,\LL_9$) contribute to anomalous TGCs.
The remaining five operators ($\LL_{4}$--$\LL_{7}$ and $\LL_{10}$)
induce anomalous quartic couplings only.

The parameter $\beta_1$ plays a special role since it multiplies a
dimension-2 operator.  It is a well-established experimental fact that
this quantity, related to the $\Delta\rho$ parameter, is small, so the
leading-order effective Lagrangian exhibits an ``isospin'' symmetry.
By definition, this symmetry forbids operators that contain $\vT$
factors and thus treat $W$ and $Z$ in an asymmetric way.  At NLO, the
symmetry is broken by $g\sin\theta_w\neq 0$ and by the up-down
differences in fermion masses and couplings, so it can at best be an
approximate symmetry.  We could simplify the anomalous couplings by
assuming isospin conservation to all orders and thus eliminate the
operators $\LL_{6}$--$\LL_{10}$ altogether, but apart from the single
observation that $\Delta\rho\approx 0$ there is no compelling reason for
this.  Therefore, we will not make this assumption.

In addition to these standard dimension-2 and dimension-4 operators,
we introduce a restricted set of dimension-6 operators
\begin{subequations}
\begin{align}
  \LL^\lambda_1 &= \ii\frac{g^3}{3M_W^2}
    \tr{\vW^{\mu\nu} \vW_\nu{}^\rho \vW_{\rho\mu}}
\\
  \LL^\lambda_2 &= \ii\frac{g^2g'}{M_W^2}
    \tr{\vB^{\mu\nu} \vW_\nu{}^\rho \vW_{\rho\mu}}
\\
  \LL^\lambda_3 &= \frac{g^2}{M_W^2}
    \tr{[\vV^\mu,\vV^\nu]\vW_\nu{}^\rho \vW_{\rho\mu}}
\\
  \LL^\lambda_4 &= \frac{g^2}{M_W^2}
    \tr{[\vV^\mu,\vV^\nu]\vB_\nu{}^\rho \vW_{\rho\mu}}
\\
  \LL^\lambda_5 &= \frac{gg'}{2M_W^2}
    \tr{\vT[\vV^\mu,\vV^\nu]}\tr{\vT\vW_\nu{}^\rho \vW_{\rho\mu}}
\end{align}
\end{subequations}
with dimensionless coefficients
$\alpha^\lambda_{1}$--$\alpha^\lambda_{5}$.  The first two operators
induce further anomalous TGCs, while all five contribute anomalous
quartic couplings.  Although these operators are formally of higher
dimension, we will see below that they occur at the same order in the
expansion as the previous operators.

\section{Resonances in the {\boldmath$\TeV$} Range}
\label{sec:resonances}

We are interested in mapping the interactions of weak bosons in the
energy range where EWSB physics becomes important, roughly $E\gtrsim
1\;\TeV$ up to several $\TeV$.  This would be straightforward for a
multi-TeV collider with sufficient luminosity.  Unfortunately, no such
collider will be available soon, and even the VLHC and CLIC projects
fulfil the requirements only partially.

Therefore, for the time being we can expect few signals of this kind
of physics.  These might be striking resonances, such that their event
rates overcompensate the low parton rates of LHC at high momentum.
Otherwise, we can carry out indirect measurements that access just the
gross properties of actual amplitudes.  For these, we should specify
to which energy range they are actually sensitive.

Clearly, indirect data will be most sensitive to the low-energy rise
of amplitudes, and there is an energy limit beyond which no variation
can possibly be detected.  A straightforward and rather generic way to
formulate this is to place a resonance at that energy and check
whether its low-energy effect is visible.  This can be done
independently for each charge (weak isospin) and spin channel.

A resonance in a given scattering channel has two parameters, the
mass~$M$ and the coupling to this channel.  If we are just interested
in the sensitivity reach, we have to get rid of the arbitrariness in
the coupling.  To this end, we first note that the total resonance
width does not exceed the mass --- otherwise the notion of a resonance
is meaningless.  To be more specific, we can introduce the ratio of
width and mass as a parameter~$f\equiv\Gamma / M$.  Since the
low-energy effect of tree-level resonance exchange is proportional to
$f^2$, the ultimate sensitivity of a low-energy measurement can be
associated with the possible maximum $f\approx 1$, i.e., a resonance
that is as wide as heavy.  While narrower states have a pronounced
effect if produced on-shell, they do less influence the low-energy
range.  Actually, a resonance with $f=1$ looks like a broad continuum
that saturates unitarity in the energy range $E\approx M$.

We therefore define the sensitivity limit $\Lambda$ of a low-energy
measurement as given by a resonance with mass $M=\Lambda$ for which
tree-level exchange would induce a $1\sigma$ shift in the fit,
compared to some assumed central value.  The resonance coupling is set
such that the width is equal to the mass (precisely: $\Gamma = fM$,
where we consider $f\leq 1$), assuming that there are no other decay
channels.  This definition ensures that a real resonance with mass
$M=\Lambda$ may have a smaller, but never a larger effect on the
considered low-energy observable.  In other words, the observable is
insensitive to anything in the high-energy amplitude beyond
$E=\Lambda$.

Looking at resonances that couple to vector boson pairs, we can limit
ourselves to spin $J=0,1,2$ and isospin $I=0,1,2$, since these are the
possible quantum numbers of a pair of spin-1, mixed-isospin (1/0)
bosons.  If isospin was conserved exactly, the only accessible $(I,J)$
combinations would be $(0,0)$, $(0,2)$, $(2,0)$, and $(1,1)$.
However, isospin is broken by the $B$ gauge boson (hypercharge) and by
the fermion couplings, therefore we should not rely on isospin
conservation.  Still, there is one combination that we can leave out,
$(I,J)=(2,1)$, since due to the Landau-Yang theorem an isospin-2
vector state does not couple to $W^+W^+$ or $W^-W^-$ pairs and is thus
indistinguishable from a vector with mixed $I=1/0$.

Along with the couplings to vector bosons, for all states considered
here we evaluate the partial width for the decay into a vector boson
pair.  If the resonance is sufficiently heavy (this is the case for
any state that is not directly accessible at the ILC), due to the
Goldstone-boson equivalence theorem this width is well approximated by
the partial width for the decay into two (unphysical) Goldstone
bosons.  This gives us a lower limit $\Gamma_{VV}$ for the total
resonance width $\Gamma$.  From the upper limit on the total width,
$\Gamma\approx M$, we can infer an upper bound for the resonance
coupling, and thus for the scattering amplitude itself.  Integrating
out the resonance gives rise to a shift in the low-energy scattering
amplitude, which is therefore also bounded in magnitude.  In the end,
these bounds have to be compared with the achievable accuracy in the
determination of the low-energy parameters.

The method for integrating out heavy states and thus obtaining their
low-energy (tree-level) effects is well known.  Given a Lagrangian
that contains quadratic and linear terms for the resonance $\Phi$,
\begin{equation}
  \LL_\Phi = \frac{z}{2} \left[\Phi(M^2 + A)\Phi + 2\Phi J\right]
\end{equation}
where $A$ and $J$ involve light fields and (covariant) derivatives,
the tree-level low-energy expansion is
\begin{equation}
  \LL_\Phi^\eff = -\frac{z}{2M^2}JJ + \frac{z}{2M^4}JAJ + O(M^{-6}).
\end{equation}
In an actual calculation, this expression is typically manipulated further
in order to relate the resulting operators to the canonical basis as
defined in Sec.~\ref{sec:chiral}.

We do not consider loop corrections due to resonance exchange, since
after proper renormalization they generically do not alter the results
at the order we are considering.  However, in cases where a symmetry
forbids the linear coupling to $J$, and thus the effect is zero in our
framework, the loop contribution is actually the leading one, although
suppressed by powers of $1/16\pi^2$ and $1/M^2$.  This happens, for
instance, for the supersymmetric partners in the MSSM.  Furthermore,
we should keep in mind that in technicolor theories there are
non-decoupling loop corrections (which originate from the massless,
confined technicolor partons) that have a rather strong impact on the
anomalous couplings that we consider.  The shifts in the oblique
corrections due to this effect have been used to rule out some of the
simplest models.  However, it is generally assumed that in the
theories considered nowadays these corrections are rather small.

In any case, in the present paper we do not intend to actually predict
the values of anomalous couplings in certain models.  Instead, by
relating the possible shifts in low-energy observables (anomalous
couplings) to the high-energy behavior of physical scattering
amplitudes (resonances) we want to estimate the physics reach of
precision measurements and express it in terms of dimensionful
parameters $\Lambda$ (resonance masses) in a meaningful way.

\subsection{Scalar Resonances}
\label{sec:scalar}

Scalar resonances are of particular interest since the most prominent
representative, a $I=0$ scalar boson, serves as a Higgs boson if its
couplings take particular values.  In extended models with Higgs
bosons, there are also scalar resonances with higher isospin.  For
instance, in the MSSM the $(H^+,H^0,H^-)$ triplet can be viewed as an
$I=1$ triplet.  As another example, the Littlest Higgs
model~\cite{little} contains a complex triplet
$(\phi^{++},\phi^+,\phi^0)$, which under isospin decomposes into a
real $I=2$ quintet and a singlet.

After the elimination of Goldstone bosons in unitarity gauge, scalars
do not mix with vector bosons, so at tree level, the low-energy
effects of a heavy scalar resonance are confined to four-boson
couplings, i.e., the parameters $\alpha_{4,5,6,7,10}$.  This is easily
verified for the explicit representations considered here.  We should
keep in mind, however, that a resonance or an equivalent contribution
in the $I=J=0$ channel (i.e., a Higgs boson) provides a (partial)
cutoff for the logarithmic divergences in the chiral Lagrangian and
thus sets the renormalization point for the anomalous couplings.  In
this sense, the parameters $\alpha_{1}$--$\alpha_{5}$ contain a
logarithmic dependence $\ln M/(16\pi^2)$ on this resonance mass.
However, after taking this renormalization into account, the residual
mass dependence due to one-loop diagrams is of order $1/(16\pi^2 M^2)$
and thus subleading compared to the tree-level contributions that are
listed below.

\subsubsection{Scalar Singlet: \boldmath$\sigma$}

This state is the generalization of a Higgs resonance.  It has two
independent linear couplings, $g_\sigma$ and $h_\sigma$.  The latter
violates isospin.  (In the following, we always adopt a notation where
$g$ couplings conserve isospin, while $h$ and $k$ couplings violate it
by one and two units, respectively.)  Neglecting self-couplings etc.\
that do not contribute to the order we are interested in, the
Lagrangian is
\begin{equation}
  \LL_\sigma = -\frac12\left[\sigma\left(M_\sigma^2 + \pd^2\right)\sigma
                             + 2\sigma j\right]
\end{equation}
where
\begin{equation}
  \label{eq:singletscalarcurr}
  j = -\frac{g_\sigma v}{2}\tr{\vV_\mu\vV^\mu}
      -\frac{h_\sigma v}{2}\left(\tr{\vT\vV_\mu}\right)^2
\end{equation}
The Higgs boson corresponds to the special values $g_\sigma=1$ and
$h_\sigma=0$.  Given the fact that we can freely add bilinear and
higher (self-)couplings, the minimal Standard Model emerges as a
special case of the chiral Lagrangian coupled to a scalar resonance.
It should be emphasized that this is an exact equivalence: a simple
nonlinear transformation of the scalar fields, that does not affect
the $S$ matrix, transforms $\LL_\sigma$ into the SM Lagrangian in its
usual form.

Integrating out $\sigma$, we obtain the values of the anomalous
couplings $\beta_1$ and $\alpha_i$.  We get zero values for $\beta_1$
and all parameters that involve field strengths, and
\def\fact{\left(\frac{v^2}{8M_\sigma^2}\right)}
\begin{subequations}
\begin{align}
  \label{eq:sigmaalpha}
  \alpha_4 &= 0
&
  \alpha_6 &= 0
\\
  \alpha_5 &= g_\sigma^2\fact
&
  \alpha_7 &= 2 g_\sigma h_\sigma\fact
\\
           &
&
  \alpha_{10} &=  2h_\sigma^2\fact
\end{align}
\end{subequations}
In the high-mass limit, the $\sigma$ width is given by
\begin{equation}
  \label{eq:sigmawidth}
  \Gamma_\sigma = \frac{g_\sigma^2 + \frac12(g_\sigma+2h_\sigma)^2}{16\pi}
                  \left(\frac{M_\sigma^3}{v^2}\right)
\end{equation}
This includes $\sigma\to W^+W^-$ and $\sigma\to ZZ$.

Scalar resonances may couple to SM fermions.  The couplings need not
follow the pattern of SM Higgs couplings that are proportional to the
fermion masses.  Altogether, the linear couplings of a scalar $\sigma$
to SM particles take the general form
\begin{equation}
  \LL = - \sigma (j_V + j_f),
\end{equation}
where $j_V \sim v V_\mu V^\mu$ is the bosonic
current~(\ref{eq:singletscalarcurr}).  The fermionic current has the
structure
\begin{align*}
  j_f &= 
  g_Q \overline{Q}_L \Sigma Q_R 
  + g_\ell \overline{\ell}_L \Sigma \ell_R 
  +  h_Q \overline{Q}_L \Sigma T Q_R 
  + h_\ell \overline{\ell}_L \Sigma T \ell_R + \hc 
\nonumber\\
  &\quad + g^L_\nu \overline{\ell}^c_L \Sigma^* \Pp \Sigma \ell_L 
  + g^R_\nu \overline{\ell}^c_R \Pp \ell_R 
\end{align*}
with $\Ppm \equiv \frac{1\pm\sigma^3}{2}$.  The $\Sigma$ factors make
the interaction terms formally $SU(2)$-invariant.  (We are assuming
baryon-number conservation.)

Integrating out the heavy singlet $\sigma$ results in the
current-current interactions:
\begin{equation}
  - \frac{1}{2 M_\sigma^2} \Bigl\{ j_V j_V 2 j_V j_f + j_f j_f +
    \Bigr\} 
\end{equation}
The first term is the purely bosonic one considered above.  The third
term is a generic four-fermion contact interaction, while the second
one is a dimension-5 operator coupling two EW gauge bosons and two
fermions.  This term should be detectable in dedicated high-precision
analyses at ILC, but is essentially unconstrained by existing data.

Four-fermion operators mediated by scalar resonances have been
discussed (in the context of fermion compositeness) in
\cite{Gonzalez-Garcia:1998ay}. The most severe limits discussed there
come from atomic parity violation experiments.  However, they are
applicable only if CP is violated, and disappear for the case of a
purely scalar or purely pseudoscalar resonance.  Limits from precision
measurements at LEP or Tevatron are generically of the order of
$\Lambda > 200-500 \,\GeV$.

\subsubsection{Scalar Triplet: \boldmath$\pi$}

If isospin is conserved, this multiplet does not have any couplings to
vector boson pairs, and instead of a resonance we might rather expect
pair production as the dominant phenomenological effect.  Furthermore,
technipions in technicolor models, as a typical realization of
isospin-1 scalars, are actually pseudoscalars and, at first glance, do
not have linear couplings at all.  However, in our treatment the logic
is opposite: we assume an effect to be present and express it in terms
of would-be resonance parameters.  Therefore, we consider the $I=1$
triplet in the resonance mode.

Writing the field as
\begin{equation}
  \vpi = \pi^a\tau^a \equiv \sqrt2(\pi^+\tau^+ + \pi^-\tau^-) + \pi^0\tau^3 
\end{equation}
the Lagrangian is
\begin{equation}
  \LL_\pi = -\frac14\tr{\vpi(M_\pi^2 + \vD^2)\vpi + 2\vpi\vj}
\end{equation}
with
\begin{equation}
  \vj = \frac{h_\pi v}{2}\vV_\mu\tr{\vT\vV^\mu}
       +\frac{h_\pi'v}{2}\vT\tr{\vV_\mu\vV^\mu}
       +\frac{k_\pi v}{2}\vT\left(\tr{\vT\vV_\mu}\right)^2
\end{equation}
Evaluating the effective Lagrangian, the nonvanishing parameters are
\def\fact{\left(\frac{v^2}{16M_\pi^2}\right)}
\begin{subequations}
\begin{align}
  \label{eq:pialpha}
  \alpha_4 &= 0
&
  \alpha_6 &= h_\pi^2\fact
\\
  \alpha_5 &= 2h_\pi^\pp\fact
&
  \alpha_7 &= 2h_\pi'(h_\pi+ 2k_\pi)\fact
\\
           &
&
  \alpha_{10} &=  4k_\pi(h_\pi + k_\pi)\fact
\end{align}
\end{subequations}
The partial widths for the decay into vector boson pairs are different
for charged and neutral pions:
\begin{subequations}
\begin{align}
  \label{eq:piwidth}
  \Gamma_{\pi^\pm} &= 
  \frac{\frac14h_\pi^2}{16\pi}\left(\frac{M_\pi^3}{v^2}\right)
\\
  \Gamma_{\pi^0} &= 
  \frac{h_\pi^\pp + \frac12(h_\pi + h_\pi' + 2k_\pi)^2}{16\pi}
  \left(\frac{M_\pi^3}{v^2}\right)
\end{align}
\end{subequations}
If there is approximate isospin conservation we expect the total
widths to be dominated by fermion pairs and by three-boson decays,
analogous to the pions of QCD.

The fermionic couplings of a triplet scalar involve the current
\begin{equation}
  j_f^a = g_Q^\pm \overline{Q}_L \tau^a \Ppm Q_R + g^\pm_\ell 
  \overline{\ell}_L \tau^a \Ppm \ell_R + \hc 
\end{equation}
Note that Majorana terms are not possible in the triplet case. 
Integrating out the heavy triplet scalar leads to similar
fermion-coupling results as for the singlet~$\sigma$.


\subsubsection{Scalar Quintet: \boldmath$\phi$}
With the notation
$\tau^{xy}\equiv\tau^x\otimes\tau^y$, we expand an isospin-2 scalar as
\begin{equation}
  \vphi = \sqrt2\left(\phi^{++}\tau^{++} + \phi^{--}\tau^{--}\right)
         + \frac{1}{\sqrt2}\left[\phi^+(\tau^{+3}+\tau^{3+})
                                 + \phi^-(\tau^{-3}+\tau^{3-})\right]
         + \frac{1}{\sqrt3}\phi^0(\tau^{33}-\tau^{+-}-\tau^{-+})
\end{equation}
The Lagrangian takes the form
\begin{equation}
  \LL_\phi = -\frac14\tr{\vphi(M_\phi^2+\vD^2)\vphi + 2\vphi\vj}
\end{equation}
with
\begin{align}
  \vj &= -\frac{g_\phi v}{2}\vV_\mu\otimes\vV^\mu
         -\frac{h_\phi v}{4}(\vT\otimes\vV_\mu + \vV_\mu\otimes\vT)
	                    \tr{\vT\vV^\mu}
         -\frac{h_\phi'v}{2}\vT\otimes\vT \tr{\vV_\mu\vV^\mu}
\nonumber \\ &\quad
         -\frac{k_\phi v}{2}\vT\otimes\vT\left(\tr{\vT\vV_\mu}\right)^2
\end{align}
We derive the nonvanishing parameters
\def\fact{\left(\frac{v^2}{16M_\phi^2}\right)}
\begin{subequations}
\begin{align}
  \label{eq:phialpha}
  \alpha_4 &= g_\phi^2\fact
&
  \alpha_6 &= h_\phi(2g_\phi+h_\phi)\fact
\\
  \alpha_5 &= 4h_\phi^\pp\fact
&
  \alpha_7 &= 2h_\phi'(g_\phi + 2 h_\phi+ 4 k_\phi)\fact
\\
           &
&
  \alpha_{10} &= \left(h_\phi^2 + 4k_\phi(g_\phi+2h_\phi+2k_\phi)\right)\fact
\end{align}
\end{subequations}
and the following expressions for the resonance widths:
\begin{subequations}
\begin{align}
  \label{eq:phiwidth}
  \Gamma_{\phi^{\pm\pm}} &= 
    \frac{g_\phi^2}{64\pi}\left(\frac{M_\phi^3}{v^2}\right)
\\
  \Gamma_{\phi^\pm} &=
    \frac{(g_\phi + h_\phi)^2}{64\pi}\left(\frac{M_\phi^3}{v^2}\right)
\\
  \Gamma_{\phi^0} &= \frac13\,
    \frac{(g_\phi-4h_\phi')^2 + 2(g_\phi+2h_\phi+2h_\phi'+4k_\phi)^2}
         {64\pi} \left(\frac{M_\phi^3}{v^2}\right)
\end{align}
\end{subequations}

For the scalar quintet (with doubly-charged components) no universal
coupling to a pair of SM fermions is possible.  These occur only for
the projection onto the singly charged and neutral components.


\subsection{Vector Resonances}
\label{sec:vector}

Vector resonances play an important role in the analysis of weak-boson
scattering.  QCD-like technicolor and the so-called BESS
models~\cite{BESS} predict a strong vector resonance $\rho_{\rm TC}$
in with the $\rho$ meson resonance in pion-pion scattering.  Vector 
resonances are also present in extended gauge theories, where they are
usually called $Z',W'$.  The low-energy effect of such states involves
all anomalous couplings in the effective Lagrangian.

There are various possibilities for coupling a vector resonance $\rho$
to gauge fields.  The couplings can be organized in powers of $1/M^2$.
Let us discuss a triplet vector resonance $\vrho$ (with isospin
conservation) for concreteness, the discussion of singlet resonances
and isospin violation is analogous.

The bosonic part of the Lagrangian may contain the operators
\begin{equation}
  v^2\tr{\vrho_\mu\vV^\mu},\quad
  \tr{\vrho_{\mu\nu}\vW^{\mu\nu}},\quad
  \tr{\vrho_\mu\vV_\nu\vW^{\mu\nu}},\quad
\end{equation}
if expanded up to dimension~$4$.  At dimension~$6$ there is an important
additional term,
\begin{equation}
  \tr{\vrho_{\mu\nu}\vW^{\rho\mu}\vW_\rho{}^\nu}.
\end{equation}

We follow the convention that in couplings with positive mass
dimension (except for the resonance masses themselves) we extract
explicit factors of $v$, not $M$.  Actually, there is a redundancy
associated to the weak-boson equation of motion,
\begin{align}
  0 &=
  - \frac{1}{g} D_\nu \vW^{\mu\nu} - \vj_f^\mu -
  i \frac{v^2}{4} {\bf V}^\mu  
\end{align}
that allows us to eliminate one of the three vector-resonance
couplings~\cite{Kilian:2003xt}, and justifies the extraction of the
dimensional parameter~$v^2$ if all dimensionless couplings are to have
identical scaling properties.  We use this redundancy to eliminate the
kinetic mixing term, $\tr{\vrho_{\mu\nu}\vW^{\mu\nu}}$.  This
condition also fixes the direct coupling of the vector resonance to
the fermionic current~$\vj^\mu_f$.

The fermionic coupling clearly has an impact on precision data.  Let
us now focus on the case of an isospin singlet vector $\omega_\mu$
(equivalent to a $Z'$ resonance). In contrast to scalars, a vector
current couples multiplets of like chirality, so we do not need extra
factors of $\Sigma$ for a gauge-invariant interaction.  We allow for
isospin breaking and decompose the currents into their up- and
down-type components:
\begin{equation}
  \LL = \omega_\mu j_f^\mu
\end{equation}
with
\begin{equation}
  j_f^\mu = g_{Q,L}^\pm \overline{Q}_L \gamma^\mu \Ppm Q_L +
  g_{Q,R}^\pm \overline{Q}_R \gamma^\mu \Ppm Q_R + 
  g_{\ell,L}^\pm \overline{\ell}_L \gamma^\mu \Ppm \ell_L +
  g_{\ell,R}^\pm \overline{\ell}_R \gamma^\mu \Ppm \ell_R 
\end{equation}

Integrating out the heavy vector resonance (here it is sufficient to
take the lowest order), one gets
\begin{equation}
  \label{vectorcurr}
  \frac{1}{2 M_\omega^2} \left( j_{V,\mu} j^\mu_V + 2 j_{f,\mu} j^\mu_V
  + j_{f,\mu} j^\mu_f \right).
\end{equation}
The second term is a redefinition of the fermionic currents of the SM
that can be attributed to the mixing of the new resonance with the SM
$Z$ boson.  In the vector-singlet case indicated here, this leads to
non-universal $Z$-fermion couplings since the current of the vector
resonance is not necessarily proportional to the SM hypercharge
current.  A vector-triplet resonance couples proportional to the SM
isospin current and thus preserves universality, but its presence
changes the meaning of the Fermi constant, which is defined by the
vector-triplet exchange interaction in muon decay.  The third term is
a four-fermion contact interaction, analogous to the scalar-resonance
case, but with different helicity structure.


\subsubsection{Vector Singlet: \boldmath$\omega$}

The Lagrangian is
\begin{align}
  \LL_\omega &= -\frac14\omega_{\mu\nu}\omega^{\mu\nu} 
  + \frac{M_\omega^2}{2}\omega_\mu\omega^\mu
  + \ii\frac{h_\omega v^2}{2}\omega_\mu\tr{\vT\vV^\mu}
  +  \frac{g v^2 k_\omega}{2 M_\omega^2} \omega_\mu \tr{\lbrack \vT,
    \vV_\nu \rbrack \vW^{\nu\mu}}       \nonumber \\
    & \quad 
  + \ii\frac{\ell_\omega}{M_\omega^2}\omega_{\mu\nu}
    \tr{\vT\vW^\nu{}_\rho\vW^{\rho\mu}}
\end{align}
and can be rewritten by partial integration:
\begin{equation}
  \label{eq:omega_curr}
  \LL_\omega =
  \frac12\left[\omega_\mu\left((M^2+\pd^2)g^{\mu\nu}-\pd^\nu\pd^\mu\right)
    \omega_\nu + 2\omega_\mu j^\mu\right]
\end{equation}
with
\begin{equation}
  j_\mu = \ii\frac{h_\omega v^2}{2}\tr{\vT\vV_\mu}
  +  \frac{g v^2 k_\omega}{2 M_\omega^2} \tr{\lbrack \vT,
    \vV^\nu \rbrack \vW_{\nu\mu}}   
  + \ii\frac{2\ell_\omega}{M_\omega^2}\pd_\nu
    \tr{\vT\vW^\nu{}_\rho\vW^\rho{}_\mu}
\end{equation}
Expanding up to second order and expressing the result in the
canonical operator basis, we obtain the coefficients
\def\fact{\left(\frac{v^2}{2M_\omega^2}\right)}
\begin{align}
  \beta_1 &= h_\omega^2 \frac{v^2}{2M_\omega^2}
\end{align}
\begin{subequations}
\begin{align}
  \alpha_1 &= h_\omega^2 \fact^2
&
  \alpha_2 &= h_\omega^2 \fact^2
\\
  \alpha_3 &= h_\omega k_\omega \fact^2
\\
  \alpha_4 &= h_\omega^2 \fact^2
&
  \alpha_6 &= -h_\omega^2 \fact^2
\\
  \alpha_5 &= -h_\omega^2 \fact^2
&
  \alpha_7 &= h_\omega^2 \fact^2
\\
  \alpha_8 &= -h_\omega^2 \fact^2
\\
  \alpha_9 &= -h_\omega (h_\omega + k_\omega) \fact^2
&
  \alpha_{10} &= 0
\end{align}
\end{subequations}
and
\begin{subequations}
\begin{align}
  \alpha^\lambda_1 &= -{h_\omega\ell_\omega}\fact^2
&
  \alpha^\lambda_2 &= {h_\omega\ell_\omega}\fact^2
\\
  \alpha^\lambda_3 &= 0
&
  \alpha^\lambda_4 &= 0
\\
  \alpha^\lambda_5 &= {h_\omega\ell_\omega}\fact^2
\end{align}
\end{subequations}
The $\omega$ boson can decay into $W^+W^-$ but not into $ZZ$, and the
pair decay width is
\begin{equation}
  \Gamma_\omega = \frac{h_\omega^2 + \frac12\ell_\omega^2}{48\pi}M_\omega
\end{equation}

Note that, at leading order in $v^2/M^2$, the $k_\omega$ coupling does
not enter the width formula.  This interaction involves a longitudinal
and a transversal gauge boson, which in the limit $v\ll M$ is
forbidden as an on-shell $\omega_\mu$ decay mode.  We could thus
interpret this term as a continuum property, not related to the
resonance, and allow for large values of $k_\omega$ (since the
$\Gamma\leq M$ constraint is irrelevant).  However, looking at the
equations of motion, consistent scaling requires $k_\omega$ to be of
the same order as the other dimensionless couplings.


\subsubsection{Vector Triplet: \boldmath$\rho$}

The vector triplet is written as
\begin{equation}
  \vrho_\mu = \rho_\mu^a\tau^a
  = \sqrt2\left(\rho_\mu^+\tau^+ + \rho_\mu^-\tau^-\right)
    + \rho_\mu^0\tau^3
\end{equation}
We write the generic Lagrangian up to order $1/M^2$ that includes
isospin-violating effects and anomalous magnetic moments:
\begin{align}
  \LL_\rho &= -\frac18\tr{\vrho_{\mu\nu}\vrho^{\mu\nu}}
  + \frac{M_\rho^2}{4}\tr{\vrho_\mu\vrho^\mu}
  + \frac{\Delta M_\rho^2}{8}\left(\tr{\vT\vrho_\mu}\right)^2
\nonumber\\ &\quad
  + \ii \frac{\mu_\rho}{2}g\tr{\vrho_\mu \vW^{\mu\nu}\vrho_\nu}
  + \ii \frac{\mu_\rho'}{2}g'\tr{\vrho_\mu \vB^{\mu\nu}\vrho_\nu}
\nonumber\\ &\quad
  + \ii \frac{g_\rho v^2}{2}\tr{\vrho_\mu\vV^\mu}
  + \ii \frac{h_\rho v^2}{2}\tr{\vrho_\mu\vT}\tr{\vT\vV^\mu}
\nonumber\\ &\quad
  + \frac{g' v^2 k_\rho}{2 M_\rho^2} \tr{ \vrho_\mu \lbrack \vB^{\nu\mu} ,
  \vV_\nu \rbrack} 
  + \frac{g v^2 k_\rho'}{4 M_\rho^2}
  \tr{ \vrho_\mu \lbrack \vT , \vV_\nu \rbrack}
  \tr{\vT \vW^{\nu\mu}} 
\nonumber\\ &\quad
  + \frac{g v^2 k''_\rho}{4 M_\rho^2}
  \tr{ \vT \vrho_\mu}
  \tr{\lbrack \vT, \vV_\nu \rbrack \vW^{\nu\mu}}
  + \ii\frac{\ell_\rho}{M_\rho^2}
    \tr{\vrho_{\mu\nu}\vW^\nu{}_\rho\vW^{\rho\mu}}
\nonumber\\ &\quad
  + \ii\frac{\ell_\rho'}{M_\rho^2}
    \tr{\vrho_{\mu\nu}\vB^\nu{}_\rho\vW^{\rho\mu}}
  + \ii\frac{\ell_\rho''}{M_\rho^2}
    \tr{\vrho_{\mu\nu}\vT}\tr{\vT\vW^\nu{}_\rho\vW^{\rho\mu}}
\end{align}
For the moment, we omit the mass splitting term.  Then, partial
integration transforms the Lagrangian into
\begin{align}
  \LL_\rho = \frac14\tr{\vrho_\mu\left(
    M_\rho^2 g^{\mu\nu} + \vD^2 g^{\mu\nu} - \vD^\nu\vD^\mu 
    + 2\ii\mu_\rho g\vW^{\mu\nu} + 2\ii\mu_\rho' g'\vB^{\mu\nu}\right)
    \vrho_\nu
    + 2\vrho_\mu\vj^\mu}
\end{align}
where
\begin{align}
  \vj_\mu &= \ii g_\rho v^2\vV_\mu + \ii g_\rho'v^2\vT\tr{\vT\vV_\mu}
\nonumber\\ &\quad
  + \frac{g' v^2 k_\rho}{M_\rho^2} \lbrack \vB_{\nu\mu}  , \vV^\nu
  \rbrack 
  +
  \frac{g v^2 k_\rho'}{2 M_\rho^2} 
  \lbrack \vT , \vV^\nu \rbrack
  \tr{\vT \vW_{\nu\mu}} 
  +
  \frac{g v^2 k''_\rho}{2 M_\rho^2}
  \vT 
  \tr{\lbrack \vT, \vV^\nu \rbrack \vW_{\nu\mu}}    
\nonumber\\ &\quad
  + \ii\frac{4\ell_\rho}{M_\rho^2}\vD_\nu
    \left(\vW^\nu{}_\rho\vW^{\rho\mu}\right)
  + \ii\frac{4\ell_\rho'}{M_\rho^2}\vD_\nu
    \left(\vB^\nu{}_\rho\vW^{\rho\mu}\right)
  + \ii\frac{4\ell_\rho''}{M_\rho^2}\vD_\nu
    \left(\vT\tr{\vT\vW^\nu{}_\rho\vW^{\rho\mu}}\right)
\end{align}
In reducing the effective Lagrangian, we use the fact that the
operators
\begin{equation}
   \tr{\vW_{\mu\nu}\vW^{\mu\nu}},
   \qquad
   \tr{\vB_{\mu\nu}\vB^{\mu\nu}},
   \qquad\text{and}\quad
   \tr{\vV_\mu\vV^\mu}
\end{equation}
can be dropped because they occur in the zeroth-order part of the
chiral Lagrangian.  These operators renormalize the measured values of
$g$, $g'$, and $v$ with respect to their bare values which are unknown
anyway.  Finally, we add the effect of the mass splitting $\Delta M^2$
to get the parameters
\def\fact{\left(\frac{v^2}{2M_\rho^2}\right)}
\begin{align}
  \beta_1 &= 4 h_\rho(g_\rho + h_\rho)\frac{v^2}{2M_\rho^2}
  - (g_\rho + 2h_\rho)^2\frac{v^2\Delta M_\rho^2}{2M_\rho^4}
\end{align}
and
\begin{subequations}
\begin{align}
  \alpha_1 &= (g_\rho + 2h_\rho)^2\fact^2
&
  \alpha_2 &= \left[- g_\rho (g_\rho (1 - \mu'_\rho) + 2 k_\rho) + 4h_\rho^2 
                    \right]\fact^2
\\
  \alpha_3 &= (g_\rho+2h_\rho)
              \left[ g_\rho \left(1 + {\mu_\rho} \right) + k''_\rho \right]
                \fact^2
\\
  \alpha_4 &= (g_\rho - 2h_\rho)^2\fact^2
&
  \alpha_6 &= 8g_\rho h_\rho\fact^2
\\
  \alpha_5 &= -(g_\rho - 2h_\rho)^2\fact^2
&
  \alpha_7 &= -8g_\rho h_\rho\fact^2
\\
  \alpha_8 &= -4h_\rho(g_\rho+h_\rho)\fact^2
\\
  \alpha_9 &=- \left[(2 h_\rho + k''_\rho)(g_\rho+2h_\rho) \right.
& 
\nonumber \\ &\qquad
            \left. + 2 h_\rho (k'_\rho + g_\rho{\mu_\rho})\right]\fact^2
&
  \alpha_{10} &= 0
\end{align}
\end{subequations}
The $\lambda$-type couplings are
\begin{subequations}
\begin{align}
  \alpha^\lambda_1 &=
  -\left[(g_\rho + 2h_\rho)(\ell_\rho+2\ell_\rho'')
                + 2g_\rho\ell_\rho\right]\fact^2
\\
  \alpha^\lambda_2 &=
  \left[(g_\rho+2h_\rho)(\ell_\rho+2\ell_\rho'') 
                - \frac{\cw}{\sw}g_\rho\ell_\rho'\right] \fact^2
\\
  \alpha^\lambda_3 &=
  -(g_\rho + 2h_\rho)\ell_\rho\fact^2
&
  \alpha^\lambda_4 &=
  -\frac{\cw}{\sw}(g_\rho + 2h_\rho)\ell_\rho'\fact^2
\\
  \alpha^\lambda_5 &=
  -(g_\rho - 2h_\rho)\ell_\rho''\fact^2
\end{align}
\end{subequations}

We note that $\beta_1$ (related to the $\Delta\rho$ or $\Delta T$
parameters) is of order $v^2/M_\rho^2$, while the other coefficients
are all of order $v^4/M_\rho^4$.  Experimentally, $\beta_1$ is known
to be small.  Usually, one draws the conclusion that $h_\rho\approx
0$, i.e., vector resonance interactions conserve isospin.  The above
formulas show that there are other possibilities: We could have
$h_\rho=-g_\rho$, which corresponds to a pseudo-symmetric case where
the components of the $\rho$ triplet couple with alternating sign.
Incidentally, in this case $\alpha_8$ (the $\Delta U$ parameter) also
vanishes, but the quartic couplings $\alpha_{4}$ to $\alpha_7$ are
significantly enhanced.  Furthermore, we cannot exclude a
cancellation, e.g., due to a nonvanishing isospin splitting.

A charged $\rho$ resonance can decay into $W^\pm Z$ and $W^\pm\gamma$:
\begin{subequations}
\begin{align}
  \Gamma_{\rho^\pm \to W^\pm Z} &= 
  \frac{(g_\rho + 2h_\rho)^2 
        + 2(\cw\ell_\rho + \frac12\sw\ell_\rho')^2}{48\pi} M_\rho
\\
  \Gamma_{\rho^\pm \to W^\pm\gamma} &=
  \frac{2(\sw\ell_\rho - \frac12\cw\ell_\rho')^2}{48\pi} M_\rho
\end{align}
\end{subequations}
For the neutral $\rho$, the Landau-Yang theorem forbids $ZZ$ and
$\gamma\gamma$ final states.  The total widths are
\begin{subequations}
\begin{align}
  \Gamma_{\rho^\pm} &= 
  \frac{(g_\rho + 2h_\rho)^2 
        + 2\ell_\rho^2 + \frac12\ell_\rho^\pp}{48\pi} M_\rho
\\
  \Gamma_{\rho^0} &= 
  \frac{(g_\rho - 2h_\rho)^2
        + 2(\ell_\rho + 2\ell_\rho'')^2}{48\pi} M_\rho
\end{align}
\end{subequations}

Again the operators with the $k$ coefficients do not change the
formula for the width of the heavy vector resonance at the order we
are considering because a helicity flip is needed, which is
proportional to the masses of the electroweak gauge bosons.


\subsection{Tensor Resonances}
\label{sec:tensor}

A massive tensor field $f^{\mu\nu}$ is subject to the conditions
\begin{align}
  f^{\mu\nu} &= f^{\nu\mu},
&
  f^\mu{}_\mu &= 0
&
  \pd_\mu f^{\mu\nu} = \pd_\nu f^{\mu\nu} = 0.
\end{align}
Its spin sum is given by
\begin{equation}
  \sum_\lambda\epsilon^*_\lambda{}^{\mu\nu}\epsilon_\lambda^{\rho\sigma}
  = \frac12\left(P^{\mu\rho}P^{\nu\sigma} +  P^{\mu\sigma}P^{\nu\rho}\right)
  -\frac13\left(P^{\mu\nu}P^{\rho\sigma}\right),
\end{equation}
where
\begin{equation}
  P^{\mu\nu}(k) = g^{\mu\nu} - \frac{k^\mu k^\nu}{M^2}.
\end{equation}
The free Lagrangian is
\begin{equation}
  \LL_f = \LL_{\rm kin} - \frac{M^2}{2}f_{\mu\nu} f^{\mu\nu}
\end{equation}
where we do not need the explicit form of the kinetic part as long as
we are just interested in the leading-order effective Lagrangian.

In the sequel, we discuss couplings of tensor resonances to fermions. 
Heavy tensor resonances beyond the EWSB scale have been
introduced in the context of extra dimensions as Kaluza-Klein
recurrences of the graviton. These particles usually couple to the
energy-momentum tensor of ordinary matter, which may serve as a
guideline for the construction of the current here. Since we are only
interested in the low-energy effective theory, after integrating out
the heavy tensor we remain with an interaction of two conserved
currents. Hence, we are allowed to omit terms proportional to a
derivative or a metric due to the transversality and tracelessness of
the tensor resonance. Therefore, dimension-4 couplings to fermions are
not possible. This is because one needs two Lorentz indices which have to
be symmetric, ruling out $\sigma^{\mu\nu}$ couplings. Therefore, the
lowest-order term in the current contains a derivative and is
dimension-5. Furthermore, since the $\gamma$ matrix flips the
chirality, a Majorana coupling at dimension-5 is not possible. 
\begin{equation}
  \LL = f_{\mu\nu} j^{\mu\nu}_f  \;\;\text{with} \;\; 
   j_{f,\mu\nu} = \frac{1}{\Lambda} \sum_{a=L/R}
   \sum_\psi \frac{g_{\psi,a}^\pm}{2} \overline{\psi}_a  (\gamma_\mu  
   \stackrel{\leftrightarrow}{\partial}_\nu + \gamma_\nu
   \stackrel{\leftrightarrow}{\partial}_\mu) \Ppm \psi_a.
\end{equation}
Here $\Lambda$ is the cutoff scale. Integrating out the tensor
resonance yields a dim.-8 contact interaction. Due to the presence of
the derivatives this operator is only relevant phenomenologically for
the heaviest SM fermions, so that no stringent bounds exist for these
terms.  Note that the usual bounds on tensor interactions concern the
antisymmetric tensor (magnetic moment-like operators).

\subsubsection{Tensor Singlet: \boldmath$f$}

Including interactions, we write the Lagrangian for a neutral tensor
field $f_{\mu\nu}$
\begin{equation}
  \LL_f = \LL_{\rm kin}
  -\frac{M_f^2}{2}f_{\mu\nu} f^{\mu\nu} + f_{\mu\nu} j^{\mu\nu}
\end{equation}
where $j_{\mu\nu}$ is a traceless symmetric tensor current:
\begin{align}
  j_{\mu\nu} &= -\frac{g_f v}{2}
  \left(\tr{\vV_\mu\vV_\nu} - \frac{g_{\mu\nu}}{4}\tr{\vV_\rho\vV^\rho}\right)
\nonumber\\ &\quad
  - \frac{h_f v}{2} \left(\tr{\vT\vV_\mu}\tr{\vT\vV_\nu} 
                          - \frac{g_{\mu\nu}}{4}(\tr{\vT\vV_\rho})^2\right)
\end{align}
Expanding the effective Lagrangian to leading order, we obtain the
nonvanishing parameters \def\fact{\left(\frac{v^2}{8M_f^2}\right)}
\begin{subequations}
\begin{align}
  \alpha_4 &= g_f^2\fact
&
  \alpha_6 &= 2g_f h_f \fact
\\
  \alpha_5 &= -\frac{g_f^2}{4}\fact
&
  \alpha_7 &= -\frac{g_f h_f}{2} \fact
\\
&&
  \alpha_{10} &= \frac32 h_f^2\fact
\end{align}
\end{subequations}
Similar to scalar resonances, at leading order no anomalous bilinear
or trilinear couplings are generated by tensor exchange.

The decay width of a tensor field can be evaluated using the spin sum
as introduced above.  We obtain
\begin{equation}
  \Gamma_f = \frac{g_f^2 + \frac12(g_f + 2h_f)^2}{16\pi}
  \left(\frac{M^3}{30v^2}\right)
\end{equation}
where the two terms in the numerator correspond to the $f\to W^+W^-$
and $f\to ZZ$ decays, respectively.

\subsubsection{Tensor Triplet: \boldmath$a$}

A triplet tensor field can be written as
\begin{equation}
  \va_{\mu\nu} = \sqrt2\left(a^+_{\mu\nu}\tau^+ +  a^-_{\mu\nu}\tau^-\right)
                 + a^0_{\mu\nu}\tau^3
\end{equation}
The Lagrangian is
\begin{equation}
  \LL_a = \LL_{\rm kin}
  - \frac{M_a^2}{4}\tr{\va_{\mu\nu}\va^{\mu\nu}}
  + \frac12\tr{\va_{\mu\nu}\vj^{\mu\nu}}
\end{equation}
where
\begin{align}
  \vj_{\mu\nu} &= 
  -\frac{h_a v}{4}\left(\vV_\mu\tr{\vT\vV_\nu} + \vV_\nu\tr{\vT\vV_\mu}
                        - \frac{g_{\mu\nu}}{2}\vV_\rho\tr{\vT\vV^\rho}\right)
\nonumber \\ &\quad
  -\frac{h_a' v}{2}\vT\left(\tr{\vV_\mu\vV_\nu}
                        - \frac{g_{\mu\nu}}{4}\tr{\vV_\rho\vV^\rho}\right)
\nonumber \\ &\quad
  -\frac{k_a v}{2}\vT\left(\tr{\vT\vV_\mu}\tr{\vT\vV_\nu}
                        - \frac{g_{\mu\nu}}{4}(\tr{\vT\vV_\rho})^2\right)
\end{align}
We obtain for the electroweak parameters
\def\fact{\left(\frac{v^2}{8M_a^2}\right)}
\begin{subequations}
\begin{align}
  \alpha_4 &= h_a^\pp\fact
&
  \alpha_6 &= \frac14 \left(\frac12h_a^2 + 4 h_a'(h_a+2k_a)\right) \fact
\\
  \alpha_5 &= -\frac{h_a^\pp}{4}\fact
&
  \alpha_7 &= \frac14\left(h_a^2 - h_a'(h_a+2k_a)\right) \fact
\\
&&
  \alpha_{10} &= \frac32 k_a(h_a+k_a)\fact
\end{align}
\end{subequations}
and for the widths
\begin{subequations}
\begin{align}
  \Gamma_{a^\pm} &=
  \frac{h_a^2}{64\pi}\left(\frac{M_a^3}{30v^2}\right)
\\
  \Gamma_{a^0} &=
  \frac{h_a^\pp + \frac12(h_a+h_a'+2k_a)^2}{16\pi}
  \left(\frac{M_a^3}{30v^2}\right)
\end{align}
\end{subequations}


\subsubsection{Tensor Quintet: \boldmath$t$}

This is analogous to the scalar quintet $\phi$:
\begin{equation}
  \vt_{\mu\nu}
  = \sqrt2\left(t_{\mu\nu}^{++}\tau^{++} + t_{\mu\nu}^{--}\tau^{--}\right)
         + \frac{1}{\sqrt2}\left[t_{\mu\nu}^+(\tau^{+3}+\tau^{3+})
                                 + t_{\mu\nu}^-(\tau^{-3}+\tau^{3-})\right]
         + \frac{1}{\sqrt3}t_{\mu\nu}^0(\tau^{33}-\tau^{+-}-\tau^{-+})
\end{equation}
The Lagrangian is
\begin{equation}
  \LL_t = \LL_{\rm kin} - \frac{M_t^2}{4}\tr{\vt_{\mu\nu}\vt^{\mu\nu}}
  + \frac12\tr{\vt_{\mu\nu}\vj^{\mu\nu}}
\end{equation}
where
\begin{align}
  \vj^{\mu\nu} &=
  -\frac{g_t v}{2}\left[\frac12
    \left(\vV^\mu\otimes\vV^\nu + \vV^\nu\otimes\vV^\mu\right)
    - \frac{g^{\mu\nu}}{4}\vV_\rho\otimes\vV^\rho\right]
\nonumber \\ &\quad
  -\frac{h_t v}{2} \left[\frac14
    \left(\vT\otimes\vV^\mu + \vV^\mu\otimes\vT\right)\tr{\vT\vV^\nu}
  + \frac14
    \left(\vT\otimes\vV^\nu + \vV^\nu\otimes\vT\right)\tr{\vT\vV^\mu}
\right. \nonumber\\ &\qquad\qquad\left.
  - \frac{g^{\mu\nu}}{8}\left(\vT\otimes\vV_\rho + \vV_\rho\otimes\vT\right)
    \tr{\vT\vV^\rho}\right]
\nonumber \\ &\quad
  -\frac{h_t' v}{2}\vT\otimes\vT
  \left[\tr{\vV^\mu\vV^\nu} - \frac{g^{\mu\nu}}{4}\tr{\vV_\rho\vV^\rho}\right]
\nonumber \\ &\quad
  -\frac{k_t v}{2}\vT\otimes\vT
  \left[\tr{\vT\vV^\mu}\tr{\vT\vV^\nu} - \frac{g^{\mu\nu}}{4}
    \left(\tr{\vT\vV_\rho}\right)^2\right]
\end{align}
The parameters are
\def\fact{\left(\frac{v^2}{16M_t^2}\right)}
\begin{subequations}
\begin{align}
  \alpha_4 &= \left(\frac14g_t^2 + 4h_t^\pp\right)\fact
&
  \alpha_6 &= \left(\frac12h_t(g_t + \frac12h_t)
                    + 4h_t'(\frac12 g_t + h_t+2k_t)\right)\fact
\\
  \alpha_5 &= \left(\frac12g_t^2 - h_t^\pp\right)\fact
&
  \alpha_7 &= \left(h_t(g_t + \frac12h_t)
                    - h_t'(\frac12 g_t + h_t+2k_t)\right)\fact
\\
&&
  \alpha_{10} &= 3\left(\frac14h_t^2 + k_t(g_t + 2h_t+2k_t)\right)\fact
\end{align}
\end{subequations}
The widths are
\def\fact{\left(\frac{M_t^3}{30v^2}\right)}
\begin{subequations}
\begin{align}
  \Gamma_{t^{\pm\pm}} &= \frac{g_t^2}{64\pi}\fact
\\
  \Gamma_{t^\pm} &= \frac{(g_t+h_t)^2}{64\pi}\fact
\\
  \Gamma_{t^0} &= \frac13\,\frac{(g_t-4h_t')^2 +
                                 2(g_t+2h_t+2h_t'+4k_t)^2}{64\pi}\fact
\end{align}
\end{subequations}

\subsection{Relating Observables to Resonance Parameters}

As illustrated by the above results, the leading effect of scalar and
tensor exchange on the anomalous couplings $\alpha_i$ is of the order $g^2
v^2/(16M^2)$, where $g$ is any of the couplings introduced in the
various Lagrangians.  For each resonance, the total width is limited
by the requirement $\Gamma_\text{tot}\lesssim M$, while it is bounded
from below by the partial widths that scale like
\begin{equation}
  \Gamma \sim \frac{g^2}{16\pi}\,\frac{M^3}{k_g v^2}
\end{equation}
with some numerical factor $k_g$. Combining this information, we get
an upper limit for the coupling strength, and hence for the anomalous
quartic couplings, which is of the order
\begin{equation}\label{alpha-limit}
  g^2 \lesssim 16\pi k_g\frac{v^2}{M^2}
  \qquad\Rightarrow\qquad
  |\alpha_i| \lesssim 4\pi k_i\left(\frac{v^2}{M^2}\right)^2.
\end{equation}
The numerical coefficients $k_g,k_i$ depend on the resonance channel
and on the type of coupling and can be read of from the relations
given in the previous sections.  For a scalar, $k$ is of order one,
while for a tensor, a typical value is $k=30$.  

To be at all sensitive to a given resonance mass $M$ (i.e., to the
behavior of the amplitude in the energy range $E\sim M$), the
experimental accuracy on the parameter $\alpha_i$ has to be at least
as good as required by~(\ref{alpha-limit}).  Furthermore, the presence
of radiative corrections and the necessity of counterterms imply an
inherent uncertainty on the anomalous couplings,
\begin{equation}
  \Delta\alpha_i \sim \mfrac{1}{16\pi^2},
\end{equation}
so that the effect of the resonance dominates only if
\begin{equation}\label{mass-reach}
  M \lesssim 4\pi v\,\sqrt[{\scriptstyle 4}]{\mfrac{k_i}{\pi}}.
\end{equation}
As a result, the reach of low-energy measurements as an indirect
model-independent determination of the high-energy amplitude behavior
will not exceed the energy range $E\sim M$ with $M$ given
by~(\ref{mass-reach}), even under favorable circumstances.

Naively, for a vector resonance the situation looks
better, since its width scales only with $M$, compared to $M^3$ for
scalar and tensor states.  However, the results of
Sec.~\ref{sec:vector} clearly show that, with the exception of
$\beta_1$, all anomalous couplings receive corrections only at order
$v^4/M^4$, so combining this with the bound on the coupling set by
$\Gamma_\text{tot}\lesssim M$, we again arrive at the
conclusion~(\ref{alpha-limit}).  In short, in the absence of fermionic
couplings, $\beta_1$ --- i.e., the $\rho$ parameter --- is the only
parameter that is sensitive to resonances, and thus to the high-energy
behavior of electroweak amplitudes, at order $v^2/M^2$.

One should keep in mind that fermionic interactions may play a
significant role.  If a resonance with mass $M$ couples to a fermionic
current $j_f$, the effective Lagrangian contains a contact interaction
$\frac{1}{M^2}j_f^2$, a dimension-$6$ operator, that scales with
$1/M^2$.  Limits on contact terms are therefore potentially more
sensitive to new phenomena than bosonic interactions (with the
exception of the $\rho$ parameter).

If both fermionic and bosonic currents couple to the resonance, there
are interactions of type $\frac{1}{M^2}j_fj_V$ that also scale with
$1/M^2$.  These terms shift the effective oblique parameters
$S,T,U$~\cite{STU} (which are usually defined in the absence of
fermionic currents) and modify vector-boson pair production on top of
the usual triple-gauge couplings.  For this reason, the measurement of
vector-boson pair production, in particular at the ILC, is very
sensitive to a high-mass QCD-like techni-$\rho$ resonance.  The QCD
$\rho$ meson does have, in our operator basis, a sizable fermion-pair
coupling. 

However, in the present paper we are mainly concerned with adding
independent information via the observation of quartic vector-boson
interactions.  For this reason, we will assume below that fermionic
couplings do not play a role.  The concrete analyses described in the
following sections thus ignore fermion currents, and the scaling
properties of purely bosonic operators do apply.


\section{LEP Observables and ILC Prospects}
\label{sec:LEP}

\subsection{Oblique Corrections}
\label{sec:oblique}

The radiative corrections to the masses and couplings of the gauge
bosons can be largely absorbed into three parameters where several,
basically equivalent, parameterizations are used. The relation of the
\pS,\pT,\pUp parameterization \cite{STU,Erler:2004nh,PDG} and the coupling
constants of the effective Lagrangian are given in
Appendix~\ref{app:oblique}.  The observables that enter in the
determination of \pS, \pTp, and \pUp\ have already been measured with
good precision at LEP, SLD and at the Tevatron~\cite{lepew}.  \pTp\ is
given by the normalization of the $Zf\bar f$ vertex, $\Delta\rho$, and
is thus obtained from the partial widths of the $Z$ decaying into
fermions.  The asymmetries at LEP and SLD, which are the quantities
that are measured with the best precision, can all expressed in terms
of an effective weak mixing angle $\sin^2 \theta_{\rm eff}$ which is
given be a linear combination of \pSp\ and \pT. The third independent
observable that enters the determination of \pS, \pTp, and \pUp\ is
the $W$-mass which is measured at LEP II and at the Tevatron. It is
given by a linear combination of all three parameters. In many models
no deviation of \pUp\ from the Standard Model (defined as the
Higgs-less electroweak theory with zero anomalous couplings) is
expected so that often \pUp\ is fixed to its SM value.  Since the
$W$-mass depends differently on \pSp\ and \pTp\ than $\sin^2
\theta_{\rm eff}$, the inclusion of the $W$ mass in this case shrinks
the error of \pSp and \pTp significantly.

\pS,\pT,\pUp\ are defined with the SM expectation subtracted so that
\pS\ = \pT\ = \pU\ = 0 in the SM per definition. However the values of
the Higgs and top quark mass strongly affect the SM predictions so
that they have to be specified in any determination of \pS,\pT,\pU.
From the recent data from LEP, SLD and the Tevatron one obtains ($m_H
= 117 \,\GeV,\, m_t = 177 \,\GeV$):
\begin{eqnarray*}
  S & = & -0.13 \pm  0.10\\
  T & = & -0.17 \pm  0.12\\
  U & = & \phantom{-} 0.22 \pm  0.13
\end{eqnarray*}
corresponding to
\begin{eqnarray*}
\alpha_1 & = & \phantom{-} 0.0026 \pm 0.0020\\
\beta_1 & = & - 0.00062 \pm 0.00043 \\
\alpha_8 & = & - 0.0044 \pm 0.0026.
\end{eqnarray*}
If instead $m_H = 1 \,\TeV$ is used one has to add 
\begin{eqnarray*}
\delta \alpha_1 & = & + 0.0020\\
\delta \beta_1 & = &  + 0.00069\\
\delta \alpha_8 & = & - 0.0002.
\end{eqnarray*}
It should however be noted that the parameters are strongly correlated and for
$m_H = 1 \,\TeV$ the data are inconsistent with \pS = \pT = \pU = 0 to
more than $4.5 \,\sigma$.

At the ILC one may improve the measurement of the leptonic
width of the $Z$ in the GigaZ running mode by a factor two
\cite{gigaz}. The main improvement is however possible in the
measurement of the weak mixing angle from the left-right
asymmetry. Here a factor ten is possible. The single parameter errors
on $\alpha_1$ and $\beta_1$ only get smaller by a factor two to three
determined by the improvement of the leptonic width, however the
correlation between the two parameters increases so that the small
axis of the error ellipse shrinks by a factor 10.

The precision on the $W$ mass can be brought to $6\,\MeV$ by a scan of the
$W$-threshold region, improving the current error by a factor five
\cite{megaw}. This improves the error on $\alpha_8$ by a factor three,
again increasing the correlations.

In many models one has $\alpha_8\propto U=0$, so that the oblique
parameters are often fitted with this constraint. In this case also
the W-mass measurements influences the large axis of the $\alpha_1 -
\beta_1$ error ellipse so that the expected improvement from ILC is a
factor four to six. 


\subsection{Trilinear Gauge Couplings}
\label{sec:TGC}

The trilinear gauge couplings have been measured at LEP from W-pair production
with small contributions from other processes like single W production.
At LEP no beam polarization was available, preventing the separation of the WWZ
from the WW$\gamma$ couplings. For this reason in the analyses the so-called
$SU(2)$ relations have been applied:
\begin{eqnarray*}
\Delta \kappa_Z & = & \Delta g_1^Z - \Delta \kappa_\gamma \tan^2 \theta_W \\
\lambda_Z & = & \lambda_\gamma
\end{eqnarray*} 
which is equivalent to demanding $\alpha_9 - \alpha_8 = \alpha_2^\lambda = 0$.
The errors turn out to be about $2/16\pi^2$ for 
$\alpha_3$ and $\alpha_1^\lambda$ and $6/16\pi^2$ for $\alpha_2$.

At ILC, using beam polarization, all triple gauge couplings can be measured
separately with small correlations~\cite{menges}. If all $\alpha$s are fitted
simultaneously, all errors are well below $0.1/16\pi^2$, except for $\alpha_3$
and $\alpha_9$ where the error is slightly above with a large correlation
between the two. If $\alpha_9$ is fixed in the fit, the error on $\alpha_3$
gets as small as the others.


\section{Triple Vector-Boson Production at the ILC}
\label{sec:TVB}

\begin{table}[b]
\caption{\label{tab:cross} Cross section for triple boson production at
  $\sqrt{s}=1000\,\GeV$ for different initial state polarization.  (A) 
  unpolarized, (B) $80\%$R electrons, and (C) $80\%$R electrons with 
  $60\%$L positrons.} 
\[
\begin{array}{c|c|c||c}
\multicolumn{3}{c||}{\mbox{WWZ}}&\mbox{ZZZ}\\[1ex]
\hline\hline
\mbox{no pol.}&e^-\mbox{pol.}&\mbox{both pol.}&\mbox{no pol.}\\[1ex]
59.1\mbox{~fb}&12.3\mbox{~fb}&5.57\mbox{~fb}&0.79\mbox{~fb}
\end{array}
\]
\end{table}
We consider the reactions $e^+e^-\rightarrow W^+W^-Z$ and $e^+e^-\rightarrow
ZZZ$ that are sensitive to generic quartic gauge couplings. These are
parameterized in terms of the effective Lagrangians $\CL_i$
(\ref{eq:efflagrange}) with coupling parameters $\ga_i$, for $i =
4,5,6,7,10$.  The processes also depend on some of the lower-order
couplings that induce triple-gauge couplings and oblique corrections;
however, regarding the high accuracy of the corresponding measurements
at the ILC (cf.\ the previous section), we accept these as
pre-determined and set them to zero for the current analysis.  We
expect that real ILC data will be analyzed by a global fit of \emph{all}
electroweak parameters, including bilinear, trilinear, and quartic
couplings, but this is beyond the scope of the present paper.

In this and the following section we investigate the sensitivity of
future experiments at the ILC on the coupling constants $\ga_i$ and
thus, indirectly, on the masses of any new resonances in the EWSB
sector.
\begin{figure}[t]
\begin{minipage}{0.32\textwidth}
    \epsfig{figure=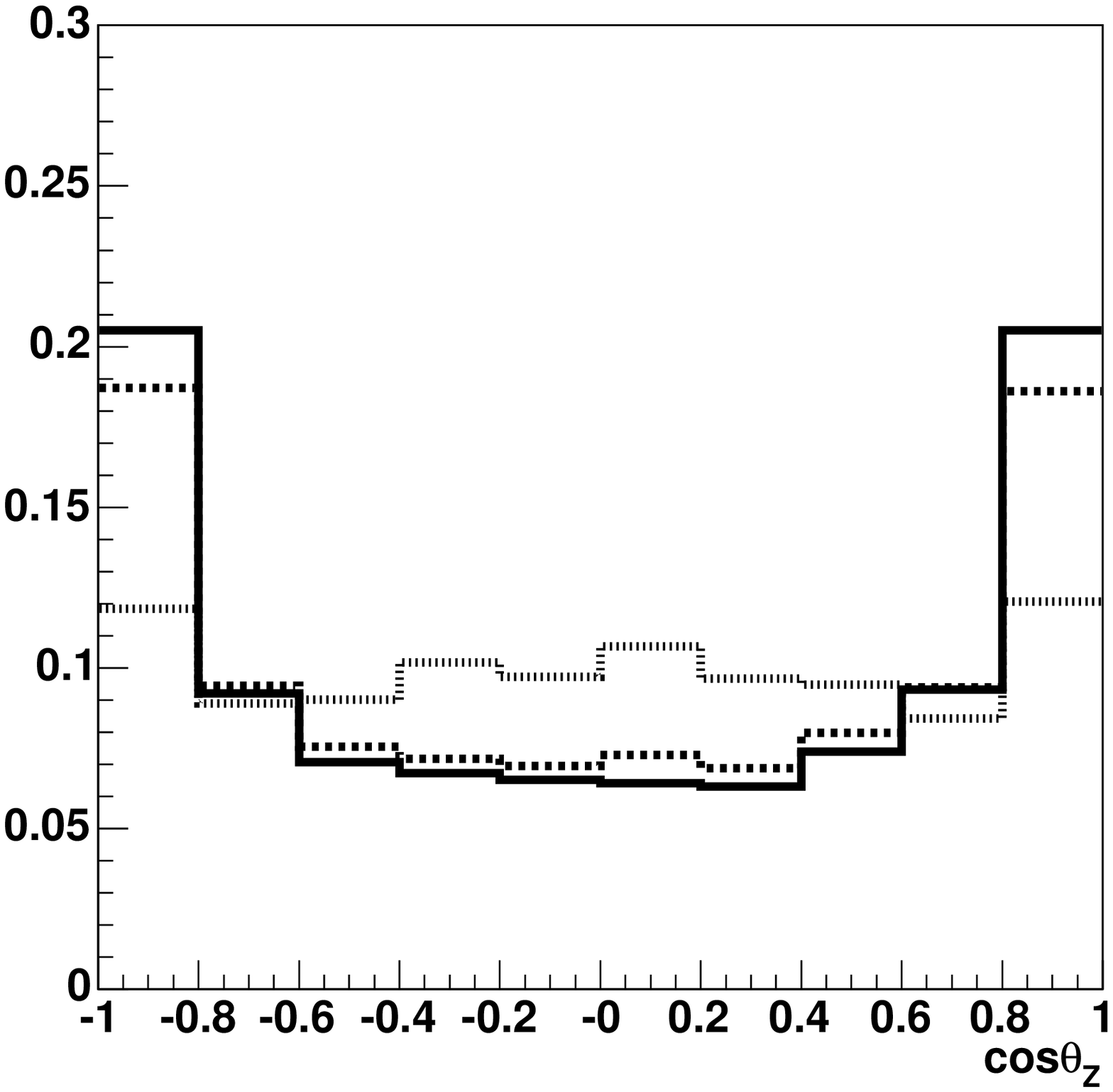,width=1.0\textwidth}
\end{minipage}\quad
\begin{minipage}{0.32\textwidth}
    \epsfig{figure=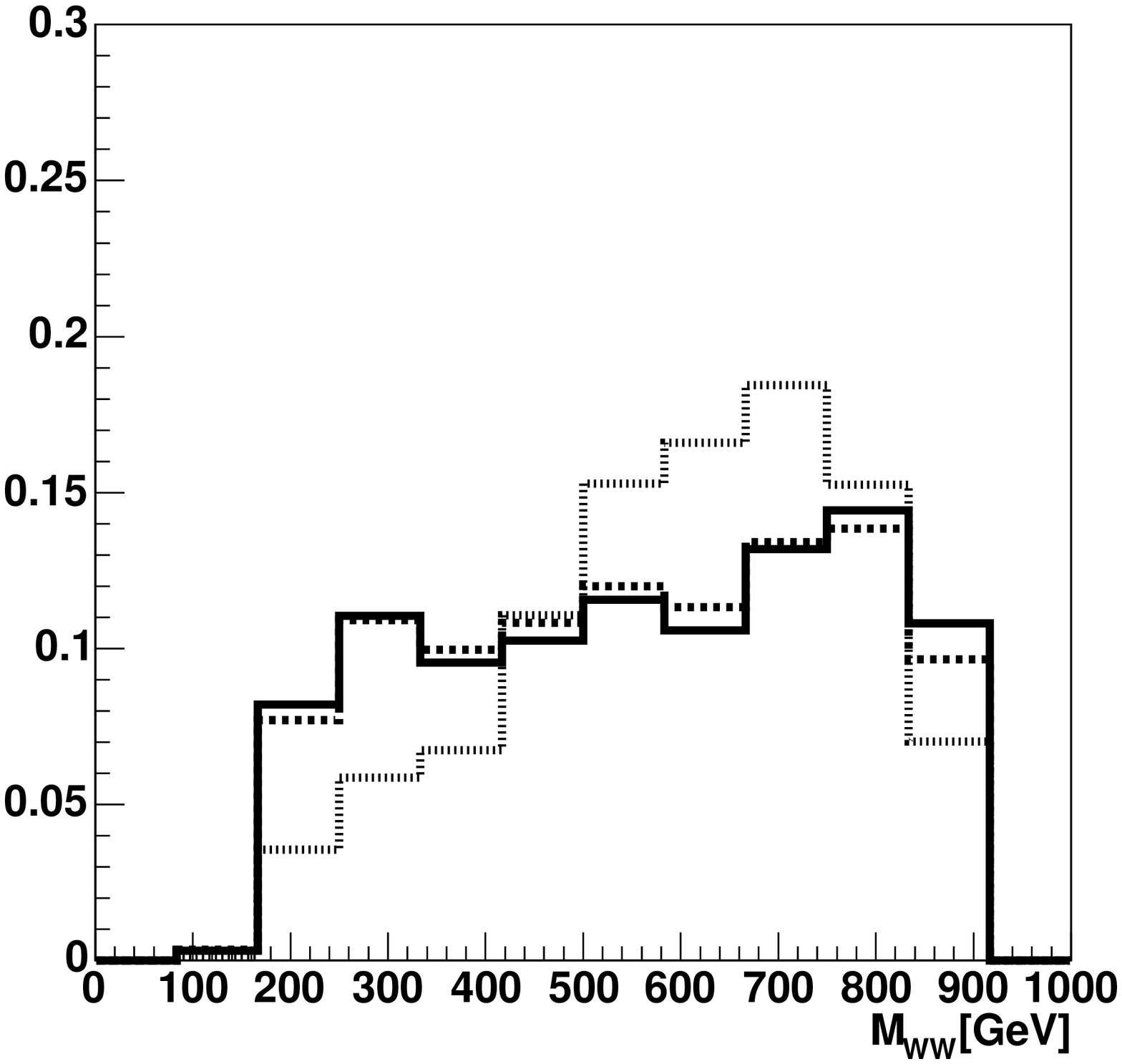,width=1.0\textwidth}
\end{minipage}\quad
\begin{minipage}{0.32\textwidth}
    \epsfig{figure=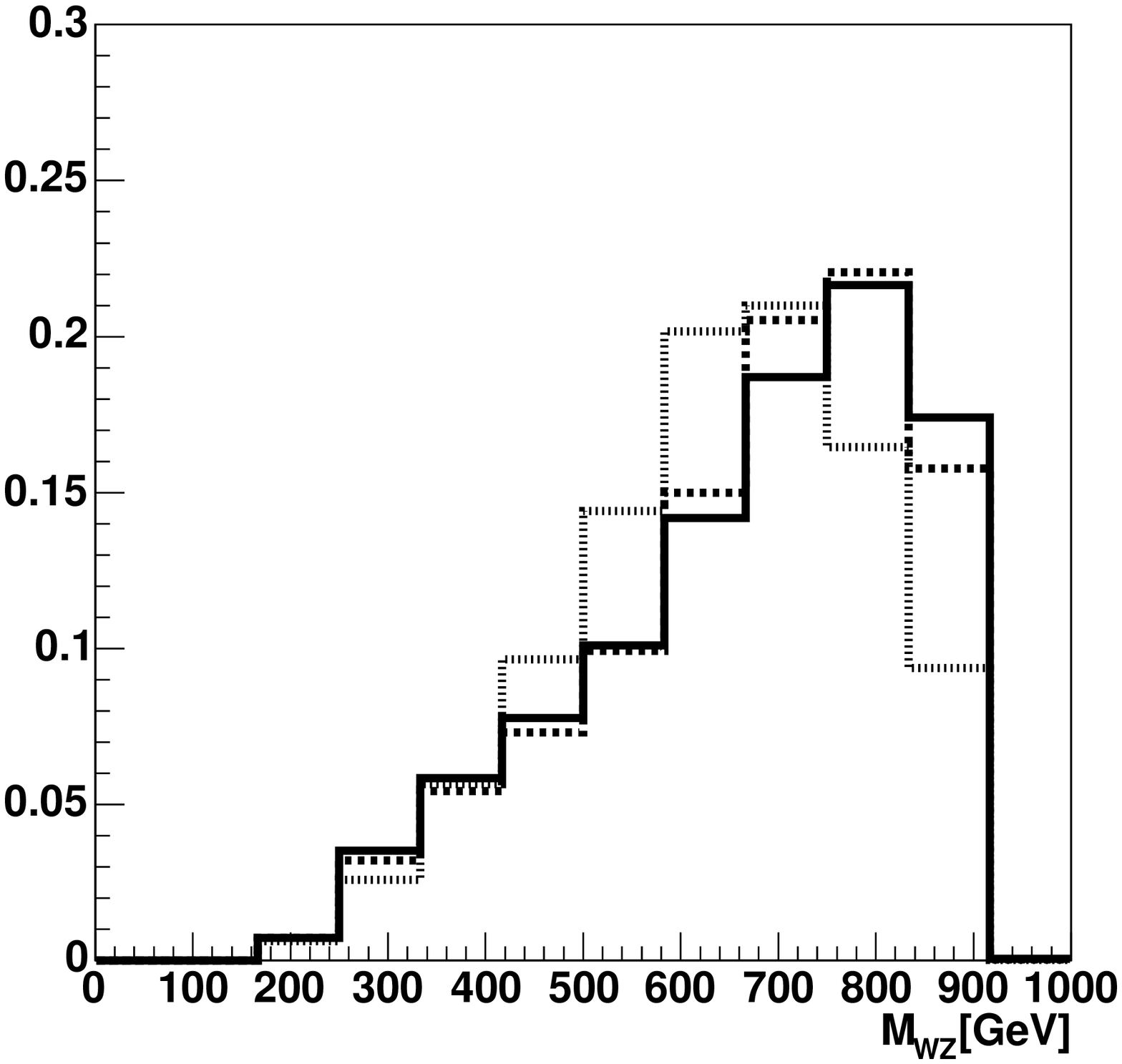,width=1.0\textwidth}
\end{minipage}
\caption{Reconstructed $\cos\theta$, $M_{WW}$, and  $M_{WZ}$  signal
 distributions for $e^+e^-\rightarrow WWZ$ and both beams polarized.
 To see the shape dependence the distributions are normalized to the
 respective total number of events for the Standard Model (solid),
 $\ga_4=1.6\pi^2 \approx 15.8$ (dashed) and $\ga_5 \approx 15.8$
 (dotted).} 
\label{fig:hist}
\end{figure}
In triple gauge-boson production processes not all anomalous couplings
can be disentangled individually.  The process $e^+e^-\rightarrow
W^+W^-Z$ depends on the $\alpha$ parameters in the two linear
combinations $\alpha_4+\alpha_6$ and $\alpha_5+\alpha_7$, while the
process $e^+e^-\rightarrow ZZZ$ depends on the single combination
$\alpha_4 + \alpha_5 + 2(\alpha_6 + \alpha_7 + \alpha_{10})$.  For the
study of triple gauge-boson production we concentrate on $\ga_4$ and
$\ga_5$ as independent couplings.

In the $WWZ$ final state the rate is dominated by a large SM
background that, however, can be substantially reduced using polarized
beams that enrich the relative appearance of longitudinal vector-boson
polarizations that are sensitive to the EWSB sector.  Hence, for $WWZ$
we investigate several running scenarios that are discussed for the
ILC: (A) unpolarized, (B) $80\%$ right-handed polarized electrons, and
(C) $80\%$ right-handed polarized electrons along with $60\%$
left-handed polarized positrons. For $ZZZ$ the SM background is much
smaller and polarization is not substantial. 

The total cross section at $\sqrt{s}=1000\;\GeV$ as calculated with
the event generator \whizard~\cite{Whizard} is given in
Table~\ref{tab:cross}.  The three-boson final state is characterized
by three four-momenta and the bosonic spins. If the bosonic spins are
not analyzed, only three kinematical variables are independent, as
follows from symmetry considerations and energy-momentum conservation.
We choose two invariant masses, $M_{WZ}^2=(p_W+p_Z)^2$,
$M_{WW}^2=(p_{W^+}+p_{W^-})^2$, and the angle $\theta$ between the
$e^-$ beam axis and the direction of the $Z$-boson.  The differential
cross section $d\sigma(M_{WW},M_{WZ},\cos\theta)$ is discretized into
bins denoted by $i,j,k$ for $M_{WZ}$, $M_{WW}$, and $\cos\theta$.
Assuming an integrated luminosity of $\int\CL=1000\;\fb^{-1}$, each bin
contains the number of events $N_{ijk}$ given by the differential
cross section.  We choose 10 bins for $\cos\theta\in[-1,1]$ and 12
bins for $M_{WZ}$ or $M_{WW}\in[0,1000]$ GeV. Since the effective
Lagrangian is linear in the anomalous couplings, $N^{\rm
theo}_{ijk}(\ga_4,\ga_5)$ is a polynomial of second order, namely
\begin{equation}
N^{\rm  theo}_{ijk}(\ga_4,\ga_5)=
N_{ijk}^{\rm sm}(1+R_{ijk}^{\rm A}\ga_4+
R_{ijk}^{\rm B}\ga_4^2+R_{ijk}^{\rm C}\ga_5+
R_{ijk}^{\rm D}\ga_5^2+R_{ijk}^{\rm E}\ga_4\ga_5)
\label{eqn:N_theo}
\end{equation}

\begin{figure}[t]
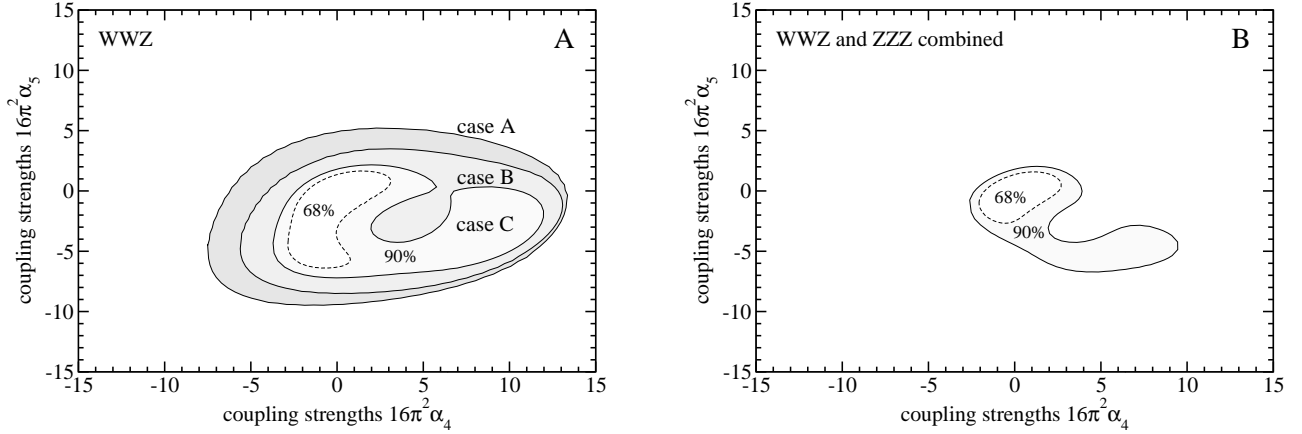

\begin{minipage}{0.47\textwidth}
    \epsfig{figure=chi2_wwz.eps,width=\textwidth}
\end{minipage}\hfill
\begin{minipage}{0.47\textwidth}
    \epsfig{figure=chi2_comb.eps,width=\textwidth}
\end{minipage}
\caption{Expected sensitivity for $\ga_4/\ga_6$ and $\ga_5/\ga_7$ at
  $\sqrt{s}=1000\,\GeV$.  Luminosity assumption 1000 fb$^{-1}$. A)
  $WWZ$-channel only, for an unpolarized beam (A) and the different
  polarizations cases, $e^-$ only polarized (B) and both beams
  polarized (C) as explained in the text. Solid lines represent $90\%$
  confidence level, the dashed line is for 68\%,
  i.e. $\Delta\chi^2=2.3$. B) Combined fit using $WWZ$ of case C and  
  $ZZZ$ production.  Lines represent $90\%$ (outer line), $68\%$
  (inner line) confidence level.}
\label{fig:result}
\end{figure}
The coefficients $R_{ijk}^{\mathrm{A} \dots\mathrm{E}}$ are determined
by reweighting, i.e., for five fixed pairs of anomalous couplings
$\ga_4$, $\ga_5$ we recalculate the respective weight, normalized to
the weight of the SM event.  By inversion, the relative weight $r_i$
of each event can then be written as an analytical function of the
anomalous couplings,
\begin{equation}
r_i = 1+ a_i \ga_4 + b_i \ga_4^2
+ c_i \ga_5 + d_i \ga_5^2 + e_i \ga_4\ga_5. 
\end{equation}
The new weights are
accumulated for each bin and finally lead to the coefficients in
(\ref{eqn:N_theo}). The kinematical variables are reconstructed as will be
explained below. Finally we calculate the $\chi^2$ contribution given
by  
\begin{equation}
\chi^2=\sum_{i,j,k}\frac{(N^{\rm exp}_{ijk}-N^{\rm theo}_{ijk}(\ga_4,\ga_5))^2}
{\gs_{ijk}^2} \quad , 
\end{equation}
where $\gs_{ijk}$ denotes the error, $i,j,k$ are the sums over bins of
$M_{WZ}$, $M_{WW}$, and $\cos\theta$. From the miminization of this
$\chi^2$ distribution we determine $\gD\ga_i$ (with all~$\ga_j=0$). The
final result is shown in Fig.~\ref{fig:result}. 

The simulation is done with the \whizard\ event
generator~\cite{Whizard} using the matrix-element generator
O'Mega~\cite{Omega,omwhiz} and the VAMP multi-channel phase space
integration package~\cite{Ohl:1998jn}.  For the study presented here,
we simulate on-shell gauge bosons and decay and hadronize the final
state using \pythia~\cite{Sjostrand:2001yu}.  Results gained from
extending this to full six-fermion matrix elements and spin
correlations will be presented in a future publication~\cite{Beyer}.
The detector is simulated using the fast simulation
\simdet~\cite{Pohl:2002vk}.  

We produce SM events corresponding to a luminosity of
$1000\;\fb^{-1}$.  Three-boson events are reconstructed via
six (hadronic) jets utilizing the YCLUS jet-finding algorithm with the
Durham recombination scheme. About 32\% of all $WWZ$ or $ZZZ$ decays
are purely hadronic. Other reconstruction channels are also possible
but presently not considered.  The dominant background is due to
$t\bar t\rightarrow b\bar b WW\rightarrow 6$ jets. We select events
with the kinematical conditions for a combination of missing energy
and transverse momentum
$E_\mathrm{mis}^2+p^2_{\perp,\mathrm{mis}}<(65\;\GeV)^2$ and
minimum jet energy $E_\mathrm{jet}^\mathrm{min}>5\;\GeV$. Two
jets are combined to form a $W$ or a $Z$ requiring
\begin{equation}
\begin{split}
-15\;\GeV &< m^\mathrm{cand} - m^\mathrm{true}_W < \delta M\\
-\delta M &< m^\mathrm{cand} - m^\mathrm{true}_Z< 15\;\GeV,
\end{split}
\end{equation}
where $\delta M= (m^\mathrm{true}_W + m^\mathrm{true}_Z)/2$, and
$m^{\rm true}$ is taken from the Particle Data Group (PDG)~\cite{PDG}.
Finally, we take the combination that minimizes the deviation from the
PDG values and do a kinematical fit of the bosonic momenta to the
total energy and momentum. The top quark is identified via a $b$-jet
that is combined with two jets from a $W$ candidate. Top-quark events
are vetoed if $ |m^\mathrm{cand}_t - m^\mathrm{true}_t|<15\;\GeV$ and
the events are consistent with the $t\bar t$ topology. The
reconstruction efficiency for $WWZ$ is about 12\%. This reflects about
36\% of all hadronic channels. The purity of the signal is about 98\%
(case A), 94\% (case B), 85\% (case C) of $WWZ$.  The reconstruction
efficiency of $ZZZ$ is about 8\%. The purity is 29\%, dominated
by the large $WWZ$ background. The reconstructed momenta are used to
determine the $\chi^2$. To minimize fluctuations in the sensitivity,
we increase statistics by factors of $5\dots100$ depending on the
process and renormalize $N_{ijk}$ accordingly. The contours and the
error intervals are calculated with \minuit~\cite{minuit}.

\begin{table}[t]
\caption{\label{tab:delta} Sensitivity of $\ga_4$ and $\ga_5$ expressed as
  $1\sigma$ errors. WWZ: two-parameter fit; ZZZ: one-parameter fit; best:
  best combination of both. }
\[
\begin{array}{c|c||r|r|r||r||r}
\hline
\multicolumn{2}{c||}{} &
\multicolumn{3}{c||}{\mbox{WWZ}}&\mbox{ZZZ}&\mbox{best}\\
\multicolumn{2}{c||}{} &
\mbox{no pol.}& e^-\mbox{ pol.}& \mbox{both pol.}&\mbox{no pol.}\\
\hline
16\pi^2\Delta\ga_4 &\gs^+&9.79  &4.21 &1.90 &3.94&1.78 \\
                   &\gs^-&-4.40  &-3.34 &-1.71 &-3.53&-1.48 \\
\hline
16\pi^2\Delta\ga_5 &\gs^+&3.05 &2.69  &1.17 &3.94&1.14 \\
                   &\gs^-&-7.10  &-6.40  &-2.19 &-3.53 &-1.64 \\
\hline
\end{array}
\]
\end{table}
Results are shown in Fig.~\ref{fig:result} and
Tab.~\ref{tab:delta}. For $WWZ$ we give in Fig.~\ref{fig:result}A the
90\% contours for the different polarization cases A, B, and C, and
for both beams polarized also the 68\% contour. The respective $\Delta
\ga_i$ are given in Tab.~\ref{tab:delta}. Note that for $WWZ$ two
parameters are independent, while for $ZZZ$ only one. Hence for $WWZ$
we get $\Delta\ga_4=\Delta\ga_6$, $\Delta\ga_5=\Delta\ga_7$, and no
sensitivity to $\ga_{10}$. For $ZZZ$ we have $\Delta\ga_4 =
\Delta\ga_5 = \frac12\Delta\ga_6 = \frac12\Delta\ga_7 =
\frac12\Delta\ga_{10}$. We find that the sensitivity strongly
increases with polarization, cf.\ the different cases A, B, and C. A
best combined fit for triple boson production is given in
Fig.~\ref{fig:result}B.  

The sensitivity could be further improved by using the information
provided by the angular distribution of jets, since the EWSB sector
mainly affects the longitudinal polarization directions of vector
bosons.  This will be covered in a future publication~\cite{Beyer}.


\section{Vector-Boson Scattering Processes at the ILC}
\label{sec:VBS}

In this section we consider those six-fermion processes in $e^+e^-$
and $e^-e^-$ collisions that depend on quartic gauge couplings via
quasi-elastic weak-boson scattering subprocesses, i.e., $VV\to VV$,
where $V=W^\pm,Z$.  We use full six-fermion matrix elements and thus
do not rely on simplifications such as the effective $W$
approximation, the Goldstone-boson equivalence theorem, or the
narrow-width approximation for vector bosons.

For the simulation we assume a c.m.\ energy of $1\;\TeV$ and a total
luminosity of $1000\;\fb^{-1}$ in the $e^+e^-$ mode. Beam polarization
of 80\% for electrons and 40\% for positrons is also assumed.  Since
the six-fermion processes under consideration contain contributions
from the triple weak-boson production processes considered in the
previous section ($ZZ$ or $W^+W^-$ with neutrinos of second and third
generation as well as a part of $\nu_e\bar \nu_eWW(ZZ)$, $e\nu_eWZ$
and $e^+e^-W^+W^-$ final states), there is no distinct separation of
signal and background. Signal processes in a separate analysis are
thus affected by all other signal processes as well as by pure
background. 

The present study extends the previous study~\cite{Chierici:2001ar}
which considered a restricted set of channels and parameters.  In
addition to the backgrounds considered there, we include single
weak-boson production in the background simulation for completeness.
We take initial-state radiation into account when generating
events. For the generation of $t\bar t$ events we use
\pythia~\cite{Sjostrand:2001yu}.  The event samples are
generated by the multi-purpose event generator
\oMega/\whizard~\cite{Omega,Whizard,omwhiz}, using exact six-fermion
tree-level matrix elements.  No flavor summation is necessary since
all possible quark final states are generated. Hadronization is done
with \pythia.  We use the \simdet~\cite{Pohl:2002vk} program to
produce the detector response of a possible ILC detector.

Table~\ref{l2ea4-t2} contains a summary of all generated processes
used for analysis and their corresponding cross sections. For pure
background processes a full $1\;\ab^{-1}$ sample is generated.  All
signal processes are generated with higher statistics. Single
weak-boson processes and $q{\bar q}$ events are generated with an
additional cut on $M(q{\bar q})> 130\;\GeV$ to reduce the number of
generated events.

\begin{table}
  \begin{center}
    \begin{tabular}{l|l|r}      
      \hline 
      Process & Subprocess & $\sigma\;[\fb]\quad$
      \\\hline\hline
      $e^+e^- \to \nu_e\bar \nu_e q \bar q q \bar q$ &
      $W^+W^- \to W^+W^-$ &
      23.19\z
      \\
      $e^+ e^- \to \nu_e \bar\nu_e q\bar qq\bar q$ &
      $W^+W^- \to ZZ$ &
      7.624 
      \\\hline 
      $e^+ e^- \to \nu \bar\nu q \bar qq\bar q$ &
      $V \to VVV$ &
      9.344 
      \\\hline 
      $e^+ e^- \to \nu e q\bar qq\bar q$ &
      $WZ \to WZ$ &
      132.3\z 
      \\
      $e^+ e^- \to e^+e^- q\bar q q \bar q$ &
      $ZZ \to ZZ$ &
      2.09\z   
      \\
      $e^+ e^- \to e^+ e^- q \bar q q \bar q$ &
      $ZZ \to W^+W^-$ &
      414.\z\z\z  
      \\\hline 
      $e^+ e^- \to b\bar b X$ &
      $e^+ e^- \to t \bar t$ & 
      331.768  
      \\\hline 
      $e^+ e^- \to q\bar q q\bar q$ &
      $e^+ e^- \to W^+ W^-$ &
      3560.108  
      \\
      $e^+ e^- \to q \bar q q \bar q$ &
      $e^+ e^- \to ZZ$ & 
      173.221  
      \\\hline 
      $e^+ e^- \to e\nu q \bar q$ &
      $e^+ e^- \to e\nu W$ & 
      279.588  
      \\
      $e^+ e^- \to e^+ e^- q \bar q$ &
      $e^+ e^- \to e^+ e^- Z$ & 
      134.935  
      \\\hline
      $e^+ e^- \to X$ &
      $e^+ e^- \to q\bar q$ &
      1637.405  \\ 
      \hline
    \end{tabular}    
    \caption{Generated processes and cross sections for signal and
    background for $\sqrt{s}=1\;\TeV$, polarization 80\% left for
    electron and 40\% right for positron beam.  For each process,
    those final-state flavor combinations are included that correspond
    to the indicated signal or background subprocess.}
    \label{l2ea4-t2}
  \end{center}  
\end{table}

The observables sensitive to the quartic couplings are the total cross
section (either reduction or increase depending on the interference
term in the amplitude and the point in parameter space), and
modification of the differential distributions in vector-boson
production angle and decay angle. This is not a full set of
observables, but some sensitive event variables, for example
transverse momentum, cannot be used since the contribution of
longitudinally polarized weak bosons is dropping faster than for
transversally polarized weak bosons with increasing transverse
momentum, and a transverse-momentum cut is unavoidable to suppress the
background in the analysis.

The event selection is done by a cut-based approach similar to
the previous analysis~\cite{Chierici:2001ar}. The general steps in the
analysis are the use of the final state $e^-$($e^+$) to tag background
(signal in $e{\nu}_{e}WZ$ case), a  cut on transverse momentum, and
\begin{table}
  \begin{center}
    \begin{tabular}{|l|l|c|c|c|c|c|}\hline 
      \textbf{$ {e^+}{e^-}\rightarrow $} & \textbf{ $
        {e^-}{e^-}\rightarrow $} 
      & \textbf{$ {\alpha}_4 $}
      & \textbf{${\alpha}_5 $}  & \textbf{${\alpha}_6 $}  & \textbf{$
        {\alpha}_7 $} 
      & \textbf{${\alpha}_{10} $}
      \\\hline 
      ${W^+}{W^-}\rightarrow {W^+}{W^-} $ & $ {W^-}{W^-}\rightarrow
        {W^-}{W^-} 
        $ & + & + & -  & -  & -   
        \\\hline 
        ${W^+}{W^-}\rightarrow ZZ $ &  & + & + & + & +  &  -
        \\\hline 
        ${W^{\pm}}Z\rightarrow {W^{\pm}}Z $ & ${W^-}Z\rightarrow
        {W^-}Z $ & + & + 
        & + & +  & -   
        \\\hline 
        $ ZZ\rightarrow ZZ $ & $ ZZ\rightarrow ZZ $ & + & + & + & +  &
        + 
        \\\hline
    \end{tabular}
    \caption{Sensitivity to quartic anomalous couplings for all
      quasi-elastic weak-boson scattering processes accessible at the
      ILC.  In addition to the $e^+e^-$ processes considered in this
      paper, we list the $e^-e^-$ processes for illustration.}
    \label{uzas}
  \end{center}
\end{table}
missing mass and energy. Realistic ZVTOP b-tagging~\cite{ztop} is used
whenever possible to improve the signal-to-background
separation. Finally, we apply cuts around the nominal masses of weak
bosons to accept only well-reconstructed events.

The extraction of quartic gauge couplings from reconstructed kinematic
variables is done by a binned likelihood fit. For each signal process,
we generate statistics much larger than the nominal $1000\;\fb^{-1}$
for $e^+e^-$ and pass the events through the detector simulation. Each
event is described by reconstructing four kinematic variables: the
event mass, the absolute value of production angle cosine, and the
absolute values of decay angle cosines for each reconstructed weak
boson. Only absolute value of the production and decay angles are used
since there is no possibility to resolve quark-antiquark and $W^+W^-$
ambiguities. 

Starting from an unweighted event sample as generated by \whizard, we
use the complete matrix elements encoded in the event generator itself
to reweight each event as a function of the quartic gauge
couplings. Each Monte-Carlo event is weighted by
\begin{equation}
  \label{eq:units1}
  R({\alpha}_i,{\alpha}_j)= 1 + A{{\alpha}_i} + B{{\alpha}_i}^2 +
  C{{\alpha}_j} + D{{\alpha}_j}^2 + E{{\alpha}_i}{{\alpha}_j}.
\end{equation}
The function $R({\alpha}_i,{\alpha}_j)$ describes the quadratic
dependence of the differential cross section on the anomalous
couplings.  It is obtained in the following way: using the generated
SM events (i.e., ${\alpha}_i\equiv 0$), we recalculate the matrix
element for each event at five different points in $\alpha_i,\alpha_j$
space and solve a set of linear equations for $A$,$B$,$C$,$D$ and
$E$. Due to the linear functional dependence of the
amplitude~\cite{ILClow} on the couplings, five points are enough to
determine the coefficients for the weighting function. The choice of
the points varies from process to process in order to fulfil the
following conditions: the distance of the point(s) from the SM value
should be large enough not to come into numerical instabilities when
solving the equations, and at the same time small enough not to come
into the region were phase space population would be significantly
different from the SM.

The obtained four-dimensional event distributions are fitted with
\minuit~\cite{minuit}, maximizing the likelihood as a function of
${\alpha}_i$,${\alpha}_j$ by taking the SM Monte-Carlo sample as
``data'':
\begin{equation}
  \label{eq:units2}
  L(\alpha_p,\alpha_q)=-\sum\limits_{i,j,k,l}{N^{SM}}(i,j,k,l)
  \ln\left( {N^{{\alpha}_p,{\alpha}_q}{}}(i,j,k,l)\right
  )+\sum\limits_{i,j,k,l} {N^{{\alpha}_p,{\alpha}_q}}(i,j,k,l)   
\end{equation}
where $i$ runs over the reconstructed event energy, $j$ over the
production angle, $k$ and $l$ over the decay
angles. ${N^{SM}}(i,j,k,l)$ are the ``data'' which correspond to the
SM Monte Carlo sample, and $ {N^{{\alpha}_p,{\alpha}_q}}(i,j,k,l) $ is
the sum of the same SM events in this bin, each reweighted by
$R({\alpha}_p,{\alpha}_q)$. Pure background events have
$R({\alpha}_p,{\alpha}_q)=1$, and for background coming from other
sensitive processes the proper weight is taken into account. After a
separate analysis for each process (see Table~\ref{uzas}), we perform
a combined fit. A small fraction of doubly-counted events that remains
after the single process analysis is uniquely assigned to one or
another set according to the distance from the nominal mass of the
weak boson pair (for example $WW$ or $ZZ$).


\section{Combined Results and Resonance Interpretation}
\label{sec:combine}

\begin{figure}
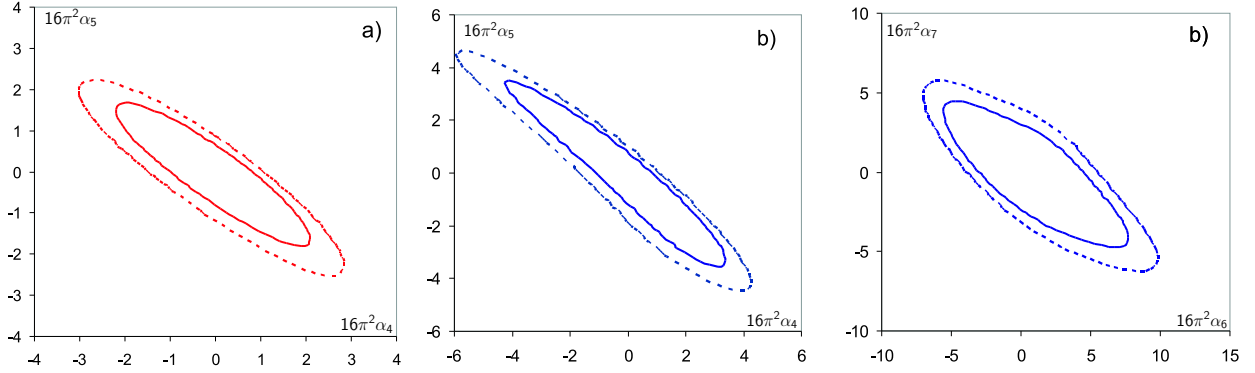

  \centering
  \includegraphics[width=55mm]{Backup_of_T1_epem_2D.eps}
  \includegraphics[width=110mm]{rezultat_zajedno_epem_5Da4a5.eps}
  \caption{Expected sensitivity (combined fit for all sensitive
  processes) to quartic anomalous couplings for a 1000 fb$\sp{-1}$
  $e^+e^-$ sample. The full line (inner one) represents 68\%, the dotted
  (outer) one 90\% confidence level.  a) conserved  $SU(2)_c$ case
  b) broken   $SU(2)_c$ case.} 
  \label{fig:kskksk} 
\end{figure}
%
%
\begin{table}[ht]
\begin{minipage}{0.45\textwidth}
\begin{center}
\begin{tabular}{|c|c|c|}

\hline \textbf{\ coupling \ } & \textbf{ $\sigma -$} & \textbf{$\sigma + $}\\

\hline \ $\alpha_4$ \ & \ -1.41 \  & \ 1.38 \ \\

\hline \ $\alpha_5$ \ & \ -1.16 \  & \ 1.09 \  \\

\hline

\end{tabular}
\caption{The expected sensitivity from 1000 fb$\sp{-1} e^+e^-$ sample
  at 1 TeV in the $SU(2)_c$ conserving case, positive and negative one
  sigma errors given separately.} 
\label{trtmrt}
\end{center}
\end{minipage}
%
\hspace{.25cm}
\begin{minipage}[t]{0.45\textwidth}
\begin{center}
\begin{tabular}{|c|c|c|}

\hline \textbf{\ coupling \ } & \textbf{ $\sigma -$} & \textbf{$\sigma + $}\\

\hline \ $\alpha_4$ \ & \ -2.72 \  & \ 2.37 \ \\

\hline \ $\alpha_5$ \ & \ -2.46 \  & \ 2.35 \  \\

\hline \ $\alpha_6$ \ & \ -3.93 \  & \ 5.53 \ \\

\hline \ $\alpha_7$ \ & \ -3.22 \  & \ 3.31 \  \\

\hline \ $\alpha_{10}$ \ & \ -5.55 \  & \ 4.55 \ \\

\hline

\end{tabular}
\caption{The expected sensitivity from 1000 fb$\sp{-1} e^+e^-$ sample
  at 1 TeV in the broken $SU(2)_c$ case, positive and negative 1
  sigma errors given separately.}  
\label{krk-t4}
\end{center}
\end{minipage}
\end{table}

%
In Table~\ref{trtmrt} and Table~\ref{krk-t4} we combine our results
for the measurement of anomalous electroweak couplings
for an integrated luminosity of $1000\;\fb^{-1}$ in the ${e^+}{e^-}$
mode, assuming $SU(2)_c$ conservation and non-conservation,
respectively.  In Fig.~\ref{fig:kskksk}, the results are displayed in
graphical form, projecting the multi-dimensional exclusion region in
$\alpha$ space around the reference point $\alpha_i\equiv 0$ onto
the two-dimensional subspaces $(\alpha_4,\alpha_5)$ and $(\alpha_6,\alpha_7)$

In order to get a more intuitive physical interpretation in terms of
a new-physics scale, in this section we transform anomalous couplings
into resonance parameters, as described in Sec.~\ref{sec:resonances}.
To this end, we also include the expected ILC results for triple gauge
couplings and oblique corrections in the fit.   Assuming one
particular resonance at a time, for each measured value of some $\alpha$
parameter, we may deduce the properties of the resonance that
would result in this particular value.  Inserting the values that
correspond to the sensitivity bound obtained by the experimental
analysis, we get a clear picture on the possible sensitivity to
resonance-like new physics in the high-energy region.

\subsection{$J=0$ Channel}

\subsubsection{Scalar Singlet: \boldmath$\sigma$}

(i) 
  We first consider the isospin conserving case, $h_{\sigma}=0$,
  which leads to $\alpha_7=\alpha_{10}=0$. Since
  $\alpha_4=\alpha_6=0$, there is only a dependence on ${\alpha}_5$ as a
  free parameter. After the fit, we get $\sigma_{\alpha_5}=0.42$ for
  the symmetric error or $-0.452<{\alpha}_5<0.397$ for the asymmetric
  ones at $1\sigma$ level.    

  \begin{figure}[t]
    \begin{center}
      \includegraphics[width=100mm]{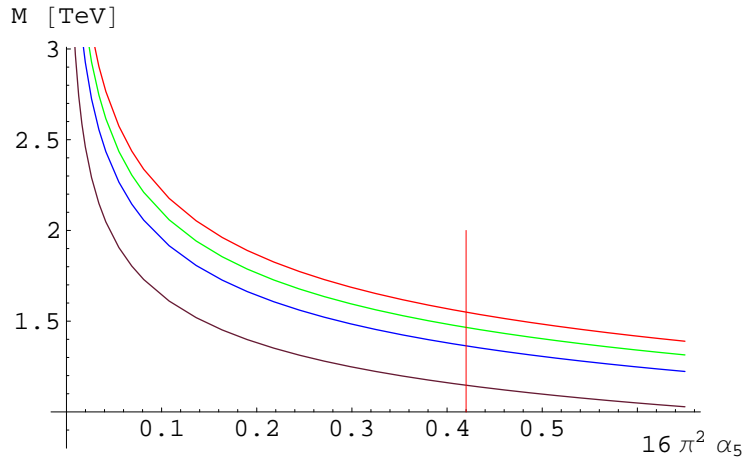}
      \caption{Mass of the scalar singlet resonance in the
      isospin-conserving case as a function of ${\alpha}_5$, with the
      resonance's width to mass ratio $f_\sigma$ equal to $1.0$ 
      in red, $0.8$ in green, $0.6$ in blue, and $0.3$ in brown,
      respectively. The vertical line in the plot is the $1\,\sigma$ limit 
      on ${\alpha}_5$.}  
      \label{fig:scalar_singlet}
    \end{center}
\end{figure}
  Expressing the width of the resonance as a fraction of its width,
  $\Gamma_\sigma = f_\sigma M_\sigma$, it is possible to solve
  Equ.~(\ref{eq:sigmawidth}) and $\alpha_5 = g_\sigma^2 \frac{v^2}{8
  M_\sigma^2}$ to obtain the resonance mass as a function of the
  quartic coupling and this fraction:  
  \begin{equation}
    \label{eq:masssigmacons}
    M_\sigma  = v  \left( \frac{4 \pi f_\sigma}{3 \alpha_5} \right)^{\frac14}
  \end{equation}
  In Fig.~\ref{fig:scalar_singlet}, we plot the mass of a scalar
  singlet resonance as a function of the coupling for a 
  given width. The vertical line in the plot is the $1\sigma$ error
  for a calculation of the mass from a given value of the width. 
  \begin{table}[b]
    \begin{center}
      \begin{tabular}{|c||c|c|c|c|}\hline
        $f_\sigma = \frac{\Gamma_\sigma}{M_\sigma}$ 
        & $1.0$ & $0.8$ & $0.6$ & $0.3$ \\\hline
        $M_\sigma \,[\TeV]$ 
        & $1.55$ & $1.46$ & $1.36$ & $1.15$ \\\hline
      \end{tabular}
      \label{tab:sigma}
      \caption{Mass reach for the scalar resonance in the
      SU(2)$\sb{c}$ conserving case depending on different
      resonance widths.} 
    \end{center}    
  \end{table}
  
\vspace{\baselineskip}\noindent
(ii)
  If we allow for isospin violation, $\alpha_4$ and $\alpha_6$ are
  still zero, leaving the three free parameters $\alpha_5$, $\alpha_7$
  and $\alpha_{10}$ for the fit. With only two independent variables,
  the system of equations (\ref{eq:sigmaalpha}), (\ref{eq:sigmawidth})
  is overconstrained with the additional relation  
  \begin{figure}[t]
    \includegraphics[width=80mm]{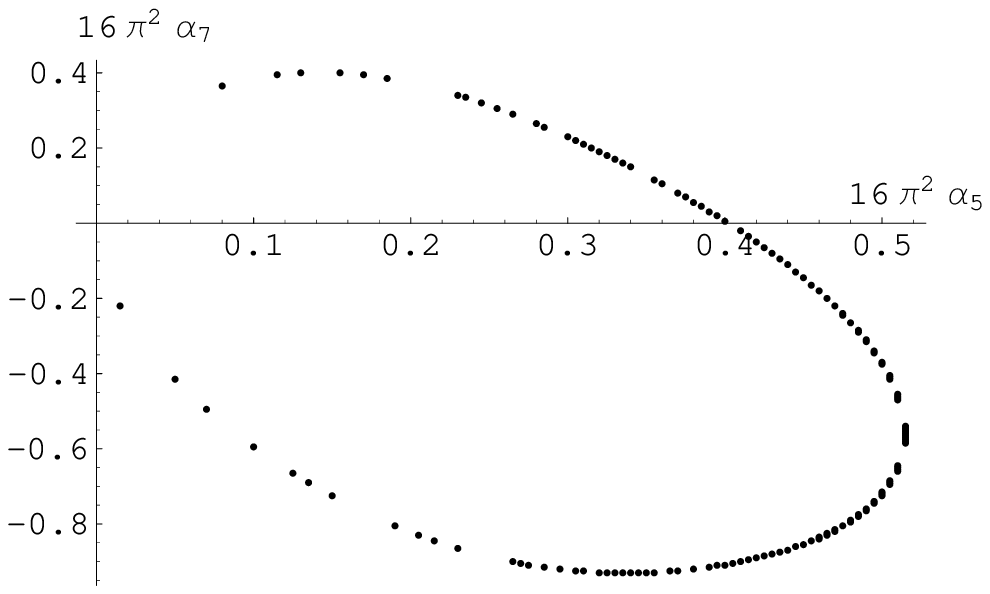}
    \includegraphics[width=80mm]{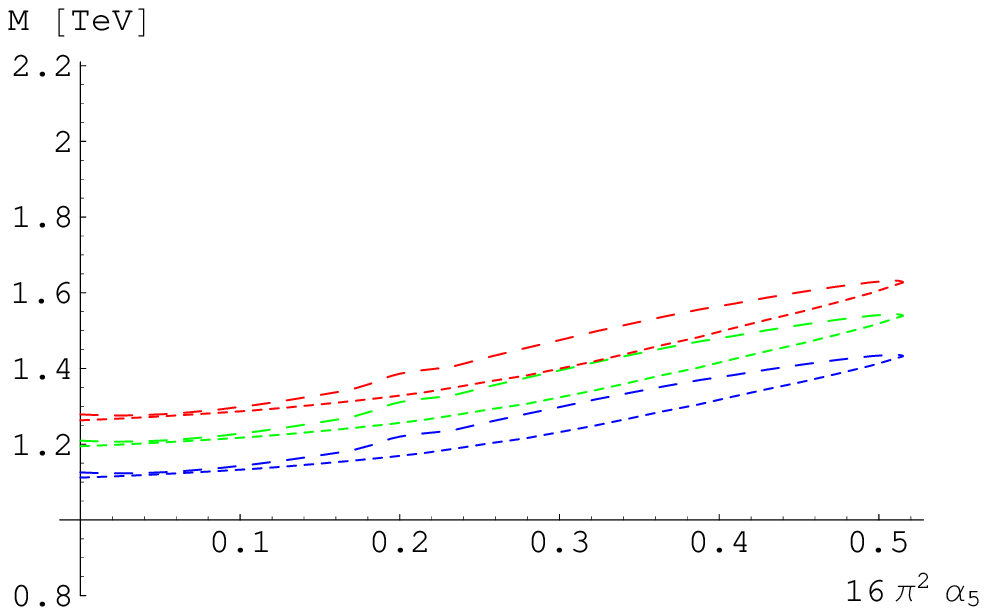}
    \caption{Scalar singlet with isospin breaking: On the left,
    $1\,\sigma$ contour in the ${\alpha}_5-{\alpha}_7$ plane. On the
    right, the dependence of the resonance mass on ${\alpha}_5$ along
    the contour for $f_\sigma = 1$ (width equal to the mass) in red,
    $f_\sigma = 0.8$ in green, and $f_\sigma = 0.6$ in blue. The
    dashed and wide-dashed lines correspond to the different branches
    of the solution of ${\alpha}_7=\mathcal{F}[\alpha_5]$,
    respectively.}     
    \label{fig:scalar_singlet_broke}
  \end{figure}

  \begin{equation}
    \label{eq:sigmarelation}
      \alpha_7^2 = 2 \alpha_5 \alpha_{10} .
  \end{equation}
  By this equation one is able to eliminate one of the couplings from
  further consideration. We will choose to eliminate
  $\alpha_{10}$. Solving the system of equations, it is now possible to
  express the mass as a function of the width, $\alpha_5$ and
  $\alpha_7$:   
  \begin{equation}
    \label{eq:masssigmanonc}
    M_\sigma = v \left( \frac{4 \pi \alpha_5 f_\sigma}{2 \alpha_5^2 +
    (\alpha_5 + \alpha_7)^2} \right)^{\frac14}.
  \end{equation}
  if we limit ourselves to the case that we vary the couplings only
  \begin{figure}
    \begin{center}
      \includegraphics[width=100mm]{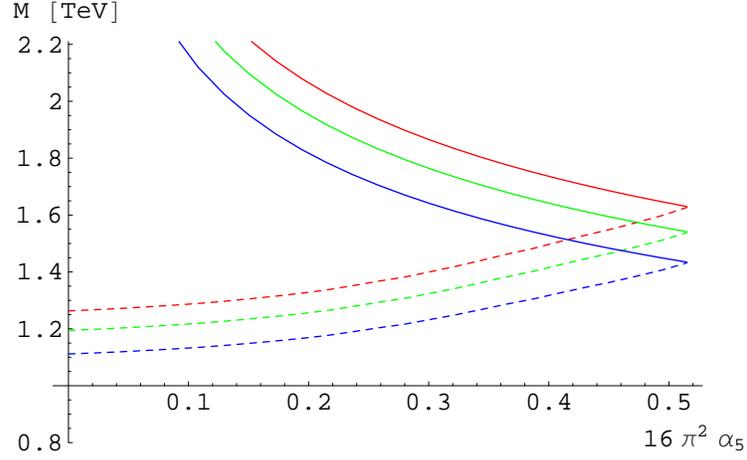}
      \caption{Allowed region for a scalar singlet resonance with
        isospin breaking as a function of ${\alpha}_5$ between the
        upper (full) and lower bound (dashed). Ratio of width to mass
        of the resonance equal to $1.0$ (red), $0.8$ (green), and $0.6$
        (blue), respectively.}    
      \label{fig:allowed_region_scalar_singlet}
    \end{center}
  \end{figure}
  along the $1\sigma$ contour in the $\alpha_5$,$\alpha_7$ plane, we
  end up with the result shown in Fig.~\ref{fig:scalar_singlet_broke}.
  \begin{table}
    \begin{center}
      \begin{tabular}{|c||c|c|c|c|}\hline
        $f_\sigma = \frac{\Gamma_\sigma}{M_\sigma}$ 
        & $1.0$ & $0.8$ & $0.6$ \\\hline
        $M_\sigma \,[\TeV]$ 
        & $1.39$ & $1.32$ & $1.23$ \\\hline
      \end{tabular}
      \label{tab:skalar_broken}
      \caption{Mass reach for the scalar singlet resonance in the case
      of isospin breaking depending on the width to mass ratio
      $f_\sigma$. We use an average mass along the contour.} 
      \end{center}     
  \end{table}
  The lower of the two dashed curves gives a lower limit on the
  allowed region. On the other hand, one can look for the maximum mass
  for the $\alpha$ parameters within the $1 \sigma$-contour. This is
  equivalent to minimizing the width (\ref{eq:sigmawidth}) by
  $g_\sigma = - 2 h_\sigma$, which yields a maximum mass (depending on
  the allowed $\alpha$ parameters) for $\alpha_5 = - \alpha_7$ of 
  \begin{equation}
    M_\sigma = v \left( \frac{2 \pi f_\sigma}{\alpha_5}
    \right)^{\frac14}.
  \end{equation}
  Fig. \ref{fig:allowed_region_scalar_singlet} shows the allowed
  region for different values of the width to mass ratio of the scalar
  singlet resonance. This means that a scalar singlet resonance
  corresponding to measured $\alpha_{5,7}$ values lying in the
  simulated $1\,\sigma$ contour cannot be heavier than the given
  upper limit.

  
\subsubsection{Scalar Triplet: \boldmath$\pi$}

(i)
    In principle, for a scalar triplet, there is no isospin-conserving
    limit, since $SU(2)$ breaking is necessary to couple a triplet to
    two identical bosonic triplets. Nevertheless, there is a case,
    where $h_\pi = k_\pi = 0$ and only $h'_\pi \neq 0$. In that case, the SM
    fields couple in an $SU(2)$ invariant way, and the isospin
    breaking resides in the coupling of the new
    resonances only. Therefore, a resonance with such a coupling would --
    to the order we are considering -- leave no trace in the
    isospin-breaking operators, but only gives a contribution to
    $\LL_5$. So experimentally, the isospin breaking is not detectable
    without direct access to the heavy resonances. Hence, again at
    leading order, there is no contribution to the width of the
    $\pi^\pm$ from electroweak gauge bosons. In this case, we regain
    the formula for the singlet case,
    \begin{equation}
      \label{eq:masspicons}
      M_{\pi^0} = v \left( \frac{  4 \pi f_{\pi^0}}{3 \alpha_5}
      \right)^{\frac14} ,
    \end{equation}
    while the charged resonance is not accessible in gauge boson
    scattering. Such a case would experimentally be indistinguishable
    from a singlet scalar. The bounds from the isospin-conserving
    singlet case also apply here. 

\vspace{\baselineskip}\noindent
(ii)
    If we allow for general isospin-breaking couplings, from
    Equ.~(\ref{eq:pialpha}) we get the constraint 
    \begin{equation}
      \label{eq:pirelation}
      \alpha_7^2 = 2 \alpha_5 (\alpha_6 + \alpha_{10}),
    \end{equation}
    which is the generalization of the singlet case. The
    correspondence between the singlet and the triplet discussed above
    goes even further. If we put $h_\pi$ to zero, the formula for the
    $\alpha$s as well as for the width between singlet and triplet
    correspond to each other with the identification: $g_\sigma
    \leftrightarrow h'_\pi$, $h_\sigma \leftrightarrow k_\pi$. In that
    case, $\alpha_6$ is zero, the charged resonance decouples at
    leading order, the two relations (\ref{eq:sigmarelation}) and
    (\ref{eq:pirelation}) are then identical, as are the formula for
    the mass of the neutral resonance as a function of
    $\alpha_{5,7}$. So for the case $h_\pi = 0$ we can reuse the
    results from the fit for the isospin-breaking singlet. 

    Allowing finally also for nonvanishing $h_\pi$ (i.e. nonvanishing
    $\alpha_6$), we again use the overconstraining to eliminate
    $\alpha_{10}$. Solving the remaining system we get: 
    \begin{subequations}
      \begin{align}
        \label{eq:masspinonc}
        M_{\pi^\pm} &=\; v \left( \frac{4 \pi f_{\pi^\pm}}{\alpha_6}
        \right)^{\frac14} \\
        M_{\pi^0}   &=\; v \left( \frac{4 \pi \alpha_5 f_{\pi^0}}{2
        \alpha_5^2 + (\alpha_5  + \alpha_7)^2}
        \right)^{\frac14} 
      \end{align}
    \end{subequations}    
    The formula for the neutral component still is the same as for the
    nonconserving singlet case, which can be understood by using the
    \begin{figure}[t]
      \begin{center}
        \includegraphics[width=80mm]{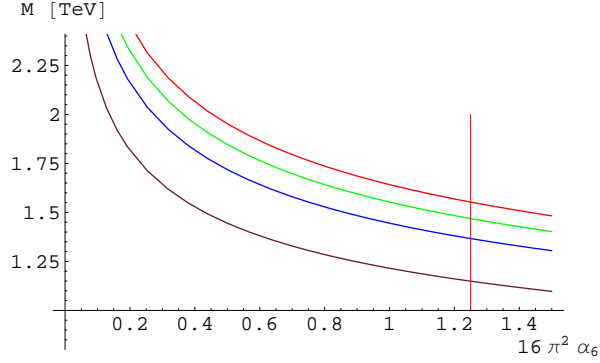}
        \caption{Dependence of the resonance mass for the charged
        scalar triplet component on ${\alpha}_6$ for different assumed
        width to mass ratios ($f_\pi = \Gamma_\pi / M_\pi = 1.0$ in
        red, $0.8$ in blue, $0.3$ in brown, respectively). The red
        vertical line represents the maximal value of ${\alpha}_6$
        along the $1\,\sigma$ surface.}    
        \label{fig:scalar_triplet_ch}
      \end{center}
    \end{figure}
    above correspondence and the replacing $2 k_\pi \to 2 k_\pi +
    h_\pi$ in the formulas for the $\alpha$s and the width. So the
    limits for the neutral component remain the same, and figures
    \ref{fig:scalar_singlet_broke} and
    \ref{fig:allowed_region_scalar_singlet} are also applicable
    here. The mass reach for the scalar triplet in the isospin
    breaking case is given in table \ref{scalar_triplet_table}. 
    \begin{table}[b]
      \begin{center}
        \begin{tabular}{|c||c|c|c|c|}\hline          
          $f_\pi = \frac{\Gamma_\pi}{M_\pi}$
          & $1.0$ & $0.8$ & $0.6$ & $0.3$ \\\hline
          $M_{\pi^0}\,[\TeV]$
          & $1.39$ & $1.32$ & $1.23$ & --- \\\hline
          $M_{\pi^\pm}\,[\TeV]$
          & $1.55$ & $1.47$ & $1.37$ & $1.15$ \\\hline
        \end{tabular}
      \end{center}
      \caption{Dependence of the mass reach for scalar triplet
        resonances on different resonance widths. For the neutral
        narrow state the mass reach is already below $1\,\TeV$.}        
      \label{scalar_triplet_table}
    \end{table}  

    A technical remark: $\alpha_5$ and $\alpha_6$ must be positive in
    order to get real solutions for the mass. The solutions for the
    mass decouple ${\alpha}_6$ and ${\alpha}_5, {\alpha}_7$ from each
    other, but we can still use the error matrix and the relation
    $\mathcal{F}({\alpha}_5,{\alpha}_6,{\alpha}_7)=1$ to fix the
    points on the $1\,\sigma$ surface.


\subsubsection{Scalar Quintet: \boldmath$\phi$}

(i)
    For isospin conservation, only $g_\phi$ and hence only $\alpha_4$
    is non-vanishing. Solving the system (\ref{eq:phialpha}),
    (\ref{eq:phiwidth}) yields 
    \begin{equation}
      \label{eq:phimasscons}
      M_\phi = v \left( \frac{4 \pi f_\phi}{\alpha_4}
      \right)^{\frac14} 
    \end{equation}    
    \begin{figure}[t]
      \includegraphics[width=80mm]{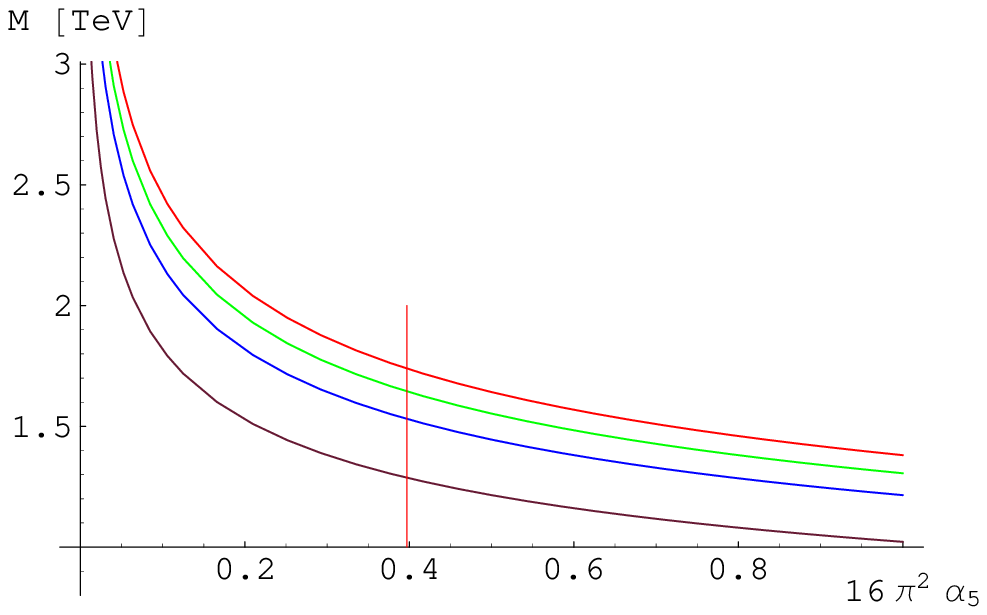}
      \includegraphics[width=80mm]{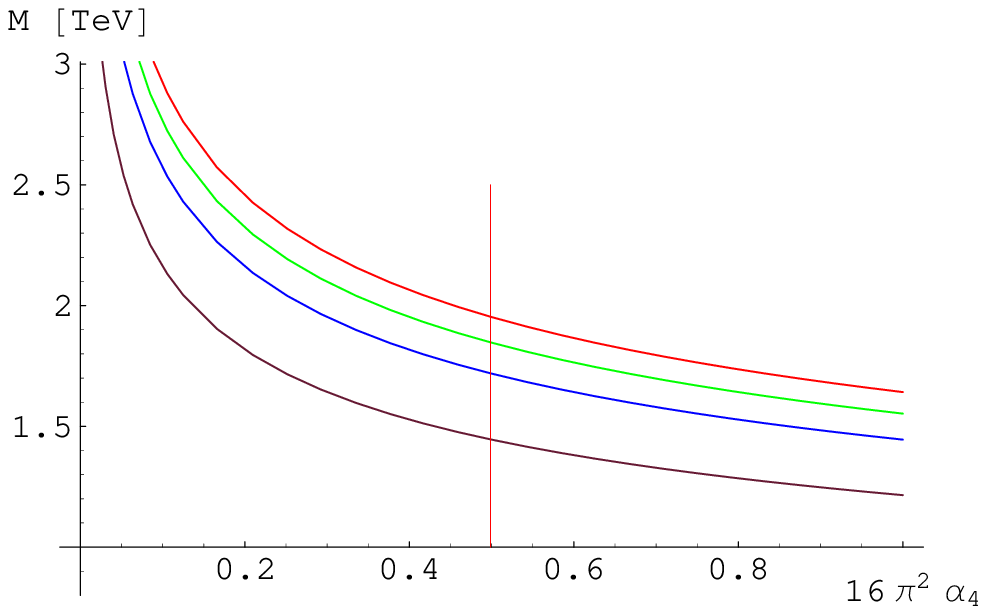}
      \caption{
        Dependence of the resonance mass for the scalar quintet 
        on the $\alpha$ parameters for different width to mass ratios,
        $f_\phi = \Gamma_\phi / M_\phi = 1.0$ (red), $0.8$ (blue),
        $0.6$ (green), and $0.3$ (brown), respectively. On the left: 
        Isospin-conserving case, degenerate mass of the whole
        multiplet as a function of $\alpha_4$.  On the right: Special
        isospin-breaking case with only $h_\phi'$ different from
        zero. Mass of the neutral component as a function of
        $\alpha_5$. The vertical red line represents the $1\,\sigma$
        limit for $\alpha_4$ and $\alpha_5$, respectively.}         
      \label{fig:scalar_quintet_cons}
    \end{figure}
    \begin{table}[b]
      \begin{center}
        \begin{tabular}{|c||c|c|c|c|}\hline
          $f_\phi = \frac{\Gamma_\phi}{M_\phi}$
          & $1.0$ & $0.8$ & $0.6$ & $0.3$ \\\hline
          $M_\phi\,[\TeV]$
          & $1.95$ & $1.85$ & $1.72$ & $1.45$ \\\hline
        \end{tabular} \qquad
        \begin{tabular}{|c||c|c|c|c|}\hline
          $f_{\phi^0} = \frac{\Gamma_{\phi^0}}{M_{\phi^0}}$
          & $1.0$ & $0.8$ & $0.6$ & $0.3$ \\\hline
          $M_{\phi^0}\,[\TeV]$
          & $2.06$ & $1.96$ & $1.82$ & $1.53$ \\\hline
        \end{tabular}
      \end{center}
      \caption{
        Mass reach for the scalar quintet depending on
        different ratios of width to mass. On the left, the 
        $SU(2)_c$ conserving case, on the right the special
        isospin-breaking case with only $h'_\phi \neq 0$.} 
      \label{scalar_quintet_table}
    \end{table}
    The results for the isospin-conserving case are shown on the left
    of Fig.~\ref{fig:scalar_quintet_cons} and
    Table~\ref{scalar_quintet_table}.  

\vspace{\baselineskip}\noindent
(ii)
      For the case of broken isospin symmetry, we first consider the
      case that only $h'_\phi \neq 0$, so that only $\alpha_5$ is
      nonvanishing. The charged and doubly-charged resonances do not
      get a contribution to their width at leading order, while
      solving for the mass of the neutral state results in
      \begin{equation}
        M_{\phi^0} = v \left( \frac {2\pi f_{\phi^0}}{\alpha_5}
        \right)^{\frac14} . 
      \end{equation}
      The fit and the $1\,\sigma$ reach are shown in the right plot of
      Fig.~\ref{fig:scalar_quintet_cons} and
      Table~\ref{scalar_quintet_table} also on the right. 

      There is a further special case when $\alpha_4 = - \alpha_6$, in
      which the charged resonance does not get a contribution to the
      width. Here also a singularity for the neutral state appears
      where the denominator for the mass of the neutral state
      vanishes. We ignore this case here.

      The next step is that we allow for nonzero $g_\phi$ {\em and}
      $h'_\phi$, which results in non-zero $\alpha_4, \alpha_5$ and
      \begin{figure}
        \begin{center}
          \includegraphics[width=80mm]{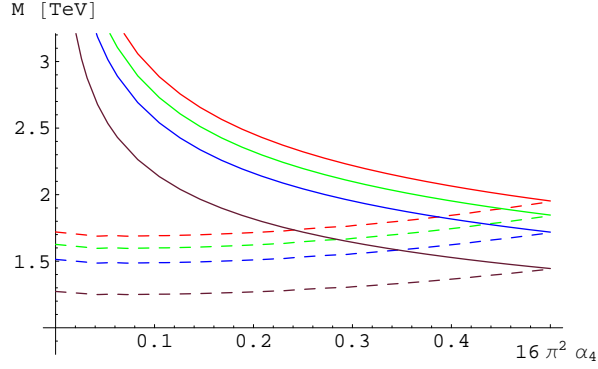}
          \caption{Dependence of the mass of the neutral component of
            a scalar quintet resonance for the case with only $g_\phi,
            h'_\phi$ non-zero, as a function of $\alpha_4$ for
            different width to mass ratios: in red $f_\phi =
            \Gamma_\phi / M_\phi = 1.0$, in green $0.8$, in blue 
            $0.6$, and in brown $0.3$. The red vertical 
            line represents the maximal value of $\alpha_4$.   }  
          \label{fig:scalar_quintet_b}
          \end{center}
      \end{figure}
      \begin{table}
        \begin{center}
          \begin{tabular}{|c||c|c|c|c|}\hline
            $f_\phi = \frac{\Gamma_\phi}{M_\phi}$
            & $1.0$ & $0.8$ & $0.6$ & $0.3$ \\\hline
            $M_{\phi0}\,[\TeV]$	
            & $1.77$ & $1.67$ & $1.55$ & $1.31$ \\\hline
          \end{tabular}
        \end{center}
        \caption{Mass reach for the neutral component of the scalar
          quintet in the case with only $g_\phi, h'_\phi$ non-zero, 
          depending on different width to mass ratios.}  
        \label{scalar_quintet_table_gh}
      \end{table}
      $\alpha_7$ with the constraint $\alpha^2_7 = \alpha_5
      \alpha_4$. Note that the term proportional to $g h'$ in 
      the width of the neutral state cancels out -- and hence the
      dependence on $\alpha_7$. In this special case, the formula for
      the masses of the charged and doubly-charged states remains the
      same as Equ.~(\ref{eq:phimasscons}). The mass reach equals the
      isospin-conserving case. In principle, isospin non-conservation
      can be detected by the different width of the neutral state. The
      solution for that component becomes     
      \begin{equation}
        \label{eq:phimasscons2}
        M_{\phi^0} = v \left( \frac{4 \pi f_{\phi^0}}{\alpha_4 + 2
        \alpha_5} \right)^{\frac14},
      \end{equation}    
      with constraints ${\alpha}_4>0$ and ${\alpha}_5>0$.

      For the completely general case of isospin breaking, the
      relation between the couplings is now a generalization of the
      triplet case, namely  
      \begin{figure}
        \begin{center}
        \includegraphics[width=80mm]{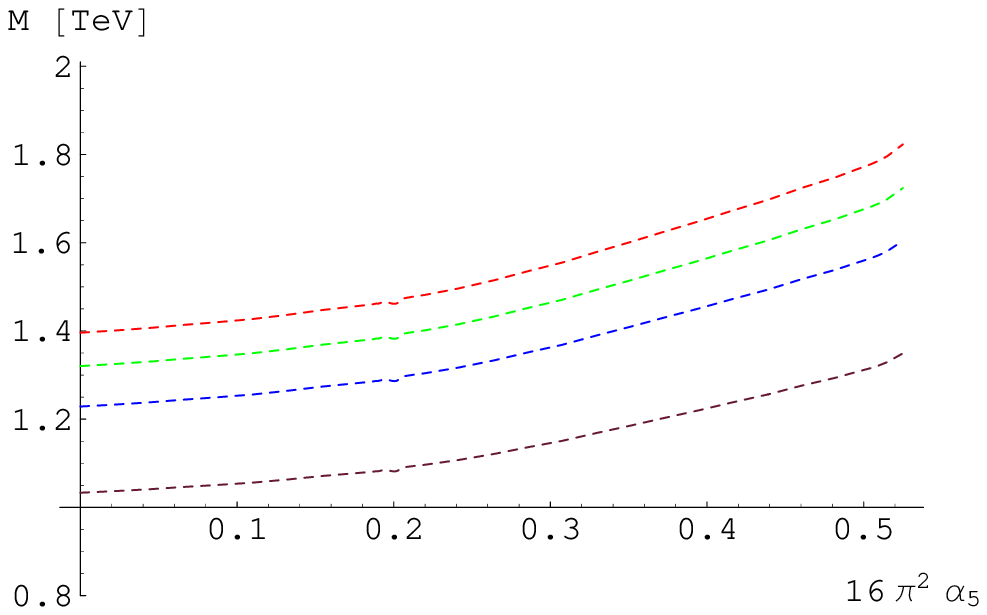}
        \includegraphics[width=80mm]{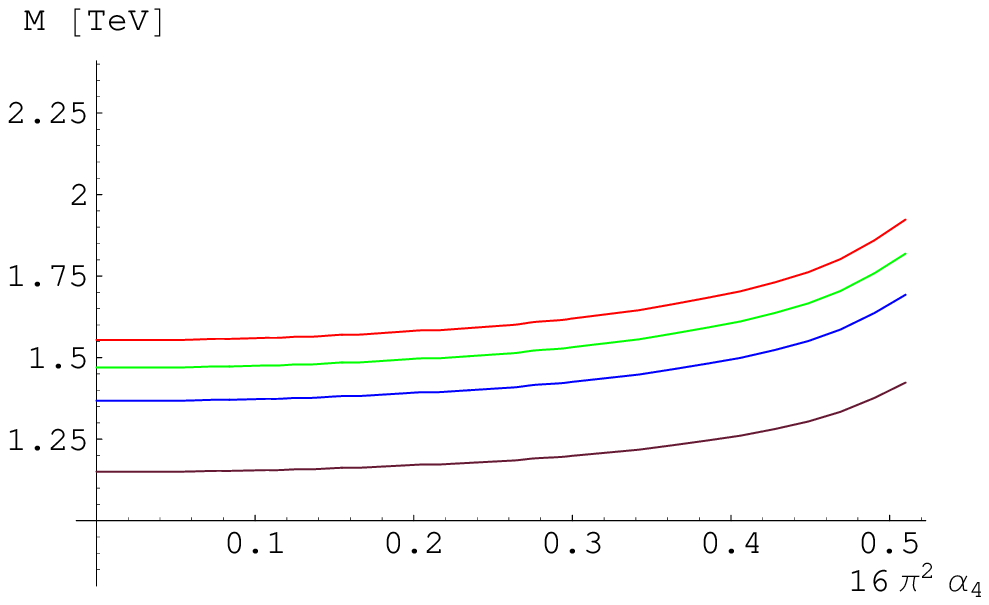}
        \caption{Dependence of the mass of the neutral (left) and the
            charged component (right) of a scalar quintet resonance
            in the completely general case, as a function of $\alpha_4$ for
            different width to mass ratios: in red $f_\phi =
            \Gamma_\phi / M_\phi = 1.0$, in green $0.8$, in blue 
            $0.6$, and in brown $0.3$. The red vertical 
            line represents the maximal value of $\alpha_4$. The
            doubly-charged component remains the same as in the
            isospin-conserving case, shown on the left of
            Fig.~\ref{fig:scalar_quintet_cons}.}
        \label{fig:scalar_quintet_full}
        \end{center}
      \end{figure}
      \begin{table}
        \begin{center}
          \begin{tabular}{|c||c|c|c|c|}\hline
            $f_\phi = \frac{\Gamma_\phi}{M_\phi}$
            & $1.0$ & $0.8$ & $0.6$ & $0.3$ \\\hline
            $M_{\phi\pm\pm}\,[\TeV]$
            & $1.95$ & $1.85$ & $1.72$ & $1.45$ \\\hline
            $M_{\phi\pm}\,[\TeV]$
            & $1.64$ & $1.55$ & $1.44$ & $1.21$ \\\hline
            $M_{\phi0}\,[\TeV]$
            & $1.55$ & $1.46$ & $1.35$ & $1.14$ \\\hline
          \end{tabular}
        \end{center}
        \label{scalar_quintet_table_full}
        \caption{Mass reach for the scalar quintet in most general
          case.$M_{\phi0}$ and $M_{\phi\pm}$ are averages over the lower 
          limit curves.}
      \end{table}      
      \begin{equation}
        \alpha_7^2 = 2 \alpha_5 \left( \frac12 \alpha_4 + \alpha_6 +
        \alpha_{10} \right),  
      \end{equation}
      to again eliminate $\alpha_{10}$. 

      We obtain the solution for the masses  
      \begin{subequations}
        \begin{align}
          \label{eq:phimassnonc}
          M_{\phi^{\pm\pm}} &=\; v \left( \frac{4\pi
              f_{\phi^{\pm\pm}}}{\alpha_4} \right)^{\frac14} 
          \\ 
          M_{\phi^\pm} &=\; v \left( \frac{4\pi
            f_{\phi^\pm}}{\alpha_4 + \alpha_6} \right)^{\frac14} \\ 
          M_{\phi^0} &=\; v \left( \frac{12\pi\alpha_5
            f_{\phi^0}}{(\sqrt{\alpha_4\alpha_5} - 2 \alpha_5)^2 + 2
              (\alpha_7 + \alpha_5)^2}\right)^{\frac14} \notag \\ 
          & \;= 
            v \left( \frac{12 \pi f_{\phi^0}}{(\sqrt{\alpha_4} - 2
              \sqrt{\alpha_5})^2 + 2 (\sqrt{\alpha_4 + 2 \alpha_6 + 2
              \alpha_{10}} + \sqrt{\alpha_5})^2} \right)^{\frac14}
        \end{align}
      \end{subequations}

      For the neutral state, the first formula is in correspondence to
      those for the non-isospin conserving singlet and triplet case,
      while the second one is better suited for taking the limit to
      the isospin-conserving case.


\subsection{$J=1$ Channel}
\subsubsection{Vector Singlet: \boldmath$\omega$}

(ii)
    For the vector singlet, isospin breaking has to be
    involved. Concerning the analysis and the fit, we ignore the
    parameter $k_\omega$ because it has no physical meaning in terms
    of the resonance mass and width, at least in the order we are
    considering. So all eight nonvanishing $\alpha$ parameters are the
    same, 
    \begin{equation}
      \alpha_1 = \alpha_2 = \alpha_4 = \alpha_7 = - \alpha_5 =
      -\alpha_6 = -\alpha_8 = -\alpha_9
    \end{equation}
    \begin{figure}
      \begin{center}
        \includegraphics[width=80mm]{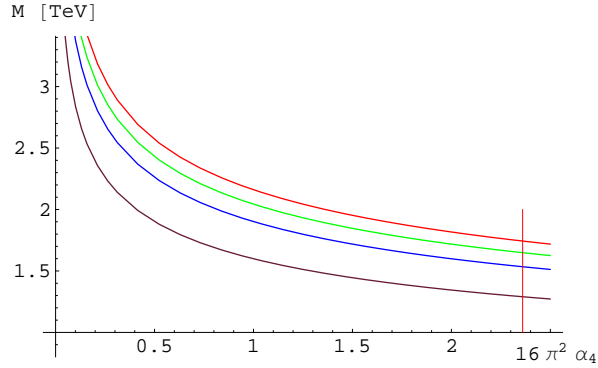}
        \caption{Dependence of the mass of a singlet vector resonance
          on ${\alpha}_4$ for different assumed $f_\pi = \Gamma_\pi /
          M_\pi = 1.0$ in red, $0.8$ in blue,  
          $0.3$ in brown, respectively. The condition $\ell_\omega=0$
          is used.}    
        \label{fig:vector_singlet_l0}
      \end{center}
    \end{figure}    
    \begin{table}
      \begin{center}
        \begin{tabular}{|c||c|c|c|c|}\hline
          $f_\omega = \frac{\Gamma_\omega}{M_\omega}$ 
          & $1.0$ & $0.8$ & $0.6$ & $0.3$ \\\hline
          $M_\omega \,[\TeV]$ 
          & $1.74$ & $1.65$ & $1.53$ & $1.29$ \\\hline
        \end{tabular}
        \caption{\label{tab:vector_l0_table}
          Mass reach for a singlet vector resonance in the case
        $\ell_\omega = 0$ for different assumed width to mass ratios:
        $f_\omega = \Gamma_\omega / M_\omega = 1.0$ (red), $0.8$
        (blue), $0.6$ (green), and $0.3$ (brown), respectively} 
      \end{center}    
    \end{table}
    Furthermore, the three non-zero $\alpha^\lambda$ parameters are
    also the same, $\alpha_2^\lambda = \alpha_5^\lambda = -
    \alpha_1^\lambda$.
    Including the parameter $k_\omega$ which could, of course, occur
    in the $\alpha$ parameters, changes the above result to
    \begin{align*}
       \alpha_1 = \alpha_2 = \alpha_4 &= \alpha_7 = - \alpha_5 =
      -\alpha_6 = -\alpha_8 \\ 
      \alpha_9 &= - (\alpha_3 + \alpha_4) \quad .
    \end{align*}
    Using only the parameters $\alpha_4$ and $\alpha_2^\lambda$
    eliminates the dependence on $k_\omega$. The mass of the singlet
    resonance is then given by
    \begin{equation}
      \label{eq:omegamass}
      M_\omega = v \left( \frac{12 \pi \alpha_4 f_\omega}{\alpha^2_4 + \frac12
      (\alpha_2^\lambda)^2} \right)^{\frac14}.
    \end{equation}
    For the fit we used the simplifying assumption $\ell_\omega = 0$, 
     \begin{figure}
       \begin{center}
         \includegraphics[width=80mm]{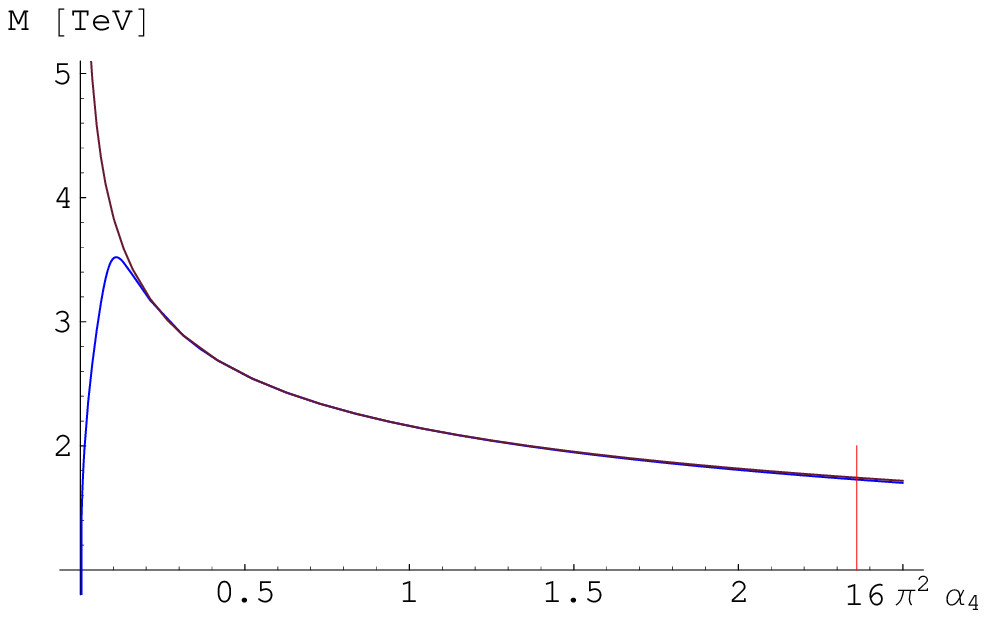}
         \includegraphics[width=80mm]{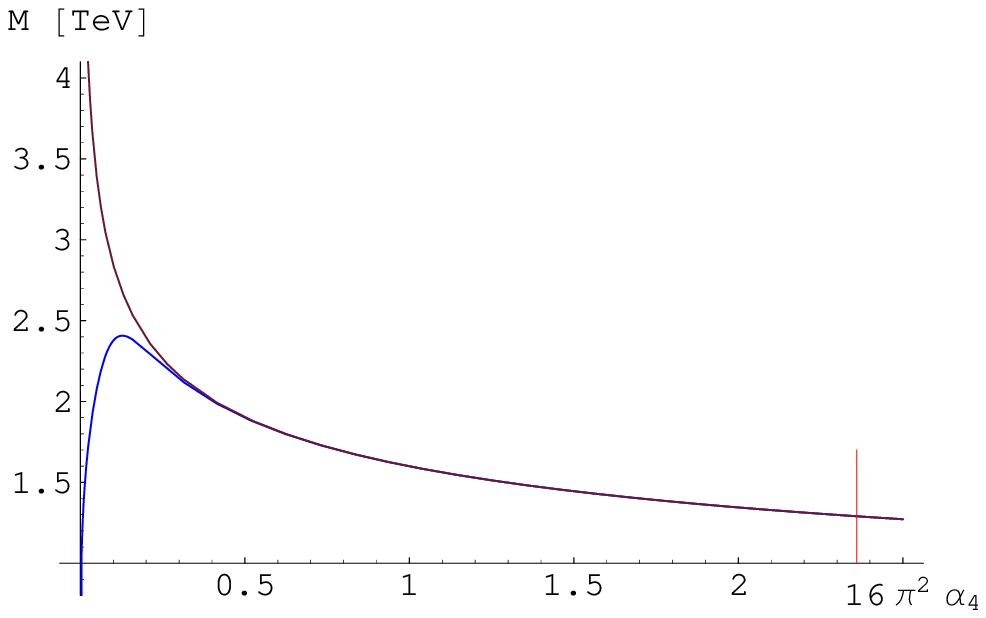}
         \caption{Dependence of the mass of a singlet vector resonance
           on ${\alpha}_4$ including the parameter $\ell_\omega$ for
           different assumed width to mass ratios: on the left 1.0, and 0.3
           on the right. The red curve is the $1\,\sigma$ upper limit
           for $\alpha_4$. The allowed resonance mass is between the
           blue and the brown curve, which are for the maximally and
           minimally allowed values of $\lambda_Z$ in \cite{menges},
           respectively.}  
         \label{fig:vector_singlet_n1}
      \end{center}
    \end{figure}
    which yields the simplified mass formula 
     \begin{equation}
      \label{eq:omegamass_simp}
      M_\omega = v \left( \frac{12 \pi f_\omega}{\alpha_4}
      \right)^{\frac14}. 
    \end{equation}   
     So this reduces to a one-parameter fit. As a cross-check, the
     error matrix for $\Delta g^Z_1, \Delta \kappa^Z, \lambda^Z $ from
     \cite{menges} has been reproduced. The limits for the vector
     singlet in the case of vanishing $\ell_\omega$ are given in
     Fig.~\ref{fig:vector_singlet_l0} and
     Table~\ref{tab:vector_l0_table}.  

     Taking also $\ell_\omega$ into account, one can solve for the
     mass of the vector resonance as a function of $\alpha_4$,
     $f_\omega$ and $\lambda_Z$. Taking the limits on the latter
     \begin{table}
       \begin{center}
         \begin{tabular}{|c||c|c|c|c|}\hline
           $f_\omega = \frac{\Gamma_\omega}{M_\omega}$
           & $1.0$ & $0.8$ & $0.6$ & $0.3$ \\\hline
           $M_\omega \,[\TeV]$
           & $2.22$ & $2.10$ & $1.95$ & $1.63$ \\\hline
         \end{tabular}
         \label{tab:vector_singlet_full}
         \caption{Mass limit for the vector singlet resonance for the
         general case with $\ell_\omega \neq 0$. The values
         in the table are average values along the lower limit} 
       \end{center}
     \end{table}
     parameter from \cite{menges}, $0 \lesssim \lambda_Z \lesssim
     0.00033$, allows one to get an allowed region region for the mass
     of a vector singlet resonance as a function of $\alpha_4$. The
     result in that case is shown in Fig.~\ref{fig:vector_singlet_n1}
     and the mass reach in Table
     \ref{tab:vector_singlet_full}. Compared to Table
     \ref{tab:vector_l0_table}, one sees that including the parameter
     $\ell_\omega$ enlarges the mass reach a bit, as one would have
     expected. 

     One point should be mentioned: a singlet vector resonance
     contributing to the electroweak sector is maximally $SU(2)_c$
     violating, and contributes significantly to $\beta_1$, i.e. $T$. 
     \begin{figure}
      \begin{center}
        \includegraphics[width=80mm]{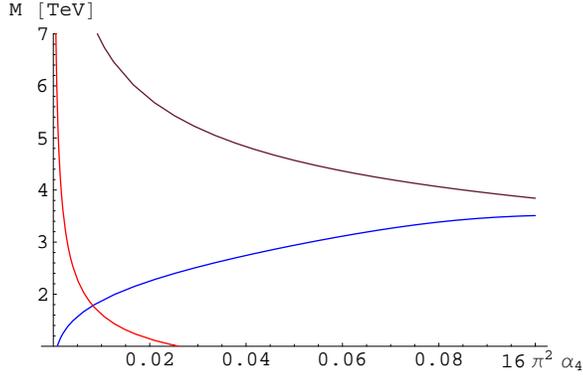}
        \caption{The allowed region for vector singlet resonances as
        function of $\alpha_4$ from the constraint $\lambda_Z$. The
        constraint from $\beta_1$, i.e. the $T$ parameter
        contribution from $\omega$ alone forces one to stay below the
        red line.} 
        \label{fig:beta}
      \end{center}
     \end{figure}
     Since this is the only constraint at order $1/M^2$, it is by far
     dominant.  As our main point is to point out what a
     measurement of the $\alpha$ parameter can do in unraveling the
     structure of electroweak symmetry breaking, we assumed that there
     is another contribution (e.g. a heavy scalar triplet) cancelling
     the $\omega$ contribution to $\beta_1$, and one is left with only
     terms of order $1/M^4$. The constraint from $T$ taken literally
     is shown in Fig.~\ref{fig:beta}, showing that most of the allowed
     parameter range is cut out.  


\subsubsection{Vector Triplet: \boldmath$\rho$}

For the analysis of the vector triplet, we assume for simplicity that
there is no mass splitting between the neutral and charged state of the
resonance. As for the vector singlet, in the sequel we ignore the
parameters $k_\rho$, $k'_\rho$, and $k''_\rho$. To the order we are 
considering they do not contribute to the widths of the resonances,
\begin{figure}[t]
  \begin{center}
    \includegraphics[width=80mm]{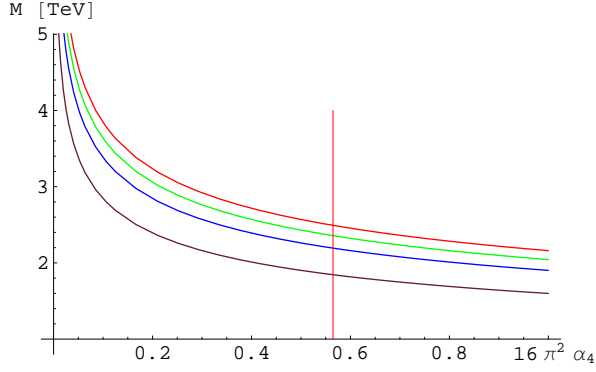}
    \caption{Dependence of the resonance mass for the vector triplet
      on ${\alpha}_4$ in the (quasi) isospin-conserving case $h_\rho =
      0$ ($g_\rho = 0$) for different assumed widths (in red
      $f_\rho = \Gamma_\rho / M_\rho = 1.0$, blue $0.8$,green
      $0.6$, brown $0.3$, respectively). All other parameters ($\mu$,
      $k$, $\ell$) are set to zero here. The red vertical
      line represents the $1\sigma$ limit for ${\alpha}_4$.}
    \label{fig:vector_triplet_llp0}
  \end{center}
\end{figure}
\begin{table}
  \begin{center}
    \begin{tabular}{|c||c|c|c|c|}\hline
      $f_\rho= \frac{\Gamma_\rho}{M_\rho}$
      & $1.0$ & $0.8$ & $0.6$ & $0.3$ \\\hline
      $M_{\rho} \ [\TeV]$
      & $2.49$ & $2.36$ & $2.19$ & $1.84$ \\\hline
    \end{tabular}
  \end{center}
\caption{Mass reach for the vector triplet if either $g_{\rho}=0$ or
  $h_{\rho}=0$ with all other parameters ($\ell$s, $k$s, $\mu$s) being
  zero, depending on different resonance widths.}
\label{vector_triplet_a}
\end{table}
and hence do not enter the electroweak fits at this stage. The same
holds in principle for the coefficients of the magnetic moment 
operators of the heavy resonances, $\mu$ and $\mu'$. As they are
quantities with a more obvious physical interpretation we try to
include them in the fits. 

\vspace{\baselineskip}\noindent
(i)
    As usual, we first consider isospin conservation, with only the
    parameters $\mu$, $g$, $\ell$ being nonzero. In principle, one
    could consider also $\mu'$, $\ell'$, and $k$ being nonzero, since in
    these terms isospin is only broken by hypercharge and not by any
    new physics effect. In this case, the relations among the
    parameters are quite simple:
    \begin{equation}
      \label{eq:rhorelationcons1}
      \alpha_1 =\; \alpha_4 = - \alpha_5 \quad \left( =  - \alpha_2
      \right) 
    \end{equation}
    The equality in parentheses holds only for $\mu' = k = 0$. For the
    $\alpha^\lambda$s we have:
    \begin{equation}
      \label{eq:rhorelationcons2}
      \alpha_1^\lambda = 3 \alpha_3^\lambda \quad \left( = - 3
      \alpha_2^\lambda \right). 
    \end{equation}
    $\alpha_2^\lambda$ gets a correction when $\ell'$ is switched on,
    and $\alpha_4^\lambda$ is not zero anymore then. 
    For the masses, we get in the pure isospin-conserving case
    \begin{equation}
      \label{eq:rhomasscons1}
      M_\rho = v \left( \frac{12 \pi \alpha_4 f_\rho}{\alpha_4^2 + 2
      (\alpha_2^\lambda)^2} \right)^{\frac14},
    \end{equation}
    while for $\ell'$ switched on:
    \begin{subequations}
      \begin{align}
        \label{eq:rhomasscons2}
        M_{\rho^\pm} &=\; v \left( \frac{12 \pi \alpha_4
          f_{\rho^\pm}}{\alpha_4^2 + 2 
          (\alpha_2^\lambda)^2 + \frac12
          \frac{\sw^2}{\cw^2}(\alpha_4^\lambda)^2} 
        \right)^{\frac14} 
        \\
        M_{\rho^0} &=\; v \left( \frac{12 \pi \alpha_4
          f_{\rho^0}}{\alpha_4^2 + 2 
          (\alpha_2^\lambda)^2} \right)^{\frac14}.
      \end{align}
    \end{subequations}
    The case $\ell_\rho = 0$ (i.e. $\alpha_2^\lambda = 0$) seems to 
    \begin{figure}
      \begin{center}
        \includegraphics[width=80mm]{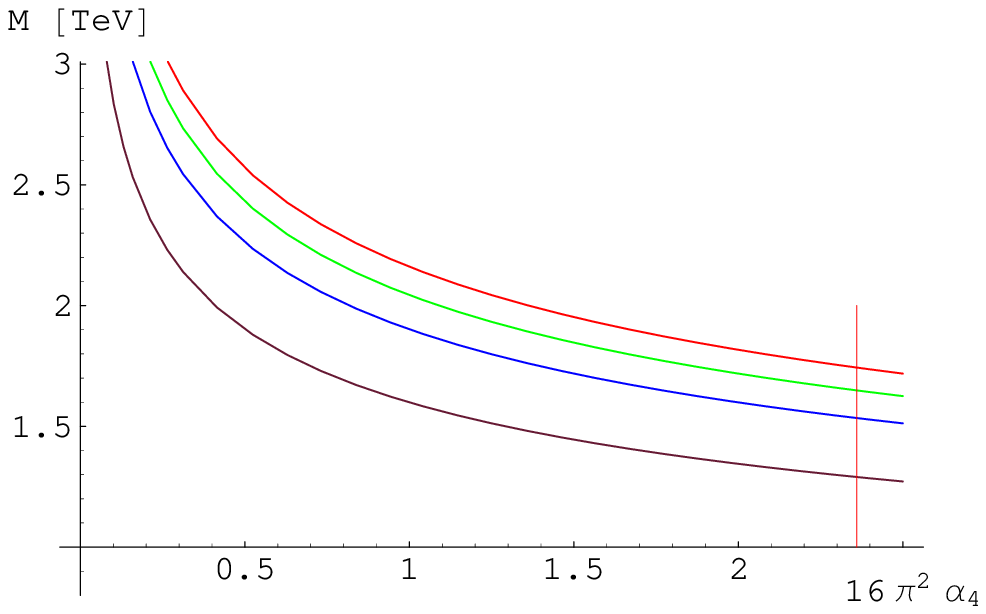}
        \includegraphics[width=80mm]{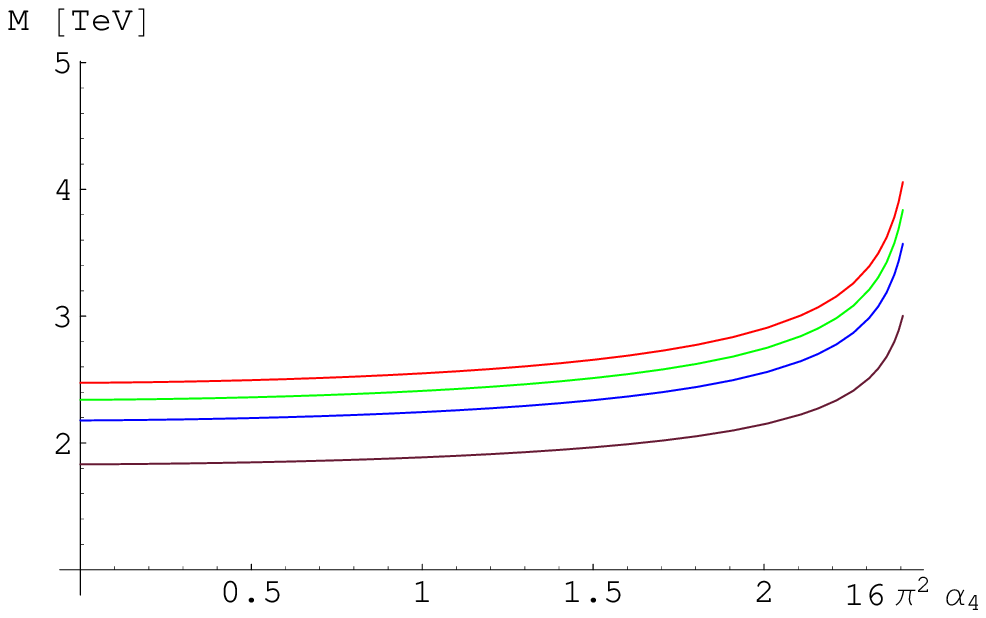}
        \caption{Dependence of the resonance mass for the vector
          triplet on ${\alpha}_4$ under the assumption $\ell_\rho =
          \ell'_\rho = \ell''_\rho =0$ for different assumed widths: in red
          $f_\rho = \Gamma_\rho / M_{\rho}= 1.0$, blue $0.8$, green
          $0.6$, brown $0.3$, respectively. On the left the neutral
          component is shown, on the right for charged one.}  
        \label{fig:vector_triplet_l02ch}
      \end{center}
      \end{figure}
    \begin{table}
      \begin{center}
        \begin{tabular}{|c||c|c|c|c|}\hline
          $f_\rho= \frac{\Gamma_\rho}{M_\rho}$
          & $1.0$ & $0.8$ & $0.6$ & $0.3$ \\\hline
          $M_{\rho^\pm} \ [\TeV]$
          & $2.67$ & $2.53$ & $2.35$ & $1.98$ \\\hline
          $M_{\rho^0} \ [\TeV]$	
          & $1.74$ & $1.65$ & $1.53$ & $1.29$ \\\hline
        \end{tabular}
      \end{center}
      \caption{Mass reach for the vector triplet under the assumption
        $\ell_{\rho}=\ell'_\rho=\ell''_\rho=0$.
        Values for the charged component are averaged over the lower limit.}  
      \label{vector_triplet_b} 
    \end{table}
    bring one back to the corresponding case for the vector
    singlet. But now the correlations among the parameters are
    different, especially $\alpha_6$ and $\alpha_7$ are zero here but
    not in the singlet case. Note also, that the assumption $g_\rho
    =0$, $h_\rho \neq 0$ leads to the same result, as the formulas for
    the width and the functional dependence of the $\alpha$s on the
    coupling change in the same manner. The mass reach for the vector
    triplet in this case is shown
    in Fig.~\ref{fig:vector_triplet_llp0} and Table~\ref{vector_triplet_a}.

\vspace{\baselineskip}\noindent
(ii)
    Taking into account isospin violation, we note that the
    following relations hold generally among the $\alpha$ and
    $\alpha^\lambda$ parameters 
    \begin{subequations}
      \begin{align}
        \label{eq:rhorelationnonc0}
        \alpha_4 &=\; -\alpha_5       &
        \alpha_6 &=\; -\alpha_7    \\
        \alpha_1 &=\; \alpha_4 + \alpha_6 & 
        \alpha_8 &=\; - \frac{\alpha_6}{2} \left( 1 + \frac{\alpha_6}{2
        (\sqrt{\alpha_1} + \sqrt{\alpha_4})^2} \right) \;\; .
      \end{align}
    \end{subequations}
    And among the $\alpha^\lambda$s:
    \begin{subequations}
      \label{eq:rhorelationnonc1}
      \begin{align}
        2 (\alpha_1^\lambda + \alpha_2^\lambda) &=\; - \left( 1 +
        \sqrt{\frac{\alpha_1}{\alpha_4}} \right) ( 2\alpha_3^\lambda +
        \alpha_4^\lambda )
      \end{align}
    \end{subequations}
    We first consider the special case $g_\rho = - h_\rho$ where the
    $\beta_1$ ($T$ parameter) vanishes (we neglect a possible $\Delta
    M_\rho$). To simplify things, we first set all the $k$s and
    $\ell$s to zero. Then, (\ref{eq:rhorelationnonc0}) simplifies
    to 
    \begin{equation}
      \label{eq:rhorelationnonc2}      
      \alpha_1 = \frac13 \alpha_2 = - \alpha_3 = \frac19 \alpha_4 = -
      \frac18 \alpha_6 = \frac12 \alpha_9 .
    \end{equation}
    \begin{figure}
      \begin{center}
        \includegraphics[width=80mm]{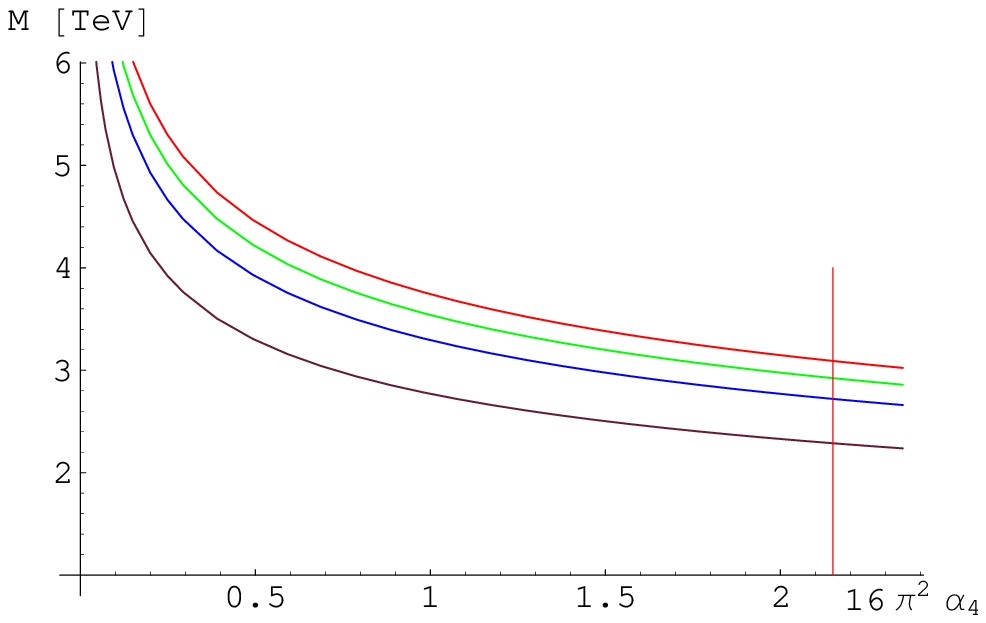}
        \includegraphics[width=80mm]{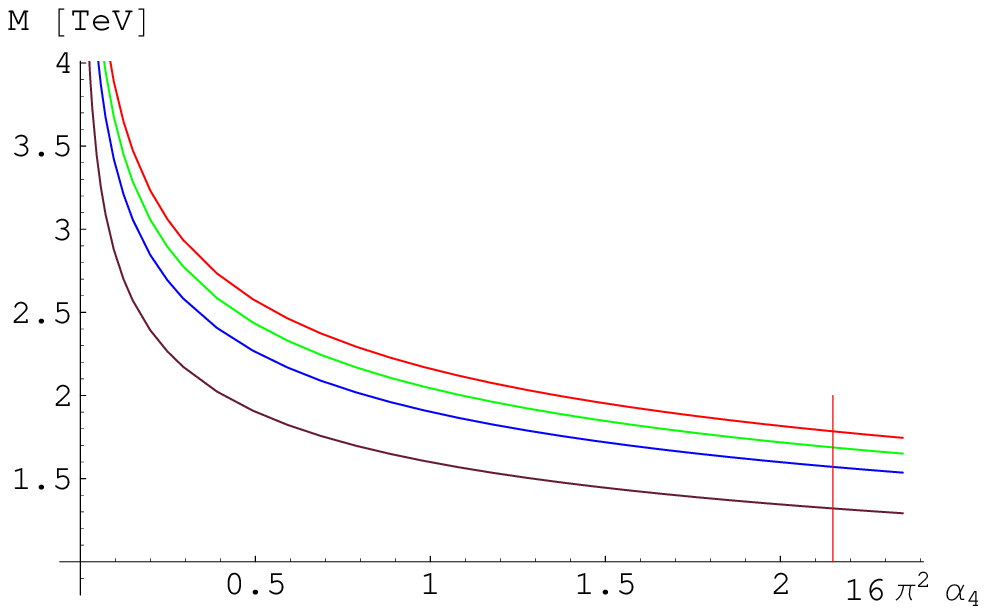}
        \caption{Dependence of the resonance mass for the vector triplet
          on $\alpha_4$ in the special isospin-violating case
          $g_\rho = - h_\rho$ for different assumed width to mass
          ratios (in red $f_\rho = \Gamma_\rho / M_{\rho^+}= 1.0$,
          blue $0.8$,green $0.6$, brown $0.3$, respectively). On the
          left is the charged resonance, on the right the neutral one.} 
        \label{fig:vector_triplet_gh}
      \end{center}
    \end{figure}
    \begin{table}
      \begin{center}
        \begin{tabular}{|c||c|c|c|c|}\hline
          $f_\rho = \frac{\Gamma_\rho}{M_\rho}$
          & $1.0$ & $0.8$ & $0.6$ & $0.3$ \\\hline
          $M_{\rho^\pm} \ [\TeV]$
          & $3.09$ & $2.92$ & $2.72$ & $2.29$ \\\hline
          $M_{\rho^0} \ [\TeV]$
          & $1.78$ & $1.69$ & $1.57$ & $1.32$ \\\hline
        \end{tabular}
      \end{center}
      \caption{Mass reach for the vector triplet in the special
      isospin-violating case $g_\rho=-h_\rho$.}
      \label{tab:vector_triplet_gh}
    \end{table}
    For this special isospin-violating case, the formulas for the
    masses of the resonances are:
    \begin{subequations}
      \label{eq:rhomassnonc0}
      \begin{align}
        M_{\rho^\pm} &=\; v \left( \frac{108 \pi 
        f_{\rho^\pm}}{\alpha_4}
        \right)^{\frac14} \\
        M_{\rho^0} &=\; v \left( \frac{12 \pi 
        f_{\rho^0}}{\alpha_4} \right)^{\frac14}  
      \end{align}
    \end{subequations}
    The dependence of the mass of the vector resonance in this case is
    shown in Fig.~\ref{fig:vector_triplet_gh}, and the mass reach in
    Table~\ref{tab:vector_triplet_gh}. Note that the difference
    between the charged and the neutral state is just the factor
    $\sqrt{3}$ from (\ref{eq:rhomassnonc0}).

    Next, we still assume $g_\rho = - h_\rho$ and hence no
    contribution to the $T$ parameter, but allow for nonzero values of 
    \begin{figure}
      \begin{center}
        \includegraphics[width=80mm]{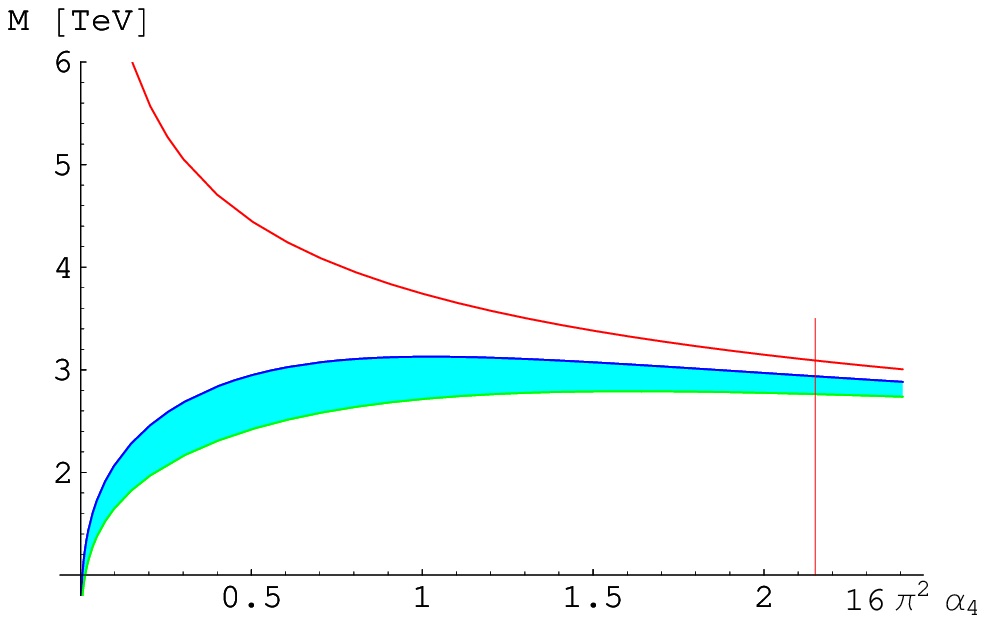}
        \includegraphics[width=80mm]{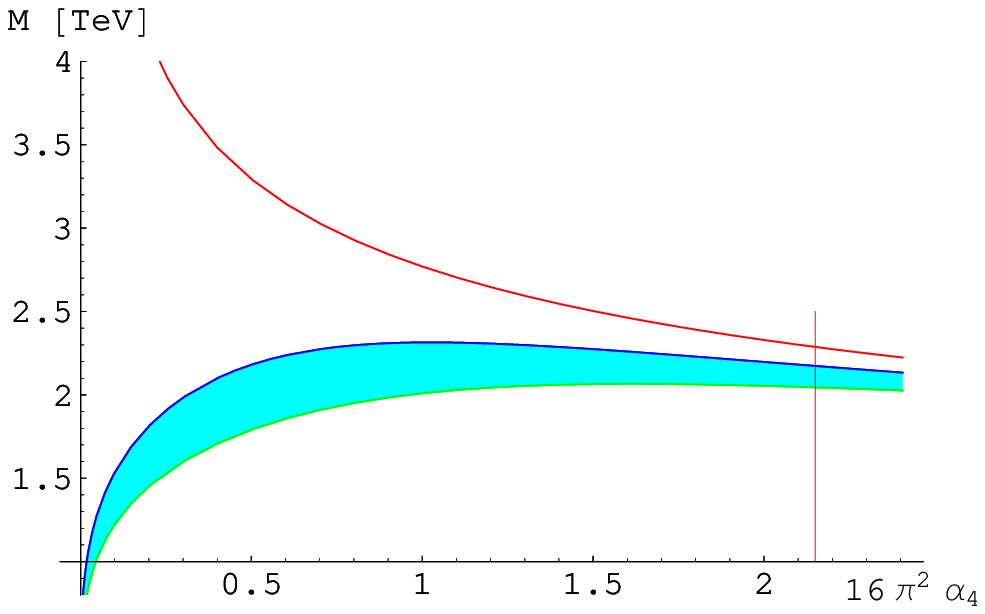}
        \caption{Dependence of the resonance mass for the charged
          component of the vector triplet 
          on ${\alpha}_4$ for different assumed widths (f=1 on the
          left and f=0.3 on the right) for $g=-h$ case, but $\ell_\rho
          , \ell'_\rho \neq 0$. The red and blue line are the lower
          and upper limit from $\lambda_Z$, respectively. Vertical red
          line: maximal allowed value of ${\alpha}_4$. The blue shaded
          area is the allowed one when $\mu_\rho, \mu'_\rho \neq 0$.}
        \label{fig:vector_triplet_gh_tr_m0_ch}
      \end{center}
    \end{figure}
    \begin{figure}
      \begin{center}
        \includegraphics[width=80mm]{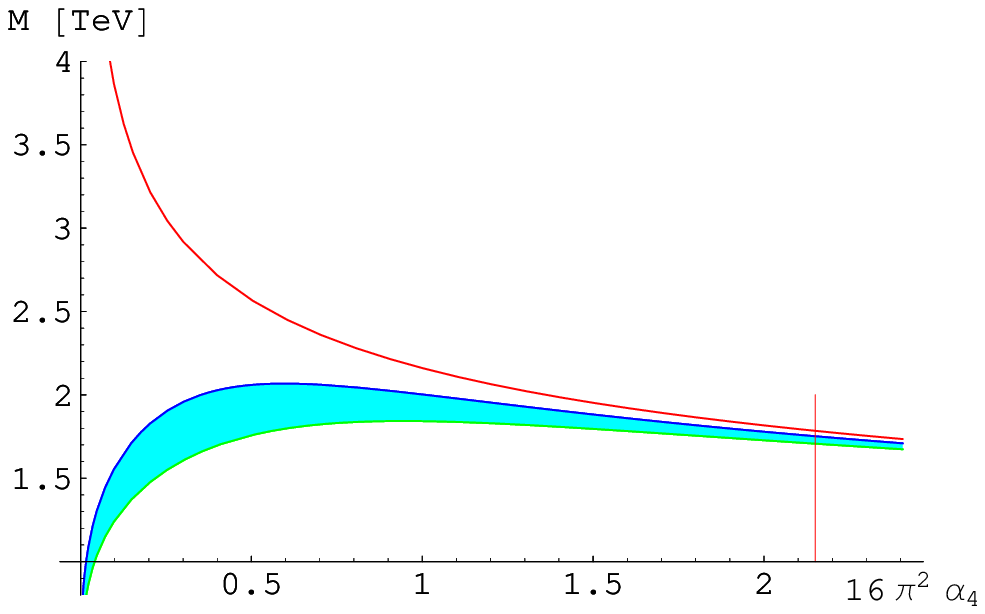}
        \includegraphics[width=80mm]{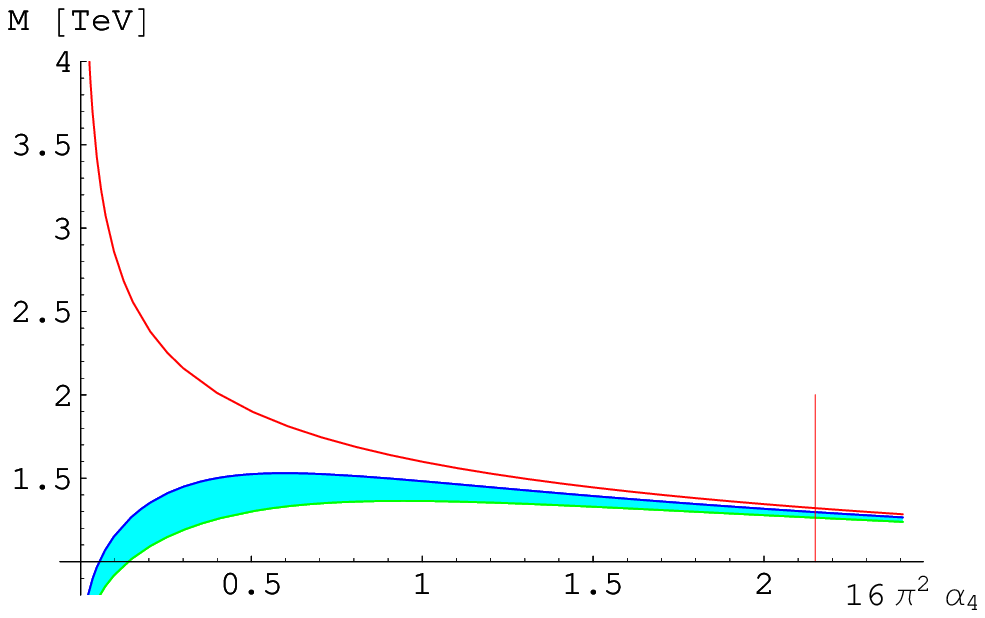}
        \caption{Same as Fig.~\ref{fig:vector_triplet_gh_tr_m0_ch},
        but for the neutral component.}
      \label{fig:vector_triplet_gh_tr_nu_m0_ch}
      \end{center}
    \end{figure}
    \begin{table}
      \begin{center}
        \begin{tabular}{|c||c|c|c|c|}\hline
          $f_\rho= \frac{\Gamma_\rho}{M_\rho}$
          & $1.0$ & $0.8$ & $0.6$ & $0.3$ \\\hline
          $M_{\rho^\pm} \ [\TeV]$
          & $2.91$ & $2.75$ & $2.56$ & $2.16$ \\\hline
          $M_{\rho^0} \ [\TeV]$
          & $1.84$ & $1.79$ & $1.66$ & $1.40$ \\\hline
        \end{tabular}
      \end{center}
      \caption{Mass reach for the vector triplet under the assumption
        $g_\rho=-h_\rho$ with nonzero $\ell_\rho$s. The values
        in the table are average values along the lower limit curve.}
      \label{vector_triplet_gh_tr_m0_ch}
    \end{table}
    $\ell_\rho$ and $\ell'_\rho$. In complete analogy to the
    discussion for the vector singlet, we now have to include the
    measurements of the triple gauge couplings to access $\lambda_Z$
    and $\lambda_\gamma$ in order to have enough equations at hand to
    solve the system, which is fulfilled if one of the $\ell_\rho$s is 
    set to zero. We take $\ell''_\rho \equiv 0$. Taking the allowed
    range of $0 \lesssim \lambda_Z \lesssim  
    0.00033$ from \cite{menges}, we again get an upper and a lower
    limiting curve for $M_\rho$ as a function of $\alpha_4$. The
    allowed range is in between. The situation is shown in
    Fig.~\ref{fig:vector_triplet_gh_tr_m0_ch} 
    for the charged state, and in
    \begin{table}
      \begin{center}
        \begin{tabular}{|c||c|c|c|c|}\hline
          $f_\rho= \frac{\Gamma_\rho}{M_\rho}$
          & $1.0$ & $0.8$ & $0.6$ & $0.3$ \\\hline
          $M_{\rho^\pm} \ [\TeV]$
          & $2.54$ & $2.41$ & $2.34$ & $1.88$ \\\hline
          $M_{\rho^0} \ [\TeV]$
          & $1.71$ & $1.62$ & $1.51$ & $1.27$ \\\hline
        \end{tabular}
      \end{center}
      \caption{Mass reach for the vector triplet with the assumptions
        $g_\rho=-h_\rho$ with nonvanishing $\ell_\rho, \ell'_\rho$ and
        nonvanishing $\mu_\rho, \mu'_\rho$. The values in the table 
        are average values along the lower limit curve.}
      \label{vector_triplet_gh_tr_m_ch}
    \end{table}
    Fig.~\ref{fig:vector_triplet_gh_tr_nu_m0_ch} for the neutral one. 
    The mass reach for this choice of parameters is given in
    Table~\ref{vector_triplet_gh_tr_m0_ch}. 

    As a next step, we still assume $g_\rho = -h_\rho$, but allow for
    nonvanishing $\mu_\rho$ and $\mu'_\rho$. This offers the
    possibility of various cancellations among the different
    parameters, especially since the $\mu$s enter linearly in the
    $\alpha$s and can have arbitrary sign. This fact completely
    cancels the gain in using a new constraint on the system, and so
    the bound for the vector resonance mass losens a bit. The allowed
    parameter regions are shown as blue shadings in
    Fig.~\ref{fig:vector_triplet_gh_tr_m0_ch} for the charged and in 
    Fig.~\ref{fig:vector_triplet_gh_tr_nu_m0_ch} for the neutral
    state, respectively. The mass reach is shown in Table
    \ref{vector_triplet_gh_tr_m_ch}. 

    Considering all isospin-violating terms, there are now (still
    ignoring the $k$s) all $\alpha$ and $\alpha^\lambda$ parameters
    nonvanishing, except for $\alpha_{10}$. The masses of the
    resonances are then:
    \begin{subequations}
      \begin{align}
        \label{eq:rhomassnonc}
        M_{\rho^\pm} &=\; v \left( \frac{12 \pi \alpha_1
        f_{\rho^\pm}}{\alpha_1^2 + 2 (\alpha^\lambda_3)^2 +
        \frac12 \frac{\sw^2}{\cw^2} (\alpha_4^\lambda)^2}
        \right)^{\frac14} \\
        M_{\rho^0} &=\; v \left( \frac{12 \pi \alpha_4
        f_{\rho^0}}{\alpha_4^2 + 2 \left( \alpha_3^\lambda
        \sqrt{\frac{\alpha_4}{\alpha_1}} + 2
        \alpha_5^\lambda \right)^2} \right)^{\frac14}  
      \end{align}
    \end{subequations}  
    Allowing for arbitrary variations of $g_\rho$ and $h_\rho$ and
    taking non-zero values for the $\ell_\rho$, $k_\rho$ and
    $\mu_\rho$ parameters into account, one again has to use the
    results from \cite{menges} to access $\lambda_\gamma$ and
    $\lambda_Z$ from the measurements of the triple gauge couplings. 
    However we found that the (independent) variation of $g_\rho $ and
    $h_\rho$ already introduces enough freedom into the system so that
    allowing for nonzero values for the other parameters does not
    extend the allowed region in $(\alpha_4,M_\rho)$ parameter space
    significantly. Hence, the allowed region shows up only as tiny
    bands below the corresponding curves in
    Figure~\ref{fig:vector_triplet_l02ch}. The limits for the mass
    reach given in Table~\ref{vector_triplet_b} do therefore not
    change significantly.

\subsection{$J=2$ Channel}
\subsubsection{Tensor Singlet: \boldmath$f$}

(i)
    For conserved isospin, $\alpha_4$ and $\alpha_5$ are non-zero, but
    related to each other by the constraint
    \begin{figure}[t]
      \begin{center}
      \includegraphics[width=80mm]{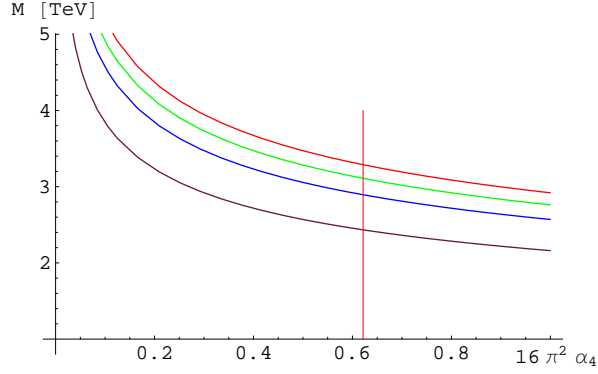}
      \caption{Dependence of the resonance mass for the tensor singlet case
        on ${\alpha}_4$ for different assumed width to mass ratios
        ($f_f = \Gamma_f / M_f = 1.0$ in red, $0.8$ in blue, $0.6$ in
        green, and $0.3$ in brown, respectively). The red vertical
        line represents the $1\,\sigma$ limit for ${\alpha}_4$.}
      \label{fig:tensor_singlet}
      \end{center}
    \end{figure}    
    \begin{equation}
      \label{eq:frelation1}
      \alpha_5 = - \frac14 \alpha_4 .
    \end{equation}
    \begin{table}[t]
      \begin{center}
        \begin{tabular}{|c||c|c|c|c|}\hline
          $f_f= \frac{\Gamma_f}{M_f}$
          & $1.0$ & $0.8$ & $0.6$ & $0.3$ \\\hline
          $M_f\ [\TeV]$
          & $3.29$ & $3.11$ & $2.89$ & $2.43$ \\\hline
        \end{tabular}
      \end{center}
      \label{tensor_singlet_table}
      \caption{Mass reach for the tensor singlet in the
        $SU(2)_c$ conserving case depending on different
        resonance widths.}
    \end{table}
    From the fit we get ${\alpha}_4=0.64369$ for the parabolic error
    and  $-0.65404<{\alpha}_4<0.62154$ for the asymmetric errors at
    $1\,\sigma$ level.  

    The mass of a singlet tensor resonance is then given by
    \begin{equation}
      \label{eq:fmasscons}
      M_f = v \left( \frac{40 \pi f_f}{\alpha_4} \right)^{\frac14} . 
    \end{equation}

\vspace{\baselineskip}\noindent
(ii)
      If we allow for isospin breaking, also $\alpha_6$, $\alpha_7$
      and $\alpha_{10}$ are non-zero, but subjected to the two
      constraints
      \begin{equation}
        \label{eq:frelation2}
        \alpha_7 = - \frac14 \alpha_6, \qquad \alpha_7^2 = - \frac23 
        \alpha_5 \alpha_{10}  \;\;\left[ \text{or} \;\;
        \alpha_6^2 = \frac83 \alpha_4 \alpha_{10} \right] ,
      \end{equation}
      \begin{figure}[t]
        \begin{center}
          \includegraphics[width=80mm]{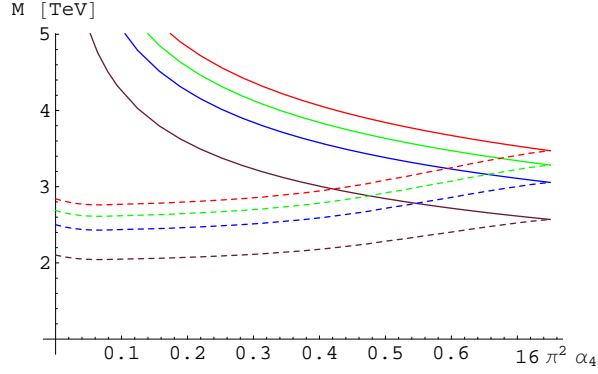}
          \caption{Dependence of the resonance mass for the tensor
            singlet with broken isospin on ${\alpha}_4$ for
            different assumed width to mass ratios: $f_f = \Gamma_f /
            M_f = 1.0$ (red), $0.8$ (blue), $0.6$ (green), and $0.3$
            (brown), respectively. Along the $1\,\sigma$ contour, the
            lower limit is given by the dashed line, while the full
            line is the upper limit.}
          \label{fig:tensor_singlet_broken}
        \end{center}
      \end{figure}
      \begin{table}
        \begin{center}
          \begin{tabular}{|c||c|c|c|c|}\hline
            $f_f= \frac{\Gamma_f}{M_f}$
            & $1.0$ & $0.8$ & $0.6$ & $0.3$ \\\hline
            $M_f\ [\TeV]$
            & $3.00$ & $2.84$ & $2.64$ & $2.22$ \\\hline
          \end{tabular}
        \end{center}
        \label{tensor_singlet_table2}
        \caption{Mass reach for the tensor singlet in the broken
          isospin case depending on different 
          resonance widths.Values in the table are average values
          along the lower limit.} 
        \end{table}
        while the former relation (\ref{eq:frelation1}) still holds. 
        We choose to take $\alpha_4$ and $\alpha_6$ as independent
        parameters. Then the mass of the tensor singlet is given by
        \begin{equation}
          \label{eq:fmassnonc}
          M_f = v \left( \frac{120 \pi \alpha_4 f_f}{2 \alpha_4^2 +
            (\alpha_4 + \alpha_6)^2} \right)^{\frac14}.
        \end{equation}
        The maximum for the resonance mass is reached when we set
        $\alpha_4 = - \alpha_6$, leaving us with a one-parameter
        fit. The maximal mass is given by $M_{f,max} = v \left( 60 \pi
        f_f/\alpha_4\right)^{\frac14}$, leading to the upper bound in
        Fig.~\ref{fig:tensor_singlet_broken}.


\subsubsection{Tensor Triplet: \boldmath$a$}

(i)
    Like for the triplet scalar, a tensor triplet as a resonance can
    only occur with the help of isospin breaking. Again, we consider
    the case $h_a = k_a = 0$, so that $h'_a$ is the only non-vanishing
    parameter. In that case, isospin breaking does not show up
    experimentally, as only $\alpha_4 = -4 \alpha_5$ is non-zero. Like
    in the scalar case, the charged resonance decouples, while for the
    mass we get the same relation as in the singlet case:
    \begin{figure}
      \begin{center}
        \includegraphics[width=80mm]{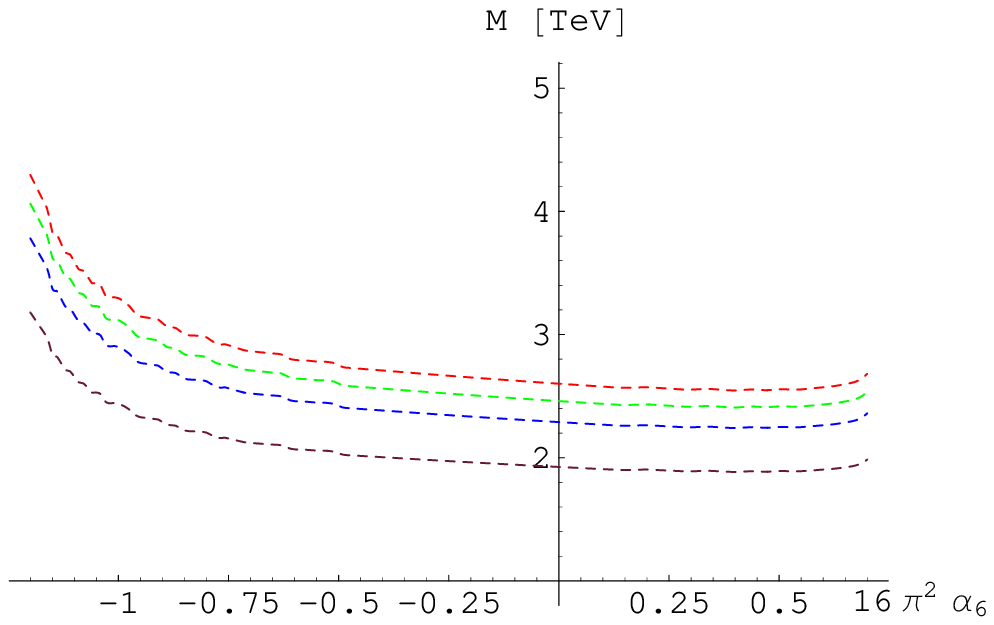}
        \includegraphics[width=80mm]{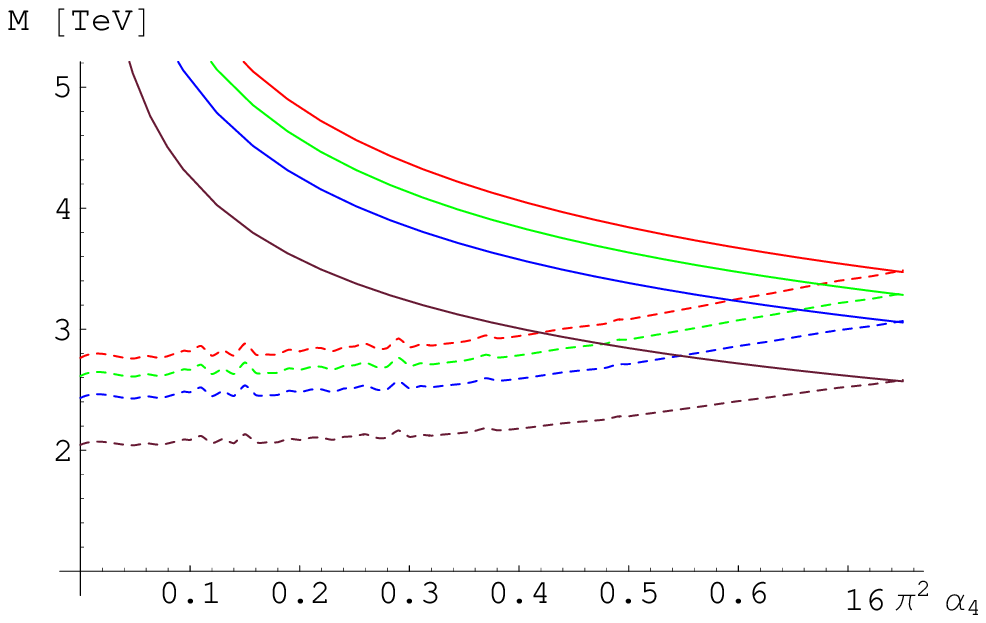}
        \caption{Mass reach for tensor triplet resonances with isospin
          breaking for different assumed width to mass ratios, $f_a =
          \Gamma_a / M_a = 1.0$ (red), $0.8$ (blue), $0.6$ (green),
          $0.3$ (brown), respectively. On the left: the charged
          components, on the right: neutral component. Full/dashed
          line: upper/lower limit within the $1\,\sigma$ contour.}           
        \label{fig:tensor_triplet}
      \end{center}
    \end{figure}
    \begin{table}
      \begin{center}
        \begin{tabular}{|c||c|c|c|c|}\hline          
          $f_a = \frac{\Gamma_a}{M_a}$
          & $1.0$ & $0.8$ & $0.6$ & $0.3$ \\\hline
          $M_{a^0}\,[\TeV]$
          & $3.01$ & $2.85$ & $2.65$ & $2.23$ \\\hline
          $M_{a^\pm}\,[\TeV]$
          & $2.81$ & $2.66$ & $2.47$ & $2.08$ \\\hline
        \end{tabular}
      \end{center}
      \label{tensor_triplet_table}
      \caption{Dependence of the mass reach for tensor triplet
        resonances on different resonance widths. For the neutral
        component, the numbers in the table are average values along
        the lower limit contour.}
    \end{table}
    \begin{equation}
      \label{eq:amasscons}
      M_{a^0} = v \left( \frac{40\pi f_{a^0}}{\alpha_4}
      \right)^{\frac14} 
    \end{equation}
    The fit (and hence the limits) is identical to the
    isospin-conserving case of the tensor singlet. 

\vspace{\baselineskip}\noindent
(ii)
    In the most general isospin-breaking case, we have five possibly
    nonvanishing parameters $\alpha_i$ for $i=4,5,6,7,10$ and two
    independent masses. There are two constraints among the
    parameters, namely 
    \begin{subequations}
      \begin{align}
        \label{eq:arelationnonc}
        \alpha_5 &= \; -\frac14 \alpha_4 \\
        (2 \alpha_6 - \alpha_7)^2  &=\; \frac92 \alpha_4 \left (
        \alpha_6 + 4 \alpha_7 + 3 \alpha_{10} \right) 
      \end{align}
    \end{subequations}
    Solving for the masses of the resonances, yields the formulas: 
    \begin{subequations}
      \begin{align}
        \label{eq:amassnonc}
        M_{a^\pm} &=\; v \left( \frac{270 \pi f_{a^\pm}}{\alpha_6 + 4
        \alpha_7} \right)^{\frac14} \\
        M_{a^0} &=\; v \left( \frac{120 \pi \alpha_4 f_{a^0}}{2
        \alpha_4^2 + (\alpha_4 + \frac89 \alpha_6 - \frac49
        \alpha_7)^2} \right)^{\frac14} 
      \end{align}
    \end{subequations}
    The denominator for the neutral component is minimized within the
    $1\,\sigma$ volume on the surface defined by $9 \alpha_4 + 8
    \alpha_6 - 4 \alpha_7 = 0$. This is equivalent to the condition
    $h_a + h'_a + 2 k_a = 0$, and maximizes the mass of the neutral state
    to become $M_{a^0} = v \left( 60 \pi f_{a^0} / \alpha_4
    \right)^{\frac14}$.


\subsubsection{Tensor Quintet: \boldmath$t$}

(i)
    For the tensor quintet, there is the case of strict isospin
    conservation, where only $\alpha_4$ and $\alpha_5$ are
    nonvanishing with the constraint $\alpha_5 = 2 \alpha_4$. This
    degeneracy is lifted as soon as the isospin breaking coupling $h'$
    is switched on. Solving for the mass yields    
    \begin{equation}
      \label{eq:tmasscons}
      M_t = v \left( \frac{30 \pi f_t}{\alpha_4} \right)^{\frac14} 
    \end{equation}

    There are four other cases, in which also only the
    isospin-conserving parameters $\alpha_{4,5}$ are non-zero and
    experimentally isospin breaking cannot be measured. This can
    either be achieved by having $h'_t \neq 0$ and all other parameters
    vanishing or $h'_t$ vanishing. In the second case, the relation
    $\alpha_5 = 2 \alpha_4$ again holds. 
    
    \begin{itemize}
      \item[a)]
        Only the coupling $h'_t$ is switched on (which is special in the
        sense that at least the SM part couples to a singlet
        invariantly). Here the charged resonances decouple, and for
        the neutral we get:
        \begin{equation}
          M_{t^0} =\; v \left( \frac{60 \pi f_{t^0}}{\alpha_4}
          \right)^{\frac14} .
        \end{equation}        

        \item[b)]
          $h'_t = h_t = 0$, $g_t = - 2 k_t$: 

        \item[c)]
          $h'_t = 0$, $g_t = k_t = -\frac12 h_t$:

        \item[d)]
          $h'_t = 0$, $g_t = 2 k_t = -\frac12 h_t$:
    \end{itemize}
    \begin{figure}[t]
      \begin{center}
        \includegraphics[width=80mm]{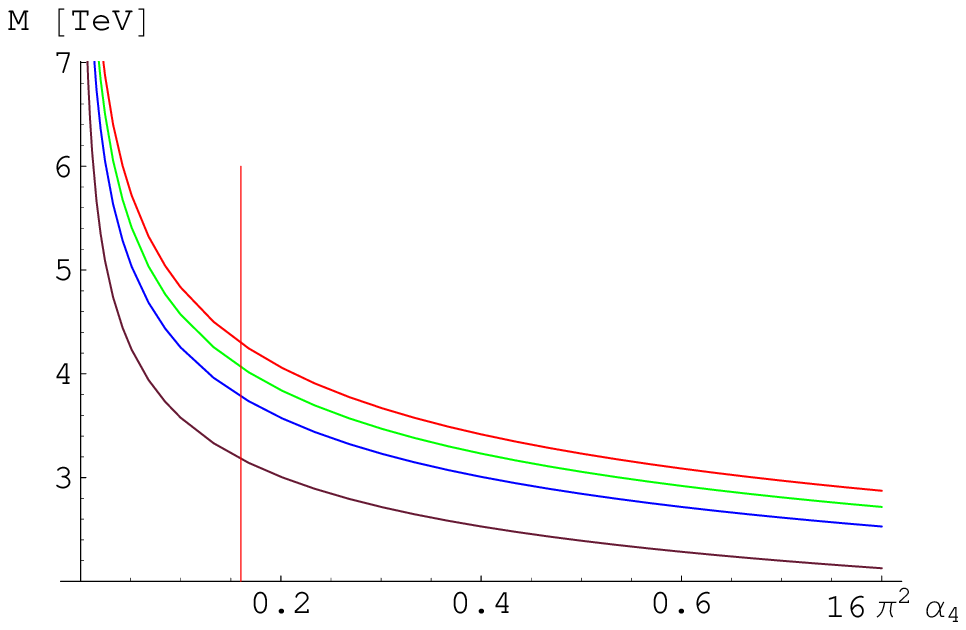}
        \includegraphics[width=80mm]{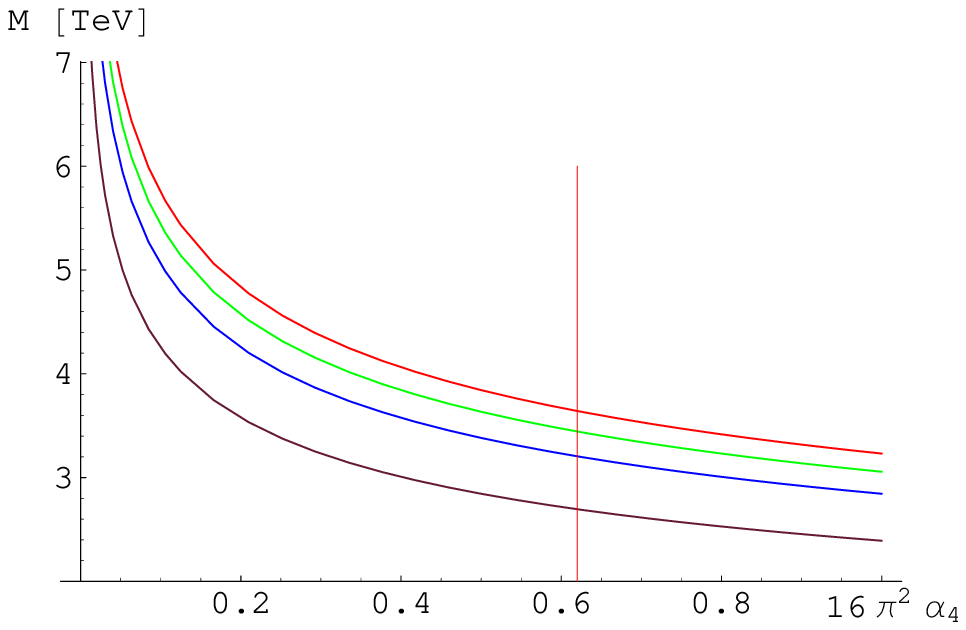}
        \qquad
      \end{center}
      \caption{Dependence of the resonance mass for the tensor quintet
        on ${\alpha}_4$ for different assumed widths (in red
        $f_{t^0} = \Gamma_{t^0} / M_{t0} = 1.0$, blue $0.8$,green
        $0.6$, brown $0.3$). The red vertical line represents the
        $1\sigma$ limit for ${\alpha}_4$. On the left:
        Isospin-conserving case and isospin-breaking cases b),c),and
        d) described in the text. On the right: isospin-breaking case a).}
      \label{fig:tensor_quintet_cons}
    \end{figure}
    \begin{table}
      \begin{center}
        \begin{tabular}{|c||c|c|c|c|}\hline
          $f= \frac{\Gamma}{M_t}$
          & $1.0$ & $0.8$ & $0.6$ & $0.3$ \\\hline
          $M_t \ [\TeV]$
          & $4.30$ & $4.06$ & $3.78$ & $3.18$ \\\hline
        \end{tabular} \qquad
        \begin{tabular}{|c||c|c|c|c|}\hline
          $f= \frac{\Gamma}{M_t}$
          & $1.0$ & $0.8$ & $0.6$ & $0.3$ \\\hline
          $M_t \ [\TeV]$
          & $3.64$ & $3.44$ & $3.20$ & $2.69$ \\\hline
        \end{tabular} \qquad        
      \end{center}
      \label{tensor_quintet_cons}
      \caption{Mass reach for the tensor quintet: On the left in the
        $SU(2)_c$ conserving case as well as for the cases b), c), and 
        d) described in the text, depending on different
        resonance widths. On the right, case a) where only $h'_t \neq
        0$.}
    \end{table}
    In all the cases b) to d), the neutral, charged and doubly charged
    resonances are degenerate in mass, and we get 
    \begin{equation}
      M_{t} =\; v \left( \frac{30 \pi f_{t}}{\alpha_4}
      \right)^{\frac14} .
    \end{equation}  
    So for the experimental sensitivity, the cases b) to d) are
    equivalent to the strictly isospin-conserving case. Here, from the
    fit we obtain $\delta\alpha_4=0.16116$ as a parabolic error and 
    $-0.17387<\alpha_4<0.15134$ as asymmetric ones at $1\,\sigma$. 

    A next case would be to consider only $g_t$ and $h'_t$ different from
    zero. In that case only $\alpha_{10}$ vanishes, while we have the
    two constraints
    \begin{equation}
       \alpha_7 = - \frac14 \alpha_6, \qquad \alpha_6^2 =
       \frac{16}{81} ( 2\alpha_4 - \alpha_5) ( \alpha_4 + 4
       \alpha_5 ) .
    \end{equation}
    So here, experimentally we can measure isospin breaking in the
    resonance sector. For the masses of the tensor resonances, there
    \begin{figure}
      \begin{center}
        \includegraphics[width=80mm]{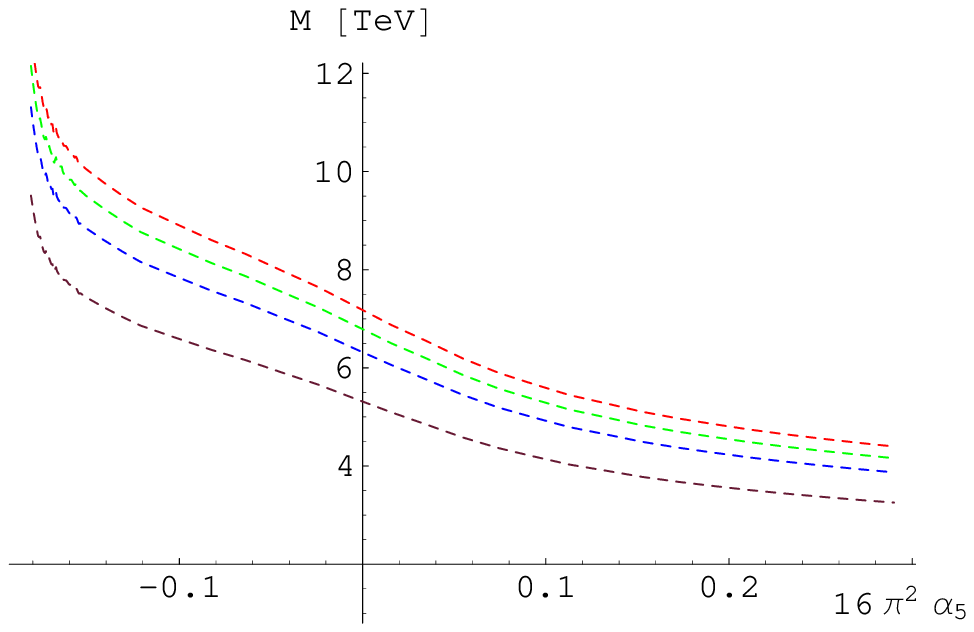}
        \includegraphics[width=80mm]{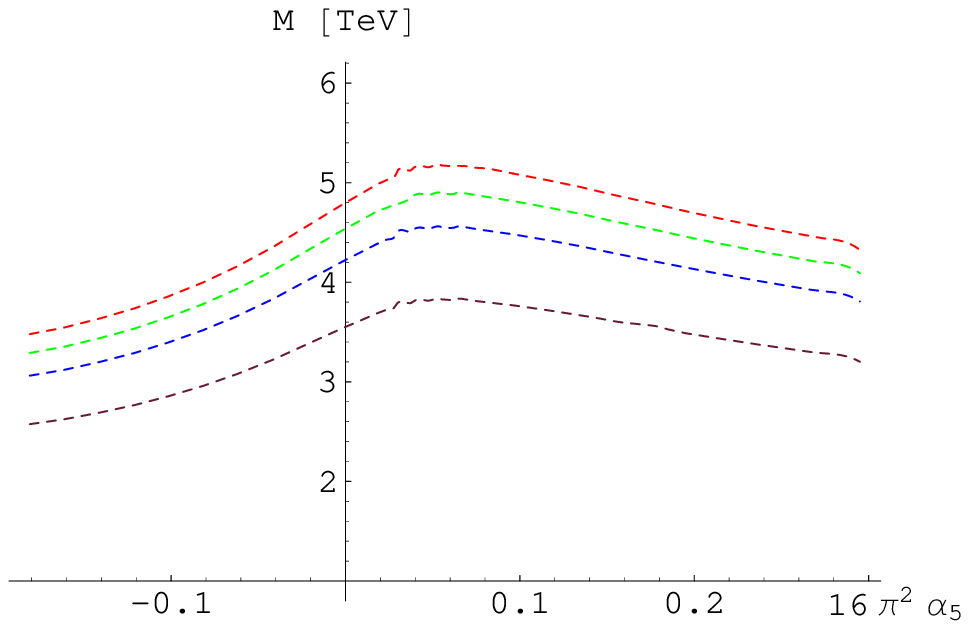}	
        \caption{Dependence of the resonance mass for the tensor
          quintet in special case $h=k=0$ for different assumed widths
          (in red $f = \Gamma / M_t = 1.0$, blue $0.8$,green
          $0.6$, brown $0.3$)}  
        \label{fig:tensor_quintet_special}
      \end{center}
    \end{figure}
    \begin{table}
      \begin{center}
        \begin{tabular}{|c||c|c|c|c|}\hline
          $f= \frac{\Gamma}{M_t}$
          & $1.0$ & $0.8$ & $0.6$ & $0.3$ \\\hline
          $M_{t^c} \ [\TeV]$
          & $6.76$ & $6.39$ & $5.95$ & $5.00$ \\\hline
 	  $M_{t^0} \ [\TeV]$	
          & $4.53$ & $4.28$ & $3.98$ & $3.35$ \\\hline
        \end{tabular}
      \end{center}
      \caption{Mass reach for the tensor quintet in the 
        $h=k=0$ case depending on different
        resonance widths.Values in the table are average over lower limit.} 
      \label{tensor_quintet_table_c}
    \end{table}
    is a splitting between the neutral and the charged ones ($t^c =
    t^\pm, t^{\pm\pm}$): 
    \begin{subequations}
      \begin{align}
        M_{t^c} &=\; v \left( \frac{270 \pi f_{t^c}}{\alpha_4 + 4 \alpha_5}
        \right)^{\frac14} \\
        M_{t^0} &=\; v \left( \frac{270 \pi f_{t^0}}{5 \alpha_4 + 2
        \alpha_5} \right)^{\frac14} 
      \end{align}
    \end{subequations}
    The mass reach in this case is shown in
    Fig.~\ref{fig:tensor_quintet_special} as well as Table
    \ref{tensor_quintet_table_c}. 

\vspace{\baselineskip}\noindent
(ii)
    For the completely general case, all couplings are non-zero, and 
    the constraint equation is
    \begin{equation}
      \label{eq:trelationnonc}
      (2 \alpha_6 - \alpha_7)^2 = (2 \alpha_4 - \alpha_5) \left( \alpha_4 +
      4 \alpha_5 + 2 \alpha_6 + 8 \alpha_7 + 6 \alpha_{10} \right).
    \end{equation}
    The masses for the tensor quintet are (we use the abbreviations
    \begin{figure}
      \begin{center}
        \includegraphics[width=80mm]{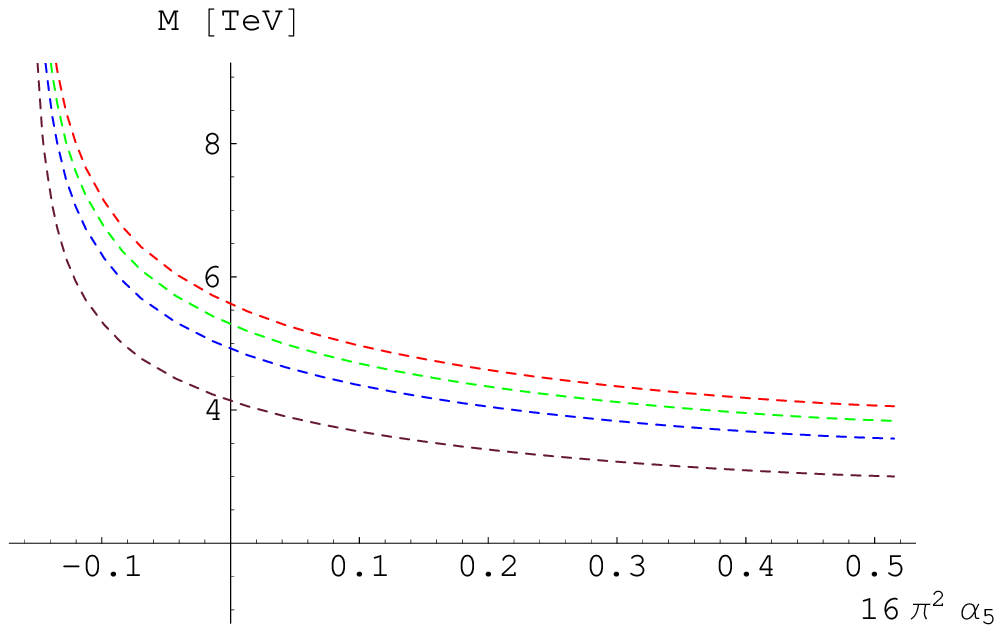}
        \includegraphics[width=80mm]{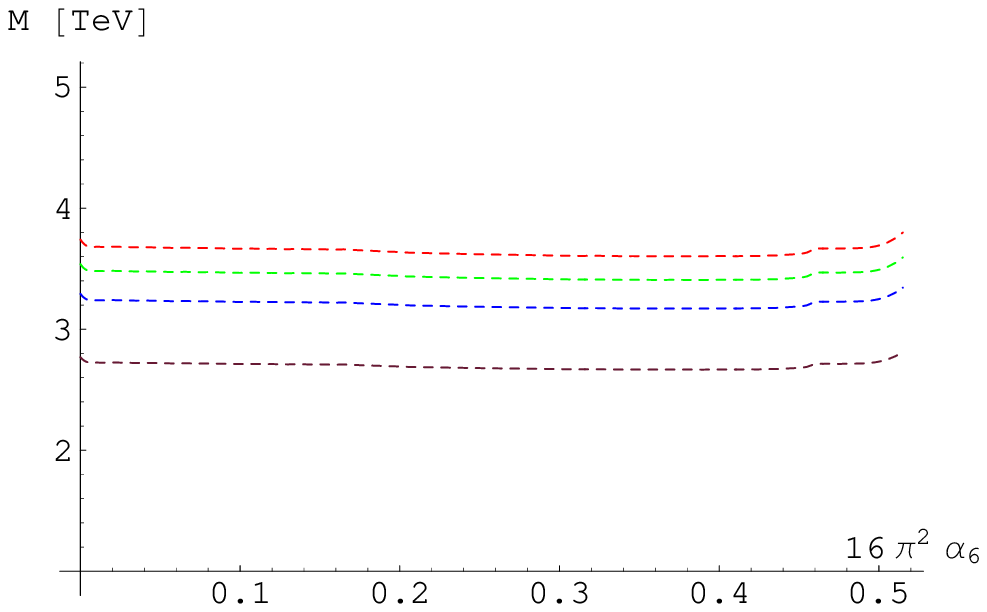}
        \includegraphics[width=80mm]{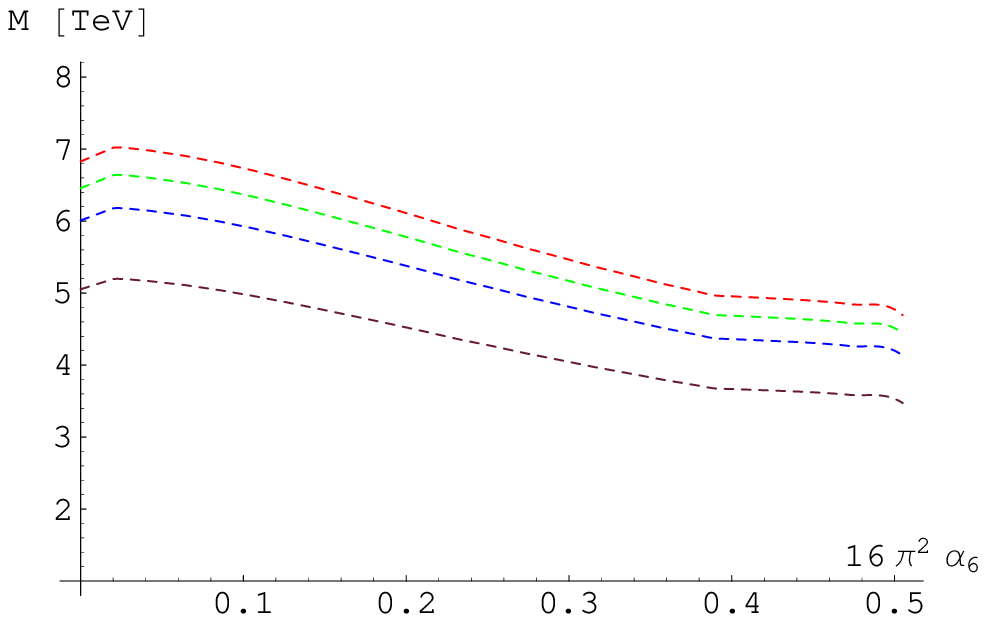}
        \caption{Dependence of the resonance mass for the tensor
          quintet in the full case for different assumed widths (in red 
          $f_t = \Gamma_t / M_t = 1.0$, blue $0.8$,green
          $0.6$, brown $0.3$, respectively). On the upper line, the doubly
          charged case is shown on the left, the charged one on the
          right, while the neutral state is in the lower line.}
        \label{fig:tensor_quintet_full}
      \end{center}
    \end{figure}
    \begin{table}
      \begin{center}
        \begin{tabular}{|c||c|c|c|c|}\hline
          $f= \frac{\Gamma}{M_t}$
          & $1.0$ & $0.8$ & $0.6$ & $0.3$ \\\hline
          $M_{t^{\pm\pm}} \ [\TeV]$
          & $5.17$ & $4.89$ & $4.55$ & $3.83$ \\\hline
          $M_{t^{\pm}} \ [\TeV]$
          & $3.64$ & $3.44$ & $3.20$ & $2.69$ \\\hline
          $M_{t^0} \ [\TeV]$
          & $5.84$ & $5.52$ & $5.14$ & $4.32$ \\\hline
        \end{tabular}
      \end{center}
      \caption{Mass reach for the tensor quintet in the
        full case depending on different
        resonance widths. The values given in the table are averaged
        over the lower limit curve.}
      \label{tensor_quintet_table_full}
      %
    \end{table}
    $\xi_{ij} = \alpha_i + 4 \alpha_j$):
    \begin{subequations}
      \begin{align}
        \label{eq:tmassnonc}
        M_{t^{\pm\pm}} &=\; v \left( \frac{270 \pi
          f_{t^{\pm\pm}}}{\xi_{45}} \right)^{\frac14} \\ 
        M_{t^\pm} &=\; v \left( \frac{270 \pi f_{t^\pm}}{\xi_{45} +
          \xi_{67}} \right)^{\frac14} \\ 
        M_{t^0} &=\; v \left( \frac{810 \pi (2\alpha_4 - \alpha_5)
          f_{t^0}}{\left[ \sqrt{(\alpha_4+4 \alpha_5)(2\alpha_4
          - \alpha_5)} - 2 (2\alpha_4 - \alpha_5) \right]^2 + 2 \left[
          2(\alpha_4 + \alpha_6) - (\alpha_5 + \alpha_7) \right]^2}
          \right)^{\frac14} \notag \\ 
          &=\;  v \left( \frac{810 \pi
          f_{t^0}}{\left[ \sqrt{\xi_{45}} - 2
          \sqrt{2\alpha_4 - \alpha_5} \right]^2 + 2 \left[ 
          \sqrt{2 \alpha_4 - \alpha_5} + \sqrt{\xi_{45} + 2 \xi_{67}
          + 6 \alpha_{10}} \right]^2}
          \right)^{\frac14}
      \end{align}
    \end{subequations}
    For the neutral component, the first formula is better suited for
    the fit, while the limit to the isospin-conserving case is easily
    visible in the second one as well as the limit to the special case
    above with $\xi_{67} \to 0$ for only $g_t$ and $h'_t$ being non-zero.


\section{Summary}

At an ILC with high energy ($1\;\TeV$) and luminosity
($1000\;\fb^{-1}$) and the possibility for both electron and positron
polarization, precise measurements of weak-boson interactions will
be feasible.  In this work we have concentrated on quartic weak-boson
couplings that enter in six-fermion processes.  Including known
results for weak-boson pair production and oblique corrections, we
have determined the possible impact on our knowledge about high-energy
weak-boson scattering amplitudes.  Our numerical results are presented
in terms of the usual set of anomalous couplings in the
chiral-Lagrangian framework.  For each spin-isospin channel, they are
conveniently re-expressed in terms of the maximal resonance mass that,
under the most favorable conditions, the measurement can be sensitive
to.

On the experimental side, the present study completes and supersedes
previous studies of weak-boson scattering and triple-boson production
in $e^+e^-$ collisions.  For weak-boson scattering processes, we have
analysed all accessible channels using an unweighted event generator
with complete six-fermion matrix elements, parton shower and
hadronization, and fast detector simulation.  The analysis uses
standard cut-based experimental techniques.  The parameters are
determined in a global multidimensional fit without implicit or
explicit assumptions of theoretical relations among them.

Triple weak-boson production provides independent information on the
parameters of interest.  While our results indicate that the ultimate
sensitivity is not as good as for the weak-boson scattering processes,
it serves as an important cross check and should be included in a
global fit of ILC data.  More details on this class of processes will
be published elsewhere~\cite{Beyer}.

In Tables~\ref{tab:final-su2}, \ref{tab:final} we combine our results
for the physics sensitivity for all spin/isospin channels.
Table~\ref{tab:final-su2} assumes $SU(2)_c$ conservation, so the
$\Delta\rho$ parameter automatically vanishes.  In this case, only
channels with $I+J$ even couple to weak-boson pairs.
Table~\ref{tab:final} shows the results without this constraint.  In
each case, a single resonance with maximal coupling (i.e., $\Gamma=M$)
was assumed to be present.  In a real situation, the particular
structure of the parameter dependence can be used to disentangle
multiple resonances.

\begin{table}
  \begin{equation*}
    \begin{array}{|c||c|c|c|}
      \hline
      \text{Spin} & I=0 & I=1 & I=2
      \\
      \hline\hline
      0 & 1.55 & - & 1.95
      \\
      1 & - & 2.49 & -
      \\
      2 & 3.29 & - & 4.30
      \\
      \hline
    \end{array}
  \end{equation*}
\caption{Accessible scale $\Lambda$ in $\TeV$ for all possible
  spin/isospin channels.  The results are derived from the analysis of
  vector-boson scattering processes at the ILC, assuming a single
  resonance with optimal properties.  Custodial $SU(2)$
  symmetry is assumed to hold.}
\label{tab:final-su2}
\end{table}

\begin{table}
  \begin{equation*}
    \begin{array}{|c||c|c|c|}
      \hline
      \text{Spin} & I=0 & I=1 & I=2
      \\
      \hline\hline
      0 & 1.39 & 1.55 & 1.95
      \\
      1 & 1.74 & 2.67 & -
      \\
      2 & 3.00 & 3.01 & 5.84
      \\
      \hline
    \end{array}
  \end{equation*}
\caption{Accesible scale $\Lambda$ in $\TeV$ for all possible
  spin/isospin channels.  The results are derived from the analysis of
  vector-boson scattering processes at the ILC, assuming a single
  resonance with optimal properties.  No constraints beyond the SM
  symmetries are assumed.}
\label{tab:final}
\end{table}

Some important properties of the relation of resonances to anomalous
couplings are worth mentioning.  First of all, we have to distinguish
resonances that (in our operator basis) couple to fermions from those
that do not.  If sizable fermion couplings are present, some anomalous
couplings scale with $1/M^2$, where $M$ is the resonance mass.
Obviously, these include four-fermion contact terms, which are thus
potentially sensitive to new-physics up to rather high scales.  The
other class of operators with $1/M^2$ scaling are mixed fermion-boson
contact terms that contribute, e.g., to vector boson pair production.
In any fixed operator basis, these operators are not related to the
triple-gauge interactions that are usually considered.  However, in
studies that deal with specific models (e.g., minimal technicolor),
they are implicitly present.  This accounts for the good physics reach
of the ILC as it has been discovered in studies of weak-boson pair
production.

In this work, we have determined the amount of information that can
possibly be gained on top of the analysis of fermionic couplings, or
otherwise if such couplings are small or absent.  In that case, the
only operator with a physical $1/M^2$ scaling corresponds to the
$\rho$ parameter, associated to custodial-$SU(2)$ violation.  Apart
from that, all $1/M^2$ effects in bosonic interactions can be absorbed
into unobservable redefinitions of the SM parameters.  Therefore, the
shifts due to heavy resonances in oblique corrections, triple-gauge
couplings, and quartic gauge couplings, all scale with $1/M^4$.  In
particular, all corrections to triple-gauge couplings
($g,\kappa,\lambda$) scale in the same way, although the operators
have formally different dimension.

Taking these considerations into account, we find limits for the
sensitivity of the ILC in the $1$ to $3~\TeV$ range, where the best
reach corresponds to the highest-spin channel.  These limits are not
as striking as possible limits from contact interactions, but agree
well with the expected direct-search limits for resonances at the LHC.
Performing global fits of all electroweak parameters, analogous to LEP
analyses, and combining data from both colliders will be important for
disentangling the contributions.  Significant knowledge about the
mechanism of electroweak symmetry breaking can thus be gained even in
`worst-case' scenarios that do not lead to striking new-physics
signatures at all.

\subsection*{Acknowledgments}

We would like to thank Peter Zerwas for useful comments and
discussions. W.K. and J.R. are supported by the
Helmholtz-Gemeinschaft, Contract No.\ VH--NG--005.

\newpage
\appendix
\section{Chiral Parameters and Anomalous Couplings}

In this section, we list the formulas that relate the operators of the
chiral Lagrangian (see Sec.~\ref{sec:chiral}) to the anomalous
couplings of vector bosons in the physical basis of $A_\mu$, $Z_\mu$,
$W^\pm_\mu$.  While standard parameterizations exist for the oblique
corrections and for the triple gauge couplings (TGC), this is not
the case for quartic anomalous couplings.

\subsection{Oblique corrections}
\label{app:oblique}

New physics that does not couple to light fermions can be parameterized
in terms of $S,T,U$.  The relations are
\begin{align}
  \Delta S &= -16\pi\alpha_1
&
  \Delta T &= 2\beta_1/{\alpha_{\rm QED}}
&
  \Delta U &= -16\pi\alpha_8
\end{align}


The oblique corrections are needed for the proper renormalization of
the SM vertices.  First, we have to specify our definition of the weak
mixing angle.  It is customary to adopt the $G_F/\alpha/M_Z$ scheme.
In this scheme, the weak mixing angle is defined by
\begin{equation}
  \sw\cw = \frac{e}{2M_Z}(\sqrt2 G_F)^{-1/2}.
\end{equation}
Furthermore, the oblique corrections renormalize the wave functions of
the vector bosons and thus affect the definition of the gauge
couplings $g$ and $g'$ in terms of $e$ and $\sw,\cw$.

A simple recipe of including the oblique corrections to the trilinear
and quartic gauge couplings is the following: (i) Expand the SM
Lagrangian in terms of physical fields
according to
\begin{equation}
  gW^3 =  eA + e\frac{\cw}{\sw}(1 + \delta_Z)Z,
\qquad\qquad
  gW^\pm  = \frac{e}{\sw}
  \left(1 + \cw^2\delta_Z - \frac{g^2}{2}\alpha_8\right)W^\pm 
\end{equation}
where
\begin{equation}
 \delta_Z =
  \frac{\beta_1+g^\pp\alpha_1}{\cw^2-\sw^2},
\end{equation}
and
(ii) switch to the $G_F/\alpha/M_Z$ scheme by the replacements
\begin{align}
  \sw &\to \sw\left(1 - \frac{\cw^2}{\cw^2-\sw^2}\beta_1
                   - \frac{e^2}{2\sw^2(\cw^2-\sw^2)}\alpha_1\right)
\\
  \cw &\to \cw\left(1 + \frac{\sw^2}{\cw^2-\sw^2}\beta_1
                   + \frac{e^2}{2\cw^2(\cw^2-\sw^2)}\alpha_1\right)
\end{align}

\subsection{Triple gauge couplings}
\label{app:triple}

We define a generic $C$ and $CP$-even triple-gauge vertex in the
standard way
\begin{align}
  \LL_{TGC} &=  \ii e\left[
    g_1^\gamma A_\mu \left(W^-_\nu W^{+\mu\nu} - W^+_\nu W^{-\mu\nu}\right)
  + \kappa^\gamma W^-_\mu W^+_\nu A^{\mu\nu}
  + \frac{\lambda^\gamma}{M_W^2}W^-_\mu{}^\nu W^+_{\nu\rho} A^{\rho\mu}
  \right]
\nonumber\\ &\quad + \ii e\frac{\cw}{\sw}\left[
    g_1^Z Z_\mu \left(W^-_\nu W^{+\mu\nu} - W^+_\nu W^{-\mu\nu}\right)
  + \kappa^Z W^-_\mu W^+_\nu Z^{\mu\nu}
  + \frac{\lambda^Z}{M_W^2}W^-_\mu{}^\nu W^+_{\nu\rho} Z^{\rho\mu}
\right]
\end{align}
The SM values are
\begin{equation}
  g_1^{\gamma,Z}=\kappa^{\gamma,Z}=1
  \quad\text{and}\quad
  \lambda^{\gamma,Z}=0
\end{equation}
The triple gauge couplings are expressed in terms of the $\alpha$
parameters as
\begin{align}
  \Delta g_1^\gamma &= 0
&
  \Delta\kappa^\gamma &= g^2(\alpha_2-\alpha_1) + g^2\alpha_3 
  + g^2(\alpha_9-\alpha_8)
\\
  \Delta g_1^Z &= \delta_Z + \frac{g^2}{\cw^2}\alpha_3
&
  \Delta\kappa^Z &= \delta_Z - g^\pp(\alpha_2-\alpha_1) + g^2\alpha_3 
  + g^2(\alpha_9-\alpha_8)
\end{align}
and
\begin{align}
  \lambda^\gamma &= -\frac{g^2}{2}
  \left(\alpha^\lambda_1 + \alpha^\lambda_2\right)
&
  \lambda^Z &= -\frac{g^2}{2}
  \left(\alpha^\lambda_1 -\frac{\sw^2}{\cw^2} \alpha^\lambda_2\right) 
\end{align}
where $\delta_Z$ is the oblique correction defined above.

This can be inverted to yield
 \begin{align}
   \alpha_2-\alpha_1 &= 
   \frac{\cw^2}{g^2}(\Delta\kappa^\gamma - \Delta\kappa^Z + \delta_Z)
 \\
   \alpha_3 &= \frac{\cw^2}{g^2}\left(\Delta g_1^Z - \delta_Z\right)
 \\
   \alpha_9-\alpha_8 &= \frac{\sw^2}{g^2}\Delta\kappa^\gamma
               + \frac{\cw^2}{g^2}(\Delta\kappa^Z - \Delta g_1^Z)
 \\
   \alpha^\lambda_1 &= 
   -\frac{2}{g^2}\left(\sw^2\lambda^\gamma + \cw^2\lambda^Z\right)
 \\
   \alpha^\lambda_2 &= -\frac{2}{g^2}\,\cw^2
   \left(\lambda^\gamma - \lambda^Z\right)
 \end{align}

\subsection{Quartic gauge couplings}
\label{app:quartic}

We define the quartic gauge couplings analogous to the TGC:
\begin{align}
  \LL_{QGC} &=
  e^2\left[ g_1^{\gamma\gamma} A^\mu A^\nu W^-_\mu W^+_\nu
           -g_2^{\gamma\gamma} A^\mu A_\mu W^{-\nu} W^+_\nu\right]
\nonumber\\ &\quad
  + e^2\frac{\cw}{\sw}\left[ g_1^{\gamma Z} A^\mu Z^\nu
                               \left(W^-_\mu W^+_\nu + W^+_\mu W^-_\nu\right)
              -2g_2^{\gamma Z} A^\mu Z_\mu W^{-\nu} W^+_\nu \right]
\nonumber\\ &\quad
  + e^2\frac{\cw^2}{\sw^2}\left[ g_1^{ZZ} Z^\mu Z^\nu W^-_\mu W^+_\nu
                  -g_2^{ZZ} Z^\mu Z_\mu W^{-\nu} W^+_\nu\right]
\nonumber\\ &\quad
  + \frac{e^2}{2\sw^2}\left[ g_1^{WW} W^{-\mu} W^{+\nu} W^-_\mu W^+_\nu
                       -g_2^{WW}\left(W^{-\mu} W^+_\mu\right)^2\right]
  + \frac{e^2}{4\sw^2\cw^4} h^{ZZ} (Z^\mu Z_\mu)^2
\end{align}
The SM values are
\begin{equation}
  g_1^{VV'}=g_2^{VV'}=1
  \quad\text{($VV'=\gamma\gamma,\gamma Z, ZZ, WW$),}\qquad\qquad
  h^{ZZ} = 0.
\end{equation}
In terms of the $\alpha$ parameters, the deviations from the SM values
are
\begin{align}
  \Delta g_1^{\gamma\gamma} &= \Delta g_2^{\gamma\gamma} = 0
                         &&= \Delta g_1^\gamma
\\
  \Delta g_1^{\gamma Z} &= \Delta g_2^{\gamma Z}
                         = \delta_Z + \frac{g^2}{\cw^2}\alpha_3
                         &&= \Delta g_1^Z
\\
  \Delta g_1^{ZZ} &= 2\Delta g_1^{\gamma Z}
                     + \frac{g^2}{\cw^4}(\alpha_4 + \alpha_6)
\\
  \Delta g_2^{ZZ} &= 2\Delta g_1^{\gamma Z}
                     - \frac{g^2}{\cw^4}(\alpha_5 + \alpha_7)
\\
  \Delta g_1^{WW} &= 2\cw^2\Delta g_1^{\gamma Z} + 2g^2(\alpha_9-\alpha_8)
                     + g^2\alpha_4
\\
  \Delta g_2^{WW} &= 2\cw^2\Delta g_1^{\gamma Z} + 2g^2(\alpha_9-\alpha_8)
                     - g^2\left(\alpha_4 + 2\alpha_5\right)
\vphantom{\frac{g^2}{\cw^2}}
\\
  h^{ZZ} &= g^2\left[\alpha_4 + \alpha_5 
                     + 2\left(\alpha_6+\alpha_7 + \alpha_{10}\right)\right]
\end{align}
There are also $\lambda$-type couplings which contain two field
strength tensors of different charge,
\begin{align}
  \LL_{QGC}^\lambda &=
  \sum_{V,V'=\gamma,Z}g_V g_{V'}\frac{\lambda^{VV'}}{M_W^2}
  V^{\mu\nu}\left[(V'_\nu W^-_\rho - V'_\rho W^-_\nu)W^{+\rho}{}_\mu
                + (V'_\nu W^+_\rho - V'_\rho W^+_\nu)W^{-\rho}{}_\mu\right] 
\nonumber\\ &\quad
  + g^2\frac{\lambda^{WW}}{M_W^2}
  (W^-_\mu W^+_\nu - W^+_\mu W^-_\nu)W^{-\nu}{}_\rho W^{+\rho\mu}
\end{align}
with
\begin{equation}
  g_\gamma = e, \qquad
  g_Z = e\cw/\sw,
\end{equation}
as well as couplings which contain two field strength tensors of equal
charge that we do not need.  Similarly, we do not consider quartic
couplings with four field strength tensors.  The SM values of
$\lambda^{VV'}$ are zero.  The quartic $\lambda$ couplings are related
to the $\alpha^\lambda$ parameters by
\begin{align}
  \lambda^{\gamma\gamma} &= 
  -\frac{g^2}{2}\left(\alpha^\lambda_1 + \alpha^\lambda_2\right)
  &&= \lambda^\gamma
\\
  \lambda^{Z\gamma} &=
  -\frac{g^2}{2}\left(\alpha^\lambda_1 - \frac{\sw^2}{\cw^2}\alpha^\lambda_2\right)
  &&= \lambda^Z
\\
  \lambda^{\gamma Z} &=
  \lambda^{\gamma\gamma} 
  -\frac{g^2}{2\cw^2}\left(\alpha^\lambda_3 + \frac12\alpha^\lambda_4\right)
\\
  \lambda^{ZZ} &=
  \lambda^{Z \gamma}
  -\frac{g^2}{2\cw^2}\left(\alpha^\lambda_3
                           - \frac{\sw^2}{2\cw^2}\alpha^\lambda_4\right)
\\
  \lambda^{WW} &=
  -\frac{g^2}{2}\left(\alpha^\lambda_1  + \alpha^\lambda_3
        + \alpha^\lambda_5\right)
\end{align}
Note that $\lambda^{\gamma\gamma}$ and $\lambda^{Z\gamma}$ are
determined by the trilinear couplings, while the other three are
independent.  The reason is that all couplings that involve the photon
field in terms of the potential $A^\mu$ directly (not via the field
strength $A^{\mu\nu}$) are connected by gauge invariance.  The same
holds for the $g^{\gamma\gamma}$ and $g^{\gamma Z}$ couplings, see
above.





\section{Chiral Lagrangian building blocks}

We define the vector field
\begin{equation}
  \vV = \Sigma(\vD\Sigma)^\dagger = -(\vD\Sigma)\Sigma^\dagger
\end{equation}
and the projection field
\begin{equation}
  \vT = \Sigma\tau^3\Sigma^\dagger
\end{equation}
Both are in the adjoint representation of $SU(2)_L$, and both are
linear combinations of Pauli matrices (this is not obvious for $\vV$),
so
\begin{align}
  \tr\vV &= 0
&
  \tr\vT &= 0
\end{align}
Their covariant derivatives are
\begin{align}
  \vD_\mu\vV_\nu &= \pd_\mu\vV_\nu + ig[\vW_\mu,\vV_\nu]
&
  \vD_\mu\vT &= [\vT,\vV_\mu]
\end{align}
Note that $\vV$ is antihermitian while $\vT$ is hermitian
\begin{align}
  \vV^\dagger &= -\vV
&
  \vT^\dagger &= \vT
\end{align}

\subsection{Unitary gauge}

In unitary gauge, these fields reduce to
\begin{align}
  \vV &\Rightarrow -ig\vW + ig'\left(B\frac{\tau^3}{2}\right)
  = -\frac{\ii g}{2}\left[\sqrt2(W^+\tau^+ + W^-\tau^-)
                           + \frac1\cw Z \tau^3\right]
\\
  \vT &\Rightarrow \tau^3
\end{align}
and we get
\begin{align}
  \tr{\vT\vV}
  &= -\frac{\ii g}{\cw} Z
\intertext{and thus}
  \tr{\vV_\mu\vV_\nu}
  &= -\frac{g^2}{2}\left( W^+_\mu W^-_\nu + W^-_\mu W^+_\nu
                          + \frac{1}{\cw^2} Z_\mu Z_\nu \right)
\\
  \tr{\vT\vV_\mu}\tr{\vT\vV_\nu}
  &= -\frac{g^2}{\cw^2}\left(Z_\mu Z_\nu \right)
\end{align}
Furthermore, we expand the field strengths in the charge eigenbasis to
obtain
\begin{align}
  \vW_{\mu\nu} &=
  \frac1{\sqrt2}\left[W^+_{\mu\nu}
    +\ii e(A_\mu W^+_\nu - A_\nu W^+_\mu)
    +\ii g\cw(Z_\mu W^+_\nu - Z_\nu W^+_\mu)\right] \tau^+
\nonumber\\ &\quad
  + \frac1{\sqrt2}\left[W^-_{\mu\nu}
    -\ii e(A_\mu W^-_\nu - A_\nu W^-_\mu)
    -\ii g\cw(Z_\mu W^-_\nu - Z_\nu W^-_\mu)\right] \tau^-
\nonumber\\ &\quad
  + \frac12\left[\sw A_{\mu\nu} + \cw Z_{\mu\nu} 
    +\ii g(W^+_\mu W^-_\nu - W^-_\mu W^+_\nu)\right] \tau^3
\\[.5\baselineskip]
  \vB_{\mu\nu} &= \frac12\left[\cw A_{\mu\nu} - \sw Z_{\mu\nu}\right]\tau^3
\\[.5\baselineskip]
  [\vV_\mu,\vV_\nu] &=  -g^2\left[
  \frac1{\cw\sqrt2}\left(Z_\mu W^+_\nu - W^+_\mu Z_\nu\right)\tau^+
  - \frac1{\cw\sqrt2}\left(Z_\mu W^-_\nu - W^-_\mu Z_\nu\right)\tau^-
  \right.
\nonumber\\ &\quad\qquad
  \left.\vphantom{\frac1{\sqrt2}}
  + \frac12\left(W^+_\mu W^-_\nu - W^-_\mu W^+_\nu\right) \tau^3
  \right]
\\[.5\baselineskip]
  \tr{\vT\vW_{\mu\nu}} &=
  \sw A_{\mu\nu} + \cw Z_{\mu\nu} 
    +\ii g\left(W^+_\mu W^-_\nu - W^-_\mu W^+_\nu\right)
\\[.5\baselineskip]
  \tr{\vT[\vV_\mu,\vV_\nu]} &=
  -g^2\left(W^+_\mu W^-_\nu - W^-_\mu W^+_\nu\right)
\end{align}

\subsection{Gaugeless limit}

Conversely, in the gaugeless limit the expansions in terms of
Goldstone fields are
\begin{align}
  \vV &\Rightarrow \frac{\ii}{v}
  \left(\pd w^k + \frac1v \epsilon^{ijk} w^i\pd w^j\right)\tau^k + O(v^{-3})
\\
  \vT &\Rightarrow \tau^3 
  + 2\sqrt2\frac{\ii}{v}\left(w^+\tau^+ - w^-\tau^-\right) + O(v^{-2})
\end{align}
Expressing both in terms of charge eigenstates, we derive the expansions
\begin{align}
  \vV &= \frac{\ii}{v}\Bigg\{\:
    \sqrt2\left[\pd w^+ + \frac{\ii}{v}\left(w^+\pd z - z\pd w^+\right)\right]
    \tau^+
\nonumber\\ &\quad\quad
   +\sqrt2\left[\pd w^- - \frac{\ii}{v}\left(w^-\pd z - z\pd w^-\right)\right]
    \tau^-
\nonumber\\ &\quad\quad
   + \left[\pd z - \frac{\ii}{v}\left(w^+\pd w^- - w^-\pd w^+\right)\right]
    \tau^3
  \Bigg\} + O(v^{-3})
\\
  \tr{\vT\vV} &= \frac{2\ii}{v}
  \left[\pd z + \frac{\ii}{v}\left(w^+\pd w^- - w^-\pd w^+\right)\right]
  + O(v^{-3})
\intertext{and thus}
  \tr{\vV_\mu\vV_\nu}
  &= -\frac{2}{v^2}\left(\pd_\mu w^+ \pd_\nu w^- + \pd_\mu w^- \pd_\nu w^+
                         + \pd_\mu z \pd_\nu z\right)
     + O(v^{-3})
\\
  \tr{\vT\vV_\mu}\tr{\vT\vV_\nu} &= 
  -\frac{4}{v^2}\left(\pd_\mu z \pd_\nu z\right)
     + O(v^{-3})
\end{align}

\subsection{Useful relations}

The following relations can be derived using the definitions and
relations above:
\begin{align}
  \tr{[\vV_\mu,\vV_\nu]^2}
  &= 2\left(\tr{\vV_\mu\vV_\nu}\right)^2
    -2\left(\tr{\vV_\mu\vV^\mu}\right)^2
\\[.5\baselineskip]
  \left(\tr{\vT[\vV_\mu,\vV_\nu]}\right)^2
  &= 4\left(\tr{\vV_\mu\vV_\nu}\right)^2
    -4\left(\tr{\vV_\mu\vV^\mu}\right)^2
\nonumber \\ &\quad
    -4\tr{\vV_\mu\vV_\nu}\tr{\vT\vV^\mu}\tr{\vT\vV^\nu}
\nonumber \\ &\quad
    +4\tr{\vV_\mu\vV^\mu}\tr{\vT\vV_\nu}\tr{\vT\vV^\nu}
\\[.5\baselineskip]
  \tr{[\vV_\mu,\vV_\nu][\vT,\vV^\mu]}
  &= -2\tr{\vV_\mu\vV_\nu}\tr{\vT\vV^\mu}
     +2\tr{\vV_\mu\vV^\mu}\tr{\vT\vV_\nu}
\\[.5\baselineskip]
  \tr{[\vT,\vV_\mu][\vT,\vV_\nu]}
  &= -4\tr{\vV_\mu\vV_\nu} + 2\tr{\vT\vV_\mu}\tr{\vT\vV_\nu}
\\[.5\baselineskip]
  \tr{\vW_{\mu\nu}[\vT,\vV^\mu]}\tr{\vT\vV^\nu}
  &= -\tr{\vW_{\mu\nu}[\vV^\mu,\vV^\nu]}
     +\tfrac12\tr{\vT\vW_{\mu\nu}}\tr{\vT[\vV^\mu,\vV^\nu]}
\end{align}
Field strength tensors:
\begin{align}
  \vD_\mu\vV_\nu - \vD_\nu\vV_\mu
  &=
  -[\vV_\mu,\vV_\nu] - \ii g\vW_{\mu\nu} + \ii g'\vB_{\mu\nu}
\\[.5\baselineskip]
  \pd_\mu\tr{\vT\vV_\nu} - \pd_\nu\tr{\vT\vV_\mu}
  &=
  \tr{\vT[\vV_\mu,\vV_\nu]} - \ii g\tr{\vT\vW_{\mu\nu}} + \ii g'B_{\mu\nu}
\\[.5\baselineskip]
  \vD_\mu(\vT\tr{\vT\vV_\nu}) - \vD_\nu(\vT\tr{\vT\vV_\mu})
  &=
  [\vT,\vV_\mu]\tr{\vT\vV_\nu} - [\vT,\vV_\nu]\tr{\vT\vV_\mu}
\nonumber\\ &\quad
  + \vT\tr{\vT[\vV_\mu,\vV_\nu]} - \ii g\vT\tr{\vT\vW_{\mu\nu}}
  + 2\ii g'\vB_{\mu\nu}
\end{align}
This is easy to see in unitary gauge:
\begin{align}
  [\vT,\vV_\mu]\tr{\vT\vV_\nu} - [\vT,\vV_\nu]\tr{\vT\vV_\mu}
  &= -2[\vV_\mu,\vV_\nu] + \vT\tr{\vT[\vV_\mu,\vV_\nu]}
\end{align}

\clearpage

\end{document}